\documentclass[a4paper,fleqn]{cas-sc}

\usepackage[numbers,sort&compress]{natbib}

\def\tsc#1{\csdef{#1}{\textsc{\lowercase{#1}}\xspace}}
\tsc{WGM}
\tsc{QE}

\usepackage{xcolor}

\graphicspath{{figures/}}

\newcommand{\Rom}{{}_{(R)}}                 

\begin{document}
\let\WriteBookmarks\relax
\def\floatpagepagefraction{1}
\def\textpagefraction{.001}

\renewcommand{\topfraction}{0.9}
\renewcommand{\bottomfraction}{0.9}
\renewcommand{\textfraction}{0.05}
\renewcommand{\floatpagefraction}{0.5}
\setcounter{topnumber}{4}
\setcounter{bottomnumber}{3}
\setcounter{totalnumber}{6}

\makeatletter
\RenewDocumentEnvironment{figure}{O{htbp}}
  {\@float{figure}[#1]}
  {\end@float}
\RenewDocumentEnvironment{figure*}{O{htbp}}
  {\@dblfloat{figure}[#1]}
  {\end@dblfloat}
\makeatother

\shorttitle{Large post-critical dynamics of an inextensible spinning fluid-conveying pinned--roller pipe}
\shortauthors{A. Fasihi et~al.}

\title[mode=title]{Large post-critical dynamics of an inextensible spinning fluid-conveying pipe with pinned--roller supports: high-order Galerkin and a modified Hencky bar-chain framework}

\author[1]{Ali Fasihi}[type=editor,
    orcid=0000-0001-7128-5208]
\ead{a.fasihi@rug.nl, ali.fasihi@dokt.p.lodz.pl}
\credit{Conceptualization, Investigation, Methodology, Validation, Software, Formal Analysis, Visualization, Writing -- Original draft, Writing -- Review \& Editing}

\author[2]{Grzegorz Kudra}[type=editor,
    orcid=0000-0003-0209-4664]
\cormark[1]
\ead{grzegorz.kudra@p.lodz.pl}
\credit{Conceptualization, Methodology, Software, Formal Analysis, Writing -- Review \& Editing, Supervision}

\author[1]{Maryam GhandchiTehrani}[type=editor,
    orcid=0000-0002-0824-4937]
\ead{m.ghandchitehrani@rug.nl}
\credit{Supervision, Investigation, Formal Analysis}

\author[2]{Jan Awrejcewicz}[type=editor,
    orcid=0000-0003-0387-921X]
\ead{jan.awrejcewicz@p.lodz.pl}
\credit{Supervision, Investigation, Formal Analysis}

\cortext[1]{Corresponding author}

\affiliation[1]{organization={Dynamics and Vibration Group, Faculty of Science and Engineering, University of Groningen},
    addressline={Nijenborgh 4},
    city={Groningen},
    postcode={9747 AG},
    state={Groningen},
    country={The Netherlands}}

\affiliation[2]{organization={Department of Automation, Biomechanics, and Mechatronics, Lodz University of Technology},
    addressline={1/15 Stefanowski},
    city={Lodz},
    postcode={90-537},
    state={Lodzkie},
    country={Poland}}

\begin{abstract}
This paper investigates the stability and large post-critical dynamics of an inextensible spinning fluid-conveying pipe with pinned--roller supports. Replacing the pinned--pinned support of the extensible counterpart with a sliding support removes the axial-stretching restoring mechanism and fundamentally changes the governing equations of motion. Derived here for this configuration, these equations contain a different set of nonlinear terms -- arising from the inextensibility constraint and the bending curvatures rather than the single axial-stretching term -- that drives a post-critical regime with large deflections. The regime is analysed with two complementary methods. The first is a Galerkin discretisation in which the bending curvatures are Taylor-expanded to ninth order, shown to be the lowest order resolving the post-critical amplitude; the standard cubic truncation overestimates the deflection significantly by missing the geometric stiffening from inextensibility. The second is a modified Hencky bar-chain model with a global angular description: a closed, $n$-independent matrix framework with exact trigonometric kinematics, directly implementable in any standard programming environment with matrix routines and adaptable to both extensible and inextensible configurations through a single boundary-condition reduction. The linearised dynamics give an ellipse-like stability boundary in the flow-velocity--rotational-speed plane with semi-axes $U=\pi$ and $\Omega=\pi^{2}$; three damping regimes are identified, including a high-rotation instability driven by rotating damping. Close agreement between the two methods across linear-stability, bifurcation, and time-history comparisons confirms the ninth-order Galerkin truncation and establishes the modified Hencky bar-chain as a reliable general-purpose discrete framework for spinning fluid-conveying pipes.
\end{abstract}




\maketitle

\section{Introduction}\label{sec:intro}
%

The dynamics of fluid-conveying pipes has been an important topic in the dynamics and stability of slender structures for several decades, because of its wide range of engineering applications and the complex dynamical behaviour these systems show. The book of P\"aidoussis~\cite{paidoussis_fluid-structure_2013} gives a broad overview of the large amount of work done on the subject, including the two main instability mechanisms -- divergence (buckling) and flutter -- caused by the Coriolis and centrifugal forces of the internal flow. Rotordynamics is a closely related field in which rotation itself produces Coriolis and centrifugal effects; an introduction is given in the book of Ishida and Yamamoto~\cite{ishida_vibrations_2012}.

The combination of these two fields -- a tubular structure that both rotates about its longitudinal axis and conveys an internal flow -- defines a more complex class of systems whose dynamics is governed by the coupled action of flow and spin. Interest in this class is recent and has been driven mainly by drill-string applications, where the interaction between the rotating drill string and the drilling fluid is one of the main sources of severe vibration and fatigue failure~\cite{ghasemloonia_review_2015}. Understanding these dynamics has direct importance for the oil and gas industry, where premature component failure due to vibration accounts for a significant part of operating cost~\cite{wu_buckling_2024,tian_dynamic_2024,ni_numerical_2025}. Beyond drill strings, similar configurations arise in rotating heat exchangers in chemical processing, centrifugal separators in the food and pharmaceutical industries, and rotating biological contactors in wastewater treatment; in all such systems the coupled dynamics of spinning and internal flow govern the onset of structural instabilities.

Early modelling work on drill-string--mud interaction was carried out by Zhang and Miska~\cite{zhang_effects_2005}, who showed that buckling and lateral vibrations of vertical drill pipes interacting with incompressible fluid have a critical flow rate at which divergence or flutter sets in, with the threshold depending on geometry, boundary conditions, axial force, and material properties. P\"aidoussis and co-workers~\cite{paidoussis_dynamics_2008} extended this by studying the effect of both internal and external mud flow, finding that the response is sensitive to the annular space between the string and the wellbore. Eftekhari and Hosseini~\cite{eftekhari_stability_2016} examined cantilevered spinning pipes of functionally graded material under fluid and thermomechanical loading. A series of studies by Liang and co-workers~\cite{liang_dynamical_2018,liang_transverse_2018,liang_vibrations_2019,liang_dynamical_2020} addressed the stability and dynamics of spinning fluid-conveying pipes with various support configurations and later examined wave-propagation and vibration-suppression mechanisms in pipes with periodic structures~\cite{liang_low-frequency_2022,liang_vibration_2022}. More recent contributions include closed-form modal analyses of multi-span spinning functionally graded pipes and the nonplanar vibration and flutter of vertically spinning cantilevered piezoelectric pipes with electromechanical coupling and gravity~\cite{fan_stability_2024,ebrahimi_nonplanar_2022}.

The analyses cited above are based on linear models that focus on eigenvalue (or frequency) calculations to determine natural frequencies and stability thresholds. Such models are limited to the stable parameter range: beyond the critical onset of buckling or flutter, linear equations of motion derived under the assumption of small strain lose validity, and their predicted amplitudes grow without bound. To capture the post-instability behaviour -- in particular the bounded post-critical equilibrium amplitudes seen in practice -- nonlinear modelling is needed, with at least third-order geometric nonlinearity. For slender pipes, the correct nonlinear formulation depends on whether the structure is treated as extensible or inextensible~\cite{semler_non-linear_1994}; nonlinear models of both types exist for various configurations of stationary fluid-conveying pipes~\cite{jiang_three-dimensional_2020}, but until recently no nonlinear study had been reported for the spinning case.

A further difference appears in the spinning case through the role of structural damping. In conservative (pinned--pinned, clamped--clamped) fluid-conveying pipes, damping does not affect the stability boundary~\cite{paidoussis_dynamic_1974}. In rotating systems, however, damping plays a more important role: it is necessary to distinguish non-rotating damping, associated with the stationary frame, from rotating damping, associated with the spinning structure itself. The former is always stabilising; the latter can be either stabilising or destabilising, depending on the operating range~\cite{dimentberg_flexural_1961}. The role of viscoelastic effects in drilling configurations has been studied recently by Galasso et al.~\cite{galasso_simulation_2025}.

A companion paper by the present authors~\cite{fasihi_nonlinear_2026} addressed both of these gaps -- the need for a nonlinear post-critical analysis of a spinning fluid-conveying pipe and the role of rotating damping -- for the \emph{extensible} pinned--pinned configuration. There, the nonlinear equations of motion are derived in PDE form, discretised by the Galerkin method, and used to map the linear-stability boundary across the flow-velocity--rotational-speed plane (with its parametric dependence on structural damping, mass ratio, and flow-profile modification) and to study the post-critical equilibrium. The present paper extends that work to the second main support type: the \emph{inextensible} pinned--roller pipe, whose downstream end is free to translate axially. With the axial-tension restoring mechanism of the extensible case removed by the sliding support, the post-critical dynamics differ fundamentally from the pinned--pinned case and call for a dedicated analysis.

The contributions of this paper are twofold. First, an inextensible Rayleigh-beam model with Kelvin--Voigt viscoelastic damping is derived for the pinned--roller spinning pipe and discretised by the Galerkin method, with the bending curvatures Taylor-expanded to ninth order so as to resolve the large post-critical deflections of the inextensible regime -- a regime in which the standard cubic truncation gives a significant overestimate by missing the geometric stiffening from inextensibility. The post-critical response is studied by direct numerical integration; the results include the linear stability map, the contributions of the individual nonlinear terms, the inertial- and rotating-frame view of the post-critical response, the global view of the post-critical surface over the $(U,\Omega)$ plane, and a comparison with the extensible pinned--pinned counterpart. Second, a modified Hencky bar-chain model is developed as a general-purpose discrete framework for spinning fluid-conveying pipes. The model uses a global angular description in which each link's orientation is expressed directly in the rotating reference frame; this yields a closed, $n$-independent matrix formulation with exact trigonometric kinematics that handles both extensible and inextensible configurations through a single boundary-condition reduction and is directly implementable in any standard programming environment with matrix routines. The two methods agree closely across linear-stability, bifurcation, and time-history comparisons, mutually confirming the ninth-order Galerkin truncation and the modified Hencky framework as reliable tools for the large-deflection regime of inextensible spinning fluid-conveying pipes.

The paper is organised as follows. Section~\ref{sec:continuous-model} derives the inextensible continuous model with pinned--roller supports. Section~\ref{sec:galerkin} presents the Galerkin discretisation. Section~\ref{sec:results} reports the linear stability and post-critical results. Section~\ref{sec:hencky} presents the modified Hencky bar-chain model and its validation.

\section{Inextensible model with pinned--roller supports}\label{sec:continuous-model}

\subsection{System configuration}\label{subsec:system}

\begin{figure}[htbp]
\centering
\includegraphics[width=0.85\textwidth]{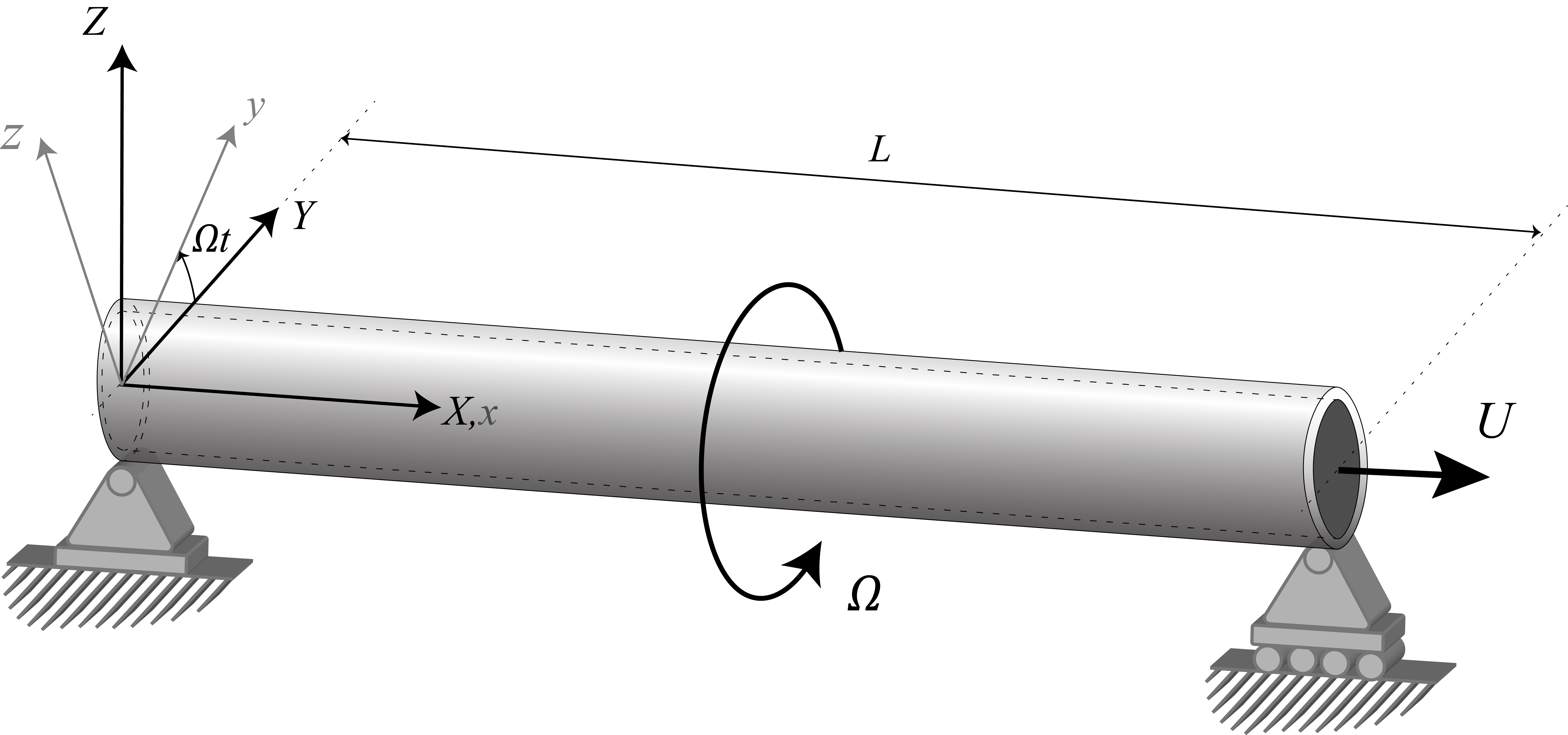}
\caption{Spinning fluid-conveying pipe with pinned--roller supports.}
\label{fig:schematic}
\end{figure}

The system under investigation consists of a slender, hollow, elastic tube of length~$L$ rotating about its longitudinal axis with angular velocity~$\Omega$ while simultaneously conveying an internal incompressible fluid at axial velocity~$U$, with pinned--roller supports as illustrated in Fig.~\ref{fig:schematic}. The pin at the upstream end permits free rotation only; the roller at the downstream end additionally permits free axial displacement. The pipe has density~$\rho$, Young's modulus~$E$, cross-sectional area~$A$, second moment of area~$I$, and Kelvin--Voigt viscoelastic damping coefficient~$\eta$; the fluid has density~$\rho_f$.

Two coordinate systems are shown in Fig.~\ref{fig:schematic}: the fixed inertial frame $(X,Y,Z)$ with unit vectors $(\mathbf{i}_x,\mathbf{i}_y,\mathbf{i}_z)$, in which Hamilton's principle is later applied, and a rotating frame $(x,y,z)$ that co-rotates with the pipe about the $X$-axis at the spin rate~$\Omega$, related to the inertial frame by the rotation angle~$\Omega t$; the rotating frame is invoked in Section~\ref{subsec:rotating-frame}, where the equations of motion are recast in co-rotating coordinates. The kinematic description of the deformed pipe is formulated in a body-attached frame following the deformed centreline, with the rotation from the inertial frame described by three successive Euler angles $(\psi,\theta,\phi)$; the explicit rotation matrices and an illustration of the deformed cross-section are given in~\cite{fasihi_nonlinear_2026}. The position of any cross-section is parametrised by the curvilinear coordinate $s$ along the pipe centreline together with the displacements $u(s,t)$, $v(s,t)$, $w(s,t)$ in the inertial $X$, $Y$, $Z$ directions, respectively. The Euler angles are related to the displacement gradients through
\begin{equation}
\sin\psi=\frac{v'}{\sqrt{(1+u')^{2}+v'^{2}}},\qquad
\sin\theta=-\frac{w'}{1+e},
\label{eq:euler-angles}
\end{equation}
where the prime denotes $\partial/\partial s$ and $e$ is the axial strain
\begin{equation}
e=\sqrt{(1+u')^{2}+v'^{2}+w'^{2}}-1.
\label{eq:axial-strain}
\end{equation}

\noindent In the same reference, the angular velocity components $\omega_{1}$, $\omega_{2}$, $\omega_{3}$ of the body-attached frame and the torsional curvature $\kappa_{1}$ together with the two bending curvatures $\kappa_{2}$, $\kappa_{3}$ of the deformed centreline are derived from these kinematic relations, and the resulting kinetic energy of the system, accounting for the pipe, the conveyed fluid and the spinning motion, takes the form

\begin{equation}
\begin{aligned}
T &= \frac{1}{2}\int_{0}^{L} m\bigl(\dot{u}^{2}+\dot{v}^{2}+\dot{w}^{2}\bigr)\,\mathrm{d}s
+ \frac{1}{2}\int_{0}^{L}\!M\Bigl[\bigl(\dot{u}+U\cos\theta\cos\psi\bigr)^{2}\\
&\quad +\bigl(\dot{v}+U\cos\theta\sin\psi\bigr)^{2}+\bigl(\dot{w}-U\sin\theta\bigr)^{2}\Bigr]\,\mathrm{d}s
+ \frac{1}{2}\int_{0}^{L}\!I_{x}\,\omega_{1}^{2}\,\mathrm{d}s
+ \frac{1}{2}\int_{0}^{L}\!I\bigl(\omega_{2}^{2}+\omega_{3}^{2}\bigr)\,\mathrm{d}s,
\end{aligned}
\label{eq:T-final}
\end{equation}
where $m$ and $M$ are the mass per unit length of the pipe and of the fluid, $I_{x}$ is the cross-section polar moment of inertia per unit length, and $I=I_{p}+I_{f}$ is the bending moment of inertia per unit length (pipe plus fluid). The elastic potential energy reads
\begin{equation}
V = \frac{1}{2}\int_{0}^{L}\bigl[EA\,e^{2} + D_{x}\,\kappa_{1}^{2} + D\bigl(\kappa_{2}^{2}+\kappa_{3}^{2}\bigr)\bigr]\,\mathrm{d}s,
\label{eq:V-final}
\end{equation}
where $D_{x}$ and $D$ are the torsional and bending rigidities. With a Kelvin--Voigt viscoelastic constitutive law for the pipe material, the variation of the nonconservative work is
\begin{equation}
\delta W_{nc} = \int_{0}^{L}\bigl(\eta\,\dot{e}\,\delta e + C_{x}\,\dot{\kappa}_{1}\,\delta\kappa_{1} + C\,\dot{\kappa}_{2}\,\delta\kappa_{2} + C\,\dot{\kappa}_{3}\,\delta\kappa_{3}\bigr)\,\mathrm{d}s,
\label{eq:Wnc-final}
\end{equation}
where $\eta$ is the material viscosity coefficient and $C_{x}$, $C$ are the corresponding torsional and bending viscous coefficients.

\subsection{Inextensibility constraint}\label{subsec:inext-constraint}

The inextensibility condition (Fig.~\ref{fig:schematic}) implies that lateral motion does not induce elongation, thereby enforcing zero axial strain ($e=0$). Under this constraint, the Euler-angle relations Eqs.~\eqref{eq:euler-angles} simplify to
\begin{align}
\sin\psi &= \frac{v'}{\sqrt{(1+u')^{2}+v'^{2}}},\notag\\
\cos\psi &= \frac{1+u'}{\sqrt{(1+u')^{2}+v'^{2}}},\notag\\
\sin\theta &= -w',\notag\\
\cos\theta &= \sqrt{(1+u')^{2}+v'^{2}}.
\label{eq:euler-angles-inext}
\end{align}
Setting the axial strain to zero in Eqs.~\eqref{eq:axial-strain}, the constraint equation becomes
\begin{equation}
\sqrt{(1+u')^{2}+v'^{2}+w'^{2}}=1,
\label{eq:inextensibility}
\end{equation}
and the axial displacement is expressed in terms of the transverse derivatives as
\begin{equation}
u'=\sqrt{1-v'^{2}-w'^{2}}-1.
\label{eq:u-in-terms-of-vw}
\end{equation}

\subsection{Equations of motion}\label{subsec:eom}

The equations of motion are derived from the extended Hamilton principle,
\begin{equation}
\int_{t_1}^{t_2} \delta(T - V + W_{nc})\,\mathrm{d}t = 0.
\label{eq:hamilton}
\end{equation}
Substituting the Taylor expansions of the kinematic quantities and retaining terms up to third order in the displacement gradients (the consequences of this truncation are addressed in Section~\ref{subsec:taylor-truncation}), each energy component is expressed in terms of the displacement variables and their derivatives. Applying Hamilton's principle and collecting the coefficients of the virtual displacements $\delta\phi$, $\delta v$, and $\delta w$ yields three coupled nonlinear partial differential equations in $\phi$, $v$, and $w$; their full forms are reported in~\ref{app:inext-details}, Eqs.~\eqref{eq:eom-phi-full}--\eqref{eq:eom-w-full}.

Due to the pipe's circular cross-section and long span, the torsional rigidity surpasses the flexural rigidity. Consequently, the natural frequency of twisting motion is much higher than that of bending motion. Bending motion is therefore incapable of exciting torsional inertia: the twist is induced mainly by bending deflections rather than by its own inertia~\cite{silva_non-linear_1988}. Setting all time-derivative terms in Eqs.~\eqref{eq:eom-phi-full} to zero gives~\cite{hosseini_free_2009}
\begin{equation}
\phi(s,t) = -\!\int_{0}^{s} v''\,w'\,\mathrm{d}s + \cdots
\label{eq:phi-quasistatic}
\end{equation}

The pipe is slender and the rotational speeds considered here are not large, so the rotary-inertia and gyroscopic groups multiplied by $I$, $I_{x}$, $D_{x}$, $C_{x}$ are dropped from the transverse-motion analysis. Substituting Eqs.~\eqref{eq:phi-quasistatic} into the two transverse equations yields, after this reduction,
\begin{align}
&(m+M)\Bigg[\ddot{v} + v''\!\int_{0}^{s}\!\!\int_{L}^{s}\!\bigl(\dot{v}'^{2}+v'\ddot{v}'+\dot{w}'^{2}+w'\ddot{w}'\bigr)\mathrm{d}s\,\mathrm{d}s \notag\\
&\qquad{}+ v'\!\int_{0}^{s}\!\bigl(\dot{v}'^{2}+v'\ddot{v}'+\dot{w}'^{2}+w'\ddot{w}'\bigr)\mathrm{d}s\Bigg] \notag\\
&{}+ M \Bigg[2U\!\Bigl(\dot{v}'+v'^{2}\dot{v}'+v'w'\dot{w}'+v''\!\int_{L}^{s}\!\bigl(v'\dot{v}'+w'\dot{w}'\bigr)\mathrm{d}s\Bigr) \notag\\
&\qquad{}+ U^{2}\!\Bigl(v''+v'^{2}v''+v'w'w''+v''\!\int_{L}^{s}\!\bigl(v'v''+w'w''\bigr)\mathrm{d}s\Bigr)\Bigg] \notag\\
&{}+ D\Big(
       v''''
       +v''^{3}
       +v''\,w''^{2}
       +v'^{2}\,v''''
       +v''\,w'\,w''' \notag\\
&\qquad{}+ v'\,w'\,w''''
       +4v'\,v''\,v'''
       +3v'\,w''\,w'''\Big) \notag\\
&{}+ C\Big(\Omega\,w''''
       +\dot{v}''''
       +v''''\,v'\,\dot{v}'
       +\tfrac{1}{2}\Omega\,w''''\,w'^{2}
       +\dot{v}'\,w'\,w'''' \notag\\
&\qquad{}+\dot{v}''\,w'\,w'''
       +3\dot{v}''\,v''^{2}
       +3v'''\,v''\,\dot{v}'
       +3\dot{v}'\,w''\,w'''
       +3\dot{v}''\,v'''\,v' \notag\\
&\qquad{}+4\dot{v}'''\,v''\,v'
       +v'''\,\dot{w}'\,w''
       +4v''\,\dot{w}''\,w''
       +2v''\,\dot{w}'\,w'''
       +3v'\,\dot{w}'''\,w'' \notag\\
&\qquad{}+3v'\,\dot{w}''\,w'''
       +v'\,\dot{w}'\,w''''
       +v'''\,w'\,\dot{w}''
       +2v''\,w'\,\dot{w}'''
       +v'\,w'\,\dot{w}'''' \notag\\
&\qquad{}+\tfrac{1}{2}\Omega\,w''''\,v'^{2}
       +\Omega\,w'\,w''\,w'''
       +\Omega\,w'''\,v''\,v'
       +\dot{v}''\,w''^{2}
       +\dot{v}''''\,v'^{2}\Big) = 0,
\label{eq:eom-v-shortened}
\end{align}

\begin{align}
&(m+M)\Bigg[\ddot{w} + w''\!\int_{0}^{s}\!\!\int_{L}^{s}\!\bigl(\dot{v}'^{2}+v'\ddot{v}'+\dot{w}'^{2}+w'\ddot{w}'\bigr)\mathrm{d}s\,\mathrm{d}s \notag\\
&\qquad{}+ w'\!\int_{0}^{s}\!\bigl(\dot{v}'^{2}+v'\ddot{v}'+\dot{w}'^{2}+w'\ddot{w}'\bigr)\mathrm{d}s\Bigg] \notag\\
&{}+ M \Bigg[2U\!\Bigl(\dot{w}'+w'^{2}\dot{w}'+w'v'\dot{v}'+w''\!\int_{L}^{s}\!\bigl(v'\dot{v}'+w'\dot{w}'\bigr)\mathrm{d}s\Bigr) \notag\\
&\qquad{}+ U^{2}\!\Bigl(w''+w'^{2}w''+w'v'v''+w''\!\int_{L}^{s}\!\bigl(v'v''+w'w''\bigr)\mathrm{d}s\Bigr)\Bigg] \notag\\
&{}+ D\Big(
       w''''
       +w''^{3}
       +w''\,v''^{2}
       +w'^{2}\,w''''
       +w''\,v'\,v''' \notag\\
&\qquad{}+ w'\,v'\,v''''
       +4w'\,w''\,w'''
       +3v''\,w'\,v'''\Big) \notag\\
&{}+ C \Big(\dot{w}''''
       -\Omega\,v''''
       -2\Omega\,v'\,w''\,w'''
       +3w'''\,w''\,\dot{w}'
       +3\dot{w}''\,w''^{2} \notag\\
&\qquad{}+\dot{w}''''\,w'^{2}
       +\dot{v}'''\,v'\,w''
       +\dot{v}''\,v''\,w''
       +4\dot{w}'''\,w''\,w'
       +3\dot{w}''\,w'''\,w' \notag\\
&\qquad{}+\dot{v}''''\,v'\,w'
       +2\dot{v}'''\,v''\,w'
       +\dot{v}''\,v'''\,w'
       +w''''\,w'\,\dot{w}'
       -\Omega\,v''^{3} \notag\\
&\qquad{}-\Omega\,v''\,w''^{2}
       -\tfrac{1}{2}\Omega\,v''''\,w'^{2}
       -\tfrac{1}{2}\Omega\,v''''\,v'^{2}
       -\Omega\,v'''\,w''\,w' \notag\\
&\qquad{}-3\Omega\,v'''\,v'\,v''\Big) = 0.
\label{eq:eom-w-shortened}
\end{align}

\subsection{Dimensionless formulation}\label{subsec:dimensionless}

The following dimensionless groups, identical to those of~\cite{fasihi_nonlinear_2026}, are introduced:
\begin{equation}
\begin{aligned}
s^{*}&=\tfrac{s}{L},\quad v^{*}=\tfrac{v}{L},\quad w^{*}=\tfrac{w}{L},\quad
t^{*}=\sqrt{\tfrac{D}{M+m}}\,\tfrac{t}{L^{2}},\\
\alpha&=\sqrt{\tfrac{D}{M+m}}\,\tfrac{\eta}{EL^{2}},\quad
\beta=\tfrac{M}{\gamma(M+m)},\quad
U^{*}=\sqrt{\tfrac{\gamma M}{D}}\,LU,\quad
\Omega^{*}=\sqrt{\tfrac{M+m}{D}}\,L^{2}\,\Omega,
\end{aligned}
\label{eq:dimensionless-groups}
\end{equation}
where $D=EI$, $M=\rho_f A_f$ is the fluid mass per unit length, $m=\rho A$ is the pipe mass per unit length, and $\gamma$ is the flow-profile modification factor of~\cite{guo_modification_2010} that accounts for the radial non-uniformity of the velocity profile in the centrifugal force term ($\gamma=4/3$ for laminar flow, $\gamma\to1$ for high-Reynolds turbulent flow).

In dimensionless form, with the asterisks dropped for compactness, the two transverse equations of motion are
\begin{align}
\ddot{v}+2\beta^{1/2}U\,\dot{v}'+\gamma U^{2}v''+v''''
&+\alpha\bigl(\dot{v}''''+\Omega\,w''''\bigr)+\mathcal{N}_{v}=0,
\label{eq:eom-v-dimless}\\[2pt]
\ddot{w}+2\beta^{1/2}U\,\dot{w}'+\gamma U^{2}w''+w''''
&+\alpha\bigl(\dot{w}''''-\Omega\,v''''\bigr)+\mathcal{N}_{w}=0,
\label{eq:eom-w-dimless}
\end{align}
where $\mathcal{N}_{v}$ and $\mathcal{N}_{w}$ collect the cubic nonlinear contributions. They decompose into five operator groups -- nonlinear mass terms~$\mathrm{NMT}$, nonlinear gyroscopic terms~$\mathrm{NGT}$, nonlinear centripetal terms~$\mathrm{NCT}$, nonlinear stiffness terms~$\mathrm{NST}$, and nonlinear damping terms~$\mathrm{NDT}$ -- as
\begin{align}
\mathcal{N}_{v}&=\mathrm{NMT}_{v}^{(1)}+\mathrm{NMT}_{v}^{(2)}
+2\beta^{1/2}U\bigl(\mathrm{NGT}_{v}^{(1)}+\mathrm{NGT}_{v}^{(2)}\bigr) \notag\\
&\quad+\gamma U^{2}\bigl(\mathrm{NCT}_{v}^{(1)}+\mathrm{NCT}_{v}^{(2)}\bigr)
+\mathrm{NST}_{v}+\alpha\,\mathrm{NDT}_{v},\label{eq:N-v-inext-decomp}\\[2pt]
\mathcal{N}_{w}&=\mathrm{NMT}_{w}^{(1)}+\mathrm{NMT}_{w}^{(2)}
+2\beta^{1/2}U\bigl(\mathrm{NGT}_{w}^{(1)}+\mathrm{NGT}_{w}^{(2)}\bigr) \notag\\
&\quad+\gamma U^{2}\bigl(\mathrm{NCT}_{w}^{(1)}+\mathrm{NCT}_{w}^{(2)}\bigr)
+\mathrm{NST}_{w}+\alpha\,\mathrm{NDT}_{w}.\label{eq:N-w-inext-decomp}
\end{align}
The explicit forms of the individual operators are given in~\ref{app:inext-details}. Each group has a distinct physical origin and a distinct dependence on the system parameters; their individual contributions to the post-critical behaviour are examined in Section~\ref{subsubsec:nonlinear-groups}.

\subsection{Complex form and rotating-frame transformation}\label{subsec:rotating-frame}

The two transverse motions are combined into a single complex variable $\rho=v+\mathrm{i}w$ and transformed to a frame co-rotating with the pipe through
\begin{equation}
\rho(s,t)=r(s,t)\,\mathrm{e}^{\mathrm{i}\Omega t},
\label{eq:rotating-transform}
\end{equation}
where $r=\Rom v + \mathrm{i}\,\Rom w$ collects the rotating-frame transverse displacements. From this point on, we drop the prefix $\Rom{}$ and use $v$ and $w$ for the rotating-frame quantities. When needed, the inertial-frame variables can be recovered through Eqs.~\eqref{eq:rotating-transform}. After substitution of Eqs.~\eqref{eq:rotating-transform} into Eqs.~\eqref{eq:eom-v-dimless}--\eqref{eq:eom-w-dimless} and separation into real and imaginary parts, the rotating-frame equations of motion take the compact form
\begin{align}
\ddot{v}-2\Omega\dot{w}-\Omega^{2}v
&+2\beta^{1/2}U(\dot{v}'-\Omega w')
+\gamma U^{2}v''+v'''' \notag\\
&\qquad+\alpha\dot{v}''''
+\mathcal{N}_{v}=0,
\label{eq:eom-v-rotating}\\[2pt]
\ddot{w}+2\Omega\dot{v}-\Omega^{2}w
&+2\beta^{1/2}U(\dot{w}'+\Omega v')
+\gamma U^{2}w''+w'''' \notag\\
&\qquad+\alpha\dot{w}''''
+\mathcal{N}_{w}=0,
\label{eq:eom-w-rotating}
\end{align}
The rotating-frame transformation introduces three additional linear terms relative to the inertial-frame Eqs.~\eqref{eq:eom-v-dimless}--\eqref{eq:eom-w-dimless}: the Coriolis terms $\mp 2\Omega\dot{w}$, $\pm 2\Omega\dot{v}$; the centrifugal terms $-\Omega^{2}v$, $-\Omega^{2}w$; and the additional flow-induced cross-coupling $\mp 2\beta^{1/2}\Omega U w'$, $\pm 2\beta^{1/2}\Omega U v'$. Conversely, the inertial-frame rotating-damping cross-coupling $\alpha\Omega w''''$ in Eqs.~\eqref{eq:eom-v-dimless} and $-\alpha\Omega v''''$ in Eqs.~\eqref{eq:eom-w-dimless} is removed by the transformation.

\section{Galerkin discretisation}\label{sec:galerkin}

The Galerkin method is now applied to reduce the rotating-frame partial differential equations of motion Eqs.~\eqref{eq:eom-v-rotating}--\eqref{eq:eom-w-rotating} to a finite-dimensional set of ordinary differential equations suitable for frequency analysis and direct numerical integration. The procedure consists of three steps: each transverse displacement field is expanded as a finite linear combination of pre-selected spatial basis functions with time-dependent generalised coordinates as the unknowns; the expansions are substituted into the equations of motion; and the resulting residuals are projected onto the basis functions through weighted integration over the pipe length. Because the equations of motion have already been transformed to the co-rotating frame in Section~\ref{subsec:rotating-frame}, the discretisation is carried out directly in that frame, so that post-critical equilibria emerge as fixed points of the discretised system and the bifurcation analysis of Section~\ref{sec:results} can be performed without an additional change of variables.

Accordingly, the transverse displacement fields in the rotating frame are expanded as
\begin{equation}
v(s,t)=\sum_{j=1}^{n}\phi_{j}(s)\,q_{jv}(t),\qquad
w(s,t)=\sum_{j=1}^{n}\phi_{j}(s)\,q_{jw}(t),
\label{eq:galerkin-expansion}
\end{equation}
where $\phi_{j}(s)$ are the spatial comparison functions, $q_{jv}(t)$ and $q_{jw}(t)$ are the time-dependent generalised coordinates of the discretised system in the rotating-frame $v$- and $w$-directions, and $n$ is the Galerkin truncation order denoting the number of modes retained. The basis functions are taken to be
\begin{equation}
\phi_{j}(s)=\sqrt{2}\,\sin(j\pi s),\qquad j=1,2,\ldots,n,
\label{eq:basis-functions}
\end{equation}
which are the mass-normalised eigenfunctions of the simply supported Euler--Bernoulli beam and satisfy both the geometric and the natural boundary conditions of the pinned--roller configuration, ensuring mode orthogonality. The truncation order $n$ is fixed at four modes throughout this paper, a choice justified by the convergence study reported in~\cite{fasihi_nonlinear_2026} for the same parameter range. Substituting Eqs.~\eqref{eq:galerkin-expansion} into the rotating-frame equations of motion Eqs.~\eqref{eq:eom-v-rotating}--\eqref{eq:eom-w-rotating}, multiplying by~$\phi_{i}(s)$, and integrating over the pipe length yields a $2n$-dimensional set of nonlinear ordinary differential equations,
\begin{equation}
\mathbf{M}\,\ddot{\mathbf{q}}
+\bigl(\mathbf{G}+\mathbf{C}\bigr)\dot{\mathbf{q}}
+\mathbf{K}\,\mathbf{q}
+\mathbf{N}(\mathbf{q},\dot{\mathbf{q}},\ddot{\mathbf{q}})=\mathbf{0},
\label{eq:galerkin-ode}
\end{equation}
in which the generalised-coordinate vector is partitioned as
\begin{equation}
\mathbf{q}=
\begin{bmatrix}\mathbf{q}_{v}\\[2pt]\mathbf{q}_{w}\end{bmatrix},\quad
\mathbf{q}_{v}=\bigl[q_{1v},\ldots,q_{nv}\bigr]^{\mathsf{T}},\quad
\mathbf{q}_{w}=\bigl[q_{1w},\ldots,q_{nw}\bigr]^{\mathsf{T}}.
\end{equation}

The mass~$\mathbf{M}$, gyroscopic~$\mathbf{G}$, damping~$\mathbf{C}$, and stiffness~$\mathbf{K}$ matrices in Eqs.~\eqref{eq:galerkin-ode} are assembled in the block form
\begin{equation}
\mathbf{M}=
\begin{bmatrix}\mathbf{A}&\mathbf{0}\\\mathbf{0}&\mathbf{A}\end{bmatrix},\qquad
\mathbf{G}=
\begin{bmatrix}2\beta^{1/2}U\,\mathbf{B}&-2\Omega\,\mathbf{A}\\[2pt]
2\Omega\,\mathbf{A}&2\beta^{1/2}U\,\mathbf{B}\end{bmatrix},
\label{eq:M-G-matrices}
\end{equation}
\begin{equation}
\mathbf{C}=
\begin{bmatrix}\alpha\,\mathbf{D}&\mathbf{0}\\\mathbf{0}&\alpha\,\mathbf{D}\end{bmatrix},\qquad
\mathbf{K}=
\begin{bmatrix}
\mathbf{D}+\gamma U^{2}\mathbf{C}-\Omega^{2}\mathbf{A} & -2\beta^{1/2}\Omega U\,\mathbf{B}\\[2pt]
2\beta^{1/2}\Omega U\,\mathbf{B} & \mathbf{D}+\gamma U^{2}\mathbf{C}-\Omega^{2}\mathbf{A}
\end{bmatrix},
\label{eq:C-K-matrices}
\end{equation}
where the building-block matrices $\mathbf{A}$, $\mathbf{B}$, $\mathbf{C}$, $\mathbf{D}$ collect the projections of the basis functions and their derivatives,
\begin{equation}
A_{ij}=\int_{0}^{1}\!\phi_{i}\phi_{j}\,\mathrm{d}s,\;
B_{ij}=\int_{0}^{1}\!\phi_{i}\phi_{j}'\,\mathrm{d}s,\;
C_{ij}=\int_{0}^{1}\!\phi_{i}'\phi_{j}'\,\mathrm{d}s,\;
D_{ij}=\int_{0}^{1}\!\phi_{i}\phi_{j}''''\,\mathrm{d}s.
\label{eq:building-blocks}
\end{equation}
The projection of the nonlinear terms in Eqs.~\eqref{eq:N-v-inext-decomp}--\eqref{eq:N-w-inext-decomp} onto $\phi_{i}$ yields the discretised nonlinear vector $\mathbf{N}$:
\begin{equation}
\mathbf{N}=\mathbf{N}_{\mathrm{NMT}}
+2\beta^{1/2}U\,\mathbf{N}_{\mathrm{NGT}}
+\gamma U^{2}\,\mathbf{N}_{\mathrm{NCT}}
+\mathbf{N}_{\mathrm{NST}}
+\alpha\,\mathbf{N}_{\mathrm{NDT}}.
\label{eq:N-decomposition}
\end{equation}
The $i$-th component in the $v$-direction reads
\begin{equation}
\begin{aligned}
\mathrm{NMT}_{v,i}&=\sum_{j,k,l=1}^{n}\Bigl(\ddot{q}_{kv}q_{lv}+\dot{q}_{kv}\dot{q}_{lv}+\ddot{q}_{kw}q_{lw}+\dot{q}_{kw}\dot{q}_{lw}\Bigr)\,q_{jv}\\
&\quad\times\Biggl[\int_{0}^{1}\!\phi_{j}''\left(\int_{0}^{s}\!\!\!\int_{1}^{s_{1}}\!\!\phi_{k}'\phi_{l}'\,\mathrm{d}s_{2}\,\mathrm{d}s_{1}\right)\phi_{i}\,\mathrm{d}s
+\int_{0}^{1}\!\phi_{j}'\left(\int_{0}^{s}\!\!\phi_{k}'\phi_{l}'\,\mathrm{d}s_{1}\right)\phi_{i}\,\mathrm{d}s\Biggr],\\[3pt]
\mathrm{NGT}_{v,i}&=\sum_{j,k,l=1}^{n}q_{jv}\bigl(q_{kv}\dot{q}_{lv}+q_{kw}\dot{q}_{lw}\bigr)
\Biggl[\int_{0}^{1}\!\phi_{j}'\phi_{k}'\phi_{l}'\,\phi_{i}\,\mathrm{d}s
+\int_{0}^{1}\!\phi_{j}''\left(\int_{1}^{s}\!\!\phi_{k}'\phi_{l}'\,\mathrm{d}s_{1}\right)\phi_{i}\,\mathrm{d}s\Biggr],\\[3pt]
\mathrm{NCT}_{v,i}&=\sum_{j,k,l=1}^{n}q_{jv}\bigl(q_{kv}q_{lv}+q_{kw}q_{lw}\bigr)
\Biggl[\int_{0}^{1}\!\phi_{j}'\phi_{k}'\phi_{l}''\,\phi_{i}\,\mathrm{d}s
+\int_{0}^{1}\!\phi_{j}''\left(\int_{1}^{s}\!\!\phi_{k}'\phi_{l}''\,\mathrm{d}s_{1}\right)\phi_{i}\,\mathrm{d}s\Biggr],\\[3pt]
\mathrm{NST}_{v,i}&=\int_{0}^{1}\!\mathrm{NST}_{v}\,\phi_{i}\,\mathrm{d}s,\qquad
\mathrm{NDT}_{v,i}=\int_{0}^{1}\!\mathrm{NDT}_{v}\,\phi_{i}\,\mathrm{d}s,
\end{aligned}
\label{eq:nonlinear-projections}
\end{equation}
with analogous expressions in the $w$-direction obtained by interchanging $v\leftrightarrow w$ in the indices of $\mathrm{NMT}$, $\mathrm{NGT}$, $\mathrm{NCT}$, and $\mathrm{NST}$.


\section{Results and discussion}\label{sec:results}

The discretised model Eqs.~\eqref{eq:galerkin-ode} is integrated numerically using a Runge--Kutta Dormand--Prince scheme with adaptive step size and relative and absolute tolerances of $10^{-10}$.

\begin{figure}[htbp]
\centering
\begin{tabular}{cc}
\includegraphics[width=0.45\textwidth]{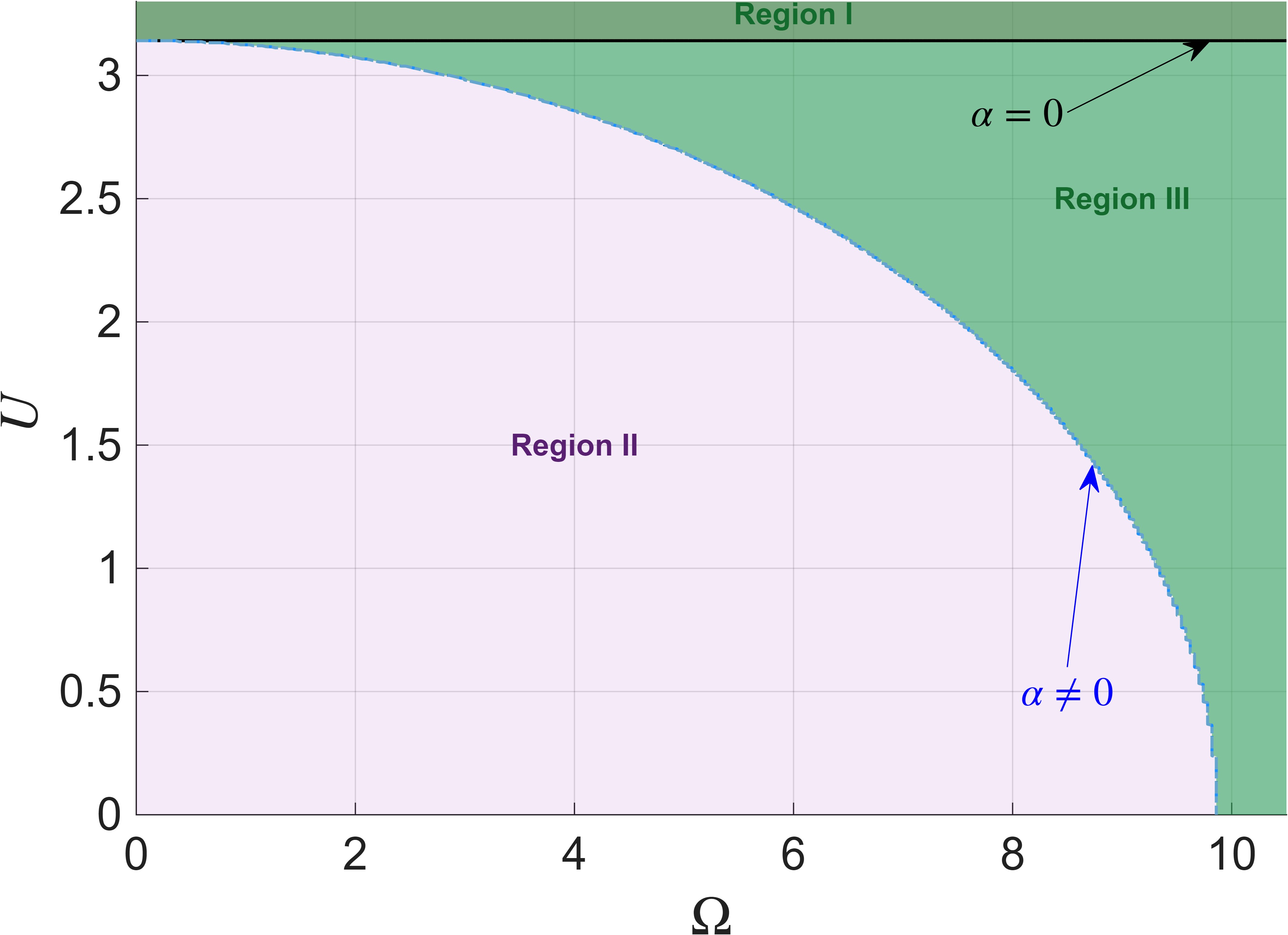} &
\includegraphics[width=0.45\textwidth]{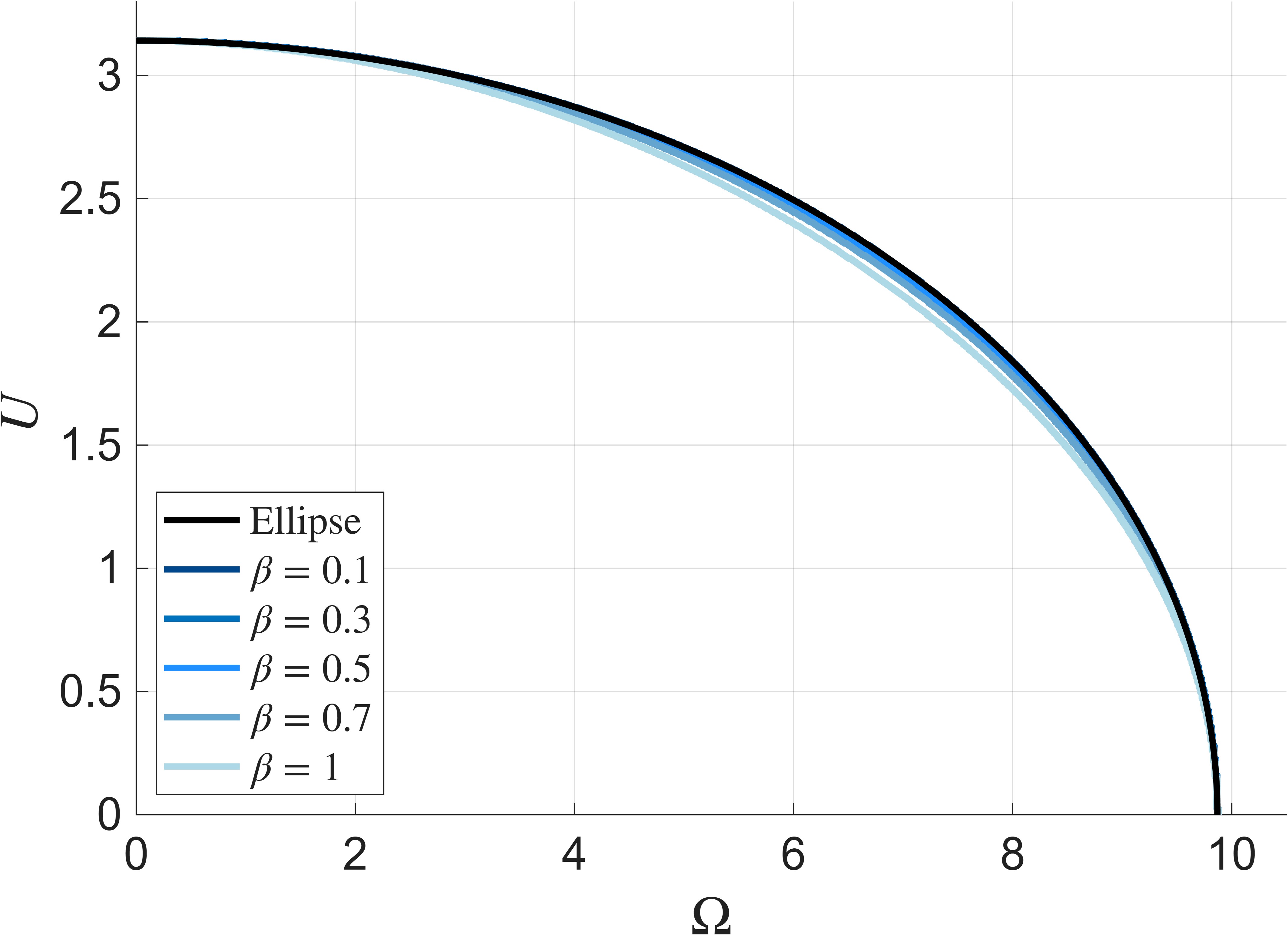} \\
(a) & (b) \\[6pt]
\multicolumn{2}{c}{\includegraphics[width=0.45\textwidth]{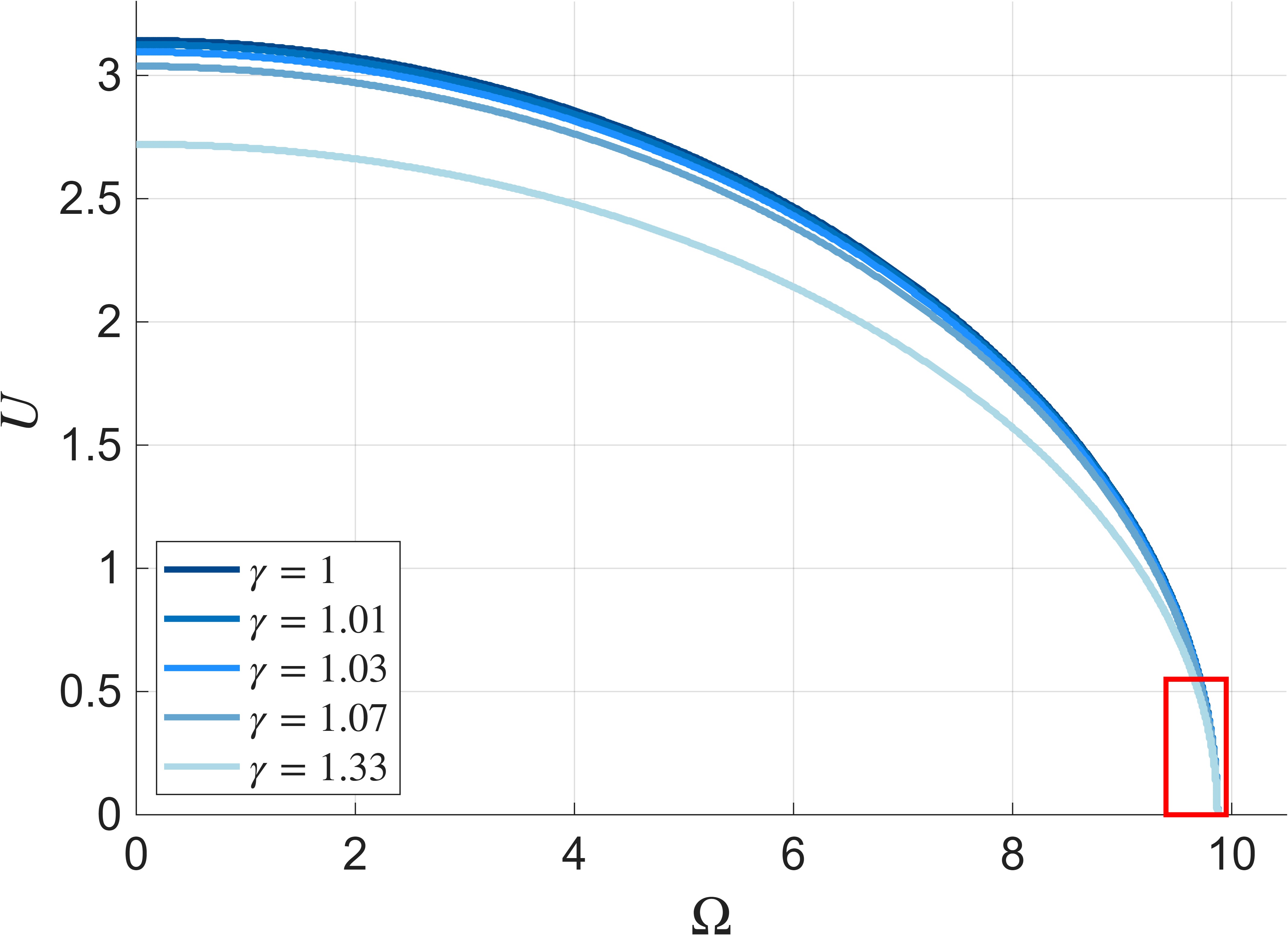}} \\
\multicolumn{2}{c}{(c)}
\end{tabular}
\caption{Linear stability boundary in the $(U,\Omega)$ plane and its parametric dependence, reproduced from~\cite{fasihi_nonlinear_2026}: (a)~three damping regimes -- $\alpha=0$; $\alpha>0$ at low $\Omega$; $\alpha>0$ at high $\Omega$; (b)~dependence on the mass ratio~$\beta$; (c)~dependence on the flow-profile modification factor~$\gamma$.}
\label{fig:linear-stability-parametric}
\end{figure}

\label{subsubsec:linear-stability-map}\label{subsubsec:swirl} The linear part of Eqs.~\eqref{eq:eom-v-rotating}--\eqref{eq:eom-w-rotating} is the same as for the extensible pinned--pinned pipe~\cite{fasihi_nonlinear_2026}. All linear-stability results therefore carry over directly. The stability boundary in the $(U,\Omega)$ plane is ellipse-like, with semi-axes $U=\pi$ and $\Omega=\pi^{2}$, and the three damping regimes ($\alpha=0$; $\alpha>0$ at low $\Omega$; $\alpha>0$ at high $\Omega$) are recovered as shown in Fig.~\ref{fig:linear-stability-parametric}(a). The boundary depends on the mass ratio~$\beta$ (Fig.~\ref{fig:linear-stability-parametric}(b)) and on the flow-profile modification factor~$\gamma$~\cite{guo_modification_2010} (Fig.~\ref{fig:linear-stability-parametric}(c)). A swirl-number argument~\cite{facciolo_study_2007} shows that pipe rotation is too weak to alter the regime within the stability region, so $\gamma$ stays fixed. The frequency analysis further reveals that only the forward-whirling mode loses stability across the unstable region; the backward-whirling mode remains stable throughout~\cite{fasihi_nonlinear_2026}.

Outside the ellipse-like stable region, the trivial equilibrium loses stability and a non-trivial post-critical state emerges. The nonlinear equations of motion are integrated throughout this post-critical region with small initial perturbations of the rotating-frame midpoint deflection. In the extensible pinned--pinned case~\cite{fasihi_nonlinear_2026}, the only source of nonlinearity is the axial stretching of the pipe. The nonlinear vector therefore contains only two terms -- a nonlinear stiffness term and the corresponding Kelvin--Voigt damping term -- both proportional to the dimensionless axial-stiffness coefficient~$\mu$. Here, with the right support free to slide, no such $\mu$-proportional axial stretching is available, and the nonlinear vector instead splits into the five distinct operator groups defined in Eqs.~\eqref{eq:N-v-inext-decomp}--\eqref{eq:N-w-inext-decomp} -- $\mathrm{NMT}$, $\mathrm{NGT}$, $\mathrm{NCT}$, $\mathrm{NST}$, and $\mathrm{NDT}$.\footnote{The cubic damping group $\mathrm{NDT}$ is set to zero throughout the paper. It arises from the Kelvin--Voigt dissipation acting on the bending curvatures and, at the cubic truncation order, is generated by the elimination of the cross-section twist angle~$\varphi$ in favour of the bending fields $v$ and $w$ via Eqs.~\eqref{eq:phi-quasistatic}, which itself follows from the kinematic condition that the torsional curvature $\kappa_{1}$ (Section~\ref{subsec:system}) vanishes in the slender, slow-spin limit. The cubic block produced by this elimination is not a faithful representation of the physical damping nonlinearity at this order: a consistent treatment would require either promoting~$\varphi$ to an independent degree of freedom or carrying the curvature expansion to higher order.} The next subsection isolates the individual contribution of each group by switching one group off at a time.

\subsection{Contributions of the nonlinear groups}\label{subsubsec:nonlinear-groups}

\begin{figure}[htbp]
\centering
\includegraphics[width=0.82\textwidth]{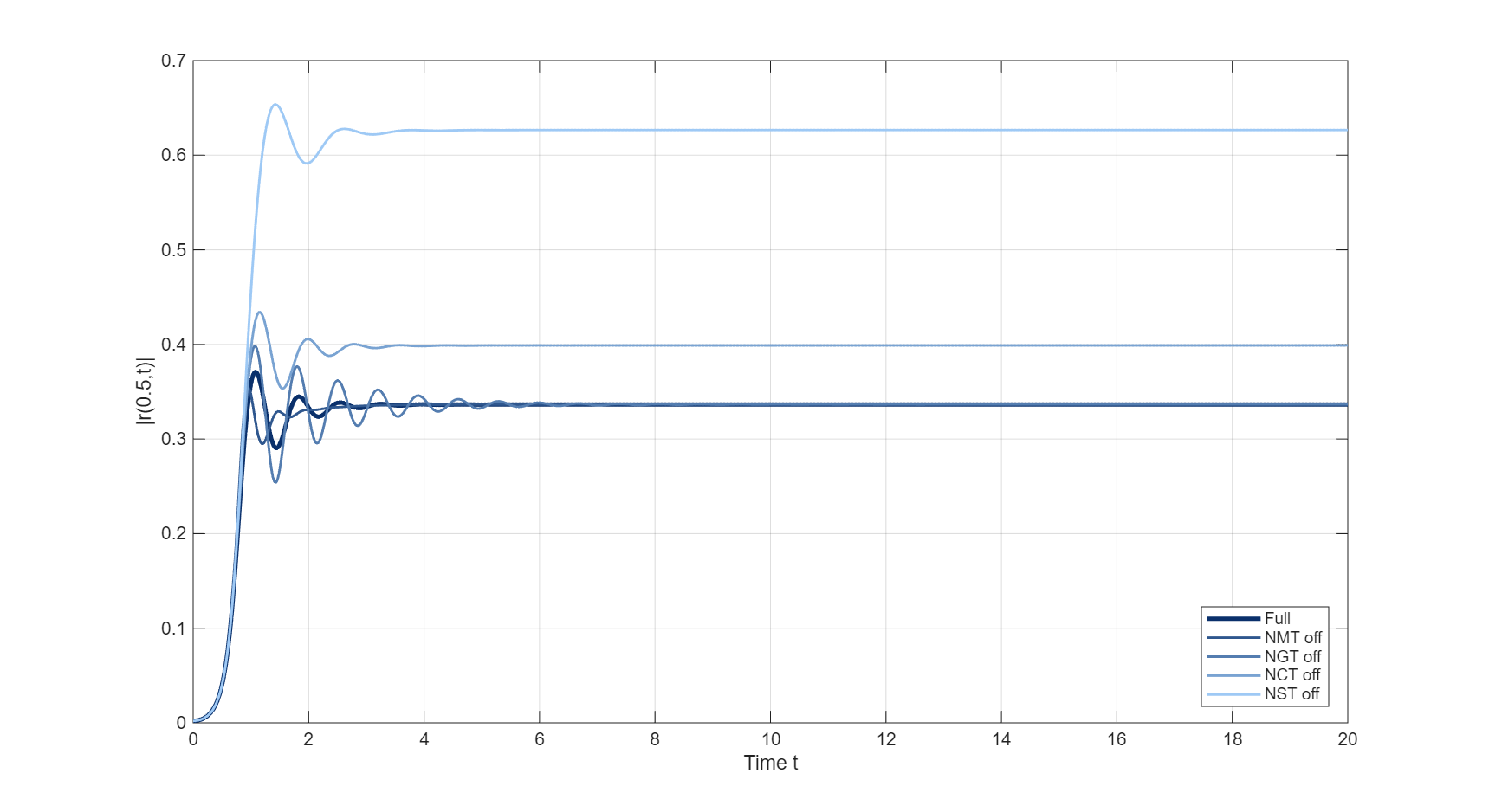}
\caption{Midpoint deflection $|r(0.5,t)|$ at $U=4$, $\Omega=4$ for the full model and for the four cases with one nonlinear group ($\mathrm{NMT}$, $\mathrm{NGT}$, $\mathrm{NCT}$, $\mathrm{NST}$) switched off. Parameters: $\alpha=0.023$, $\sqrt{\beta}=0.536925$.}
\label{fig:group-timehistory}
\end{figure}

At the post-critical equilibrium -- a fixed point in the rotating-frame coordinates, corresponding to steady forward whirling in the inertial frame~\cite{fasihi_nonlinear_2026} -- the groups $\mathrm{NMT}$ and $\mathrm{NGT}$ vanish identically, because every term they contain carries at least one generalised velocity or acceleration. They therefore affect only the transient approach to equilibrium and leave the equilibrium amplitude unchanged. The groups $\mathrm{NCT}$ and $\mathrm{NST}$, in contrast, survive at the equilibrium and set its amplitude through the balance against the linearised stiffness~$\mathbf{K}$. These roles are confirmed numerically by switching one group off at a time at a representative post-critical point, $U=4$, $\Omega=4$. The full model is integrated to its rotating-frame steady state, and the simulation is then repeated with each group switched off in turn. Figure~\ref{fig:group-timehistory} shows the resulting midpoint deflection histories $|r(0.5,t)|$. Removing $\mathrm{NMT}$ or $\mathrm{NGT}$ leaves the steady-state amplitude unchanged, as predicted, but the two groups affect the transient in opposite directions: switching off $\mathrm{NMT}$ shrinks the transient oscillations, while switching off $\mathrm{NGT}$ enlarges them.

\begin{figure}[htbp]
\centering
\includegraphics[width=0.72\textwidth]{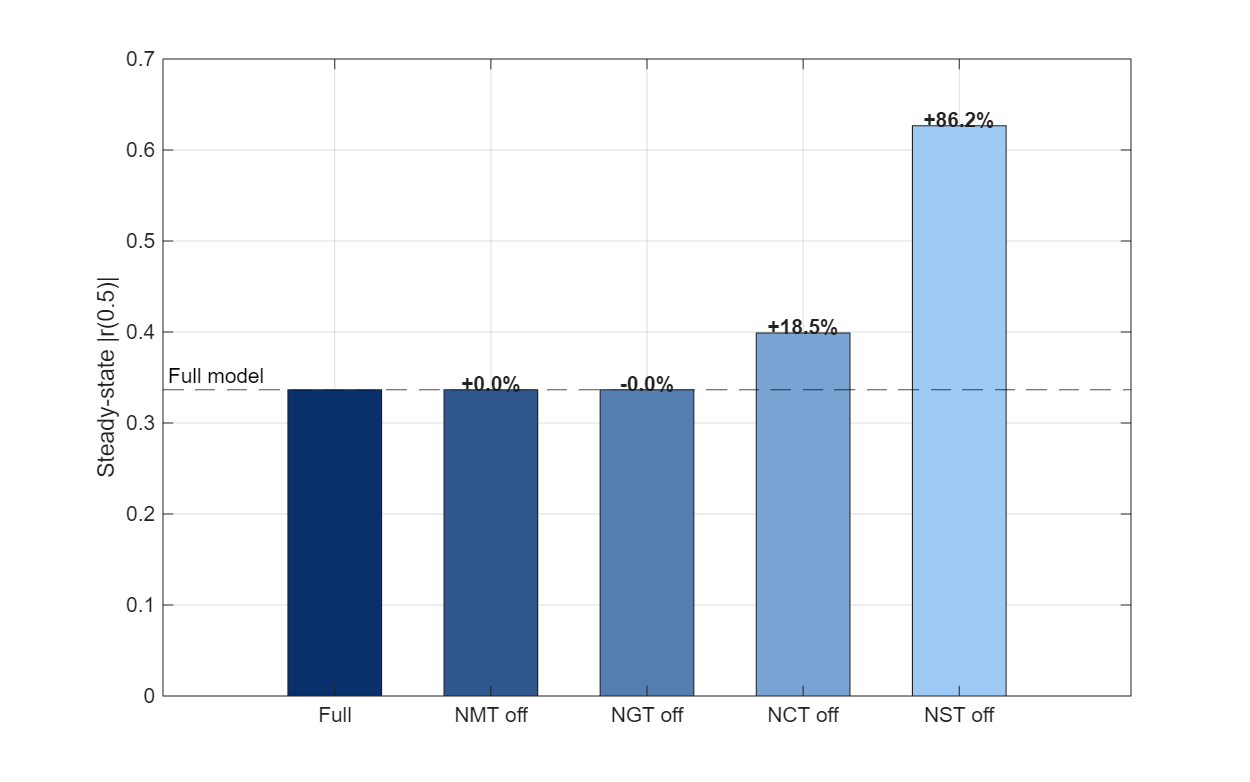}
\caption{Steady-state midpoint amplitude $|r(0.5)|$ for the full model and for each case with one group switched off, with the percentage shift relative to the full model annotated. The velocity-dependent groups $\mathrm{NMT}$, $\mathrm{NGT}$ leave the equilibrium amplitude unchanged; $\mathrm{NCT}$ and $\mathrm{NST}$ shift it, with $\mathrm{NST}$ dominant. Parameters as in Fig.~\ref{fig:group-timehistory}.}
\label{fig:group-bar}
\end{figure}

Figure~\ref{fig:group-bar} reports the resulting steady-state amplitudes with the percentage shift relative to the full model. Removing $\mathrm{NCT}$ raises the equilibrium amplitude moderately ($+18.5\%$), and removing $\mathrm{NST}$ raises it by $+86\%$, confirming that the bending-curvature stiffness $\mathrm{NST}$ is by far the dominant contribution to the post-critical equilibrium.

$\mathrm{NST}$ dominates the post-critical equilibrium and depends on the bending curvatures $\eta_{1}, \eta_{2}$ through the bending part of the elastic potential energy, $\tfrac{1}{2}\int_{0}^{1}(\eta_{1}^{2}+\eta_{2}^{2})\,\mathrm{d}s$. The truncation order of those curvatures therefore determines the equilibrium amplitude.\label{subsec:taylor-truncation} The curvatures themselves are the spinning-angle-independent components of the bending, defined kinematically via the Euler angles $\psi$, $\theta$, $\phi$ of the body-attached frame as
\begin{equation}
\eta_{1}=\psi'\sin\phi\cos\theta+\theta'\cos\phi,\qquad
\eta_{2}=\psi'\cos\phi\cos\theta-\theta'\sin\phi,
\label{eq:eta-euler}
\end{equation}
with the full Euler-angle derivation given in~\cite[Eq.~(A.21)]{fasihi_nonlinear_2026}.
and related to the body-attached bending curvatures $\kappa_{2}, \kappa_{3}$ of Section~\ref{subsec:system} by the planar rotation
\begin{equation}
\kappa_{2}=\eta_{2}\sin(\Omega t)+\eta_{1}\cos(\Omega t),\qquad
\kappa_{3}=\eta_{2}\cos(\Omega t)-\eta_{1}\sin(\Omega t).
\label{eq:eta-kappa}
\end{equation}

\begin{figure}[htbp]
\centering
\includegraphics[width=0.72\textwidth]{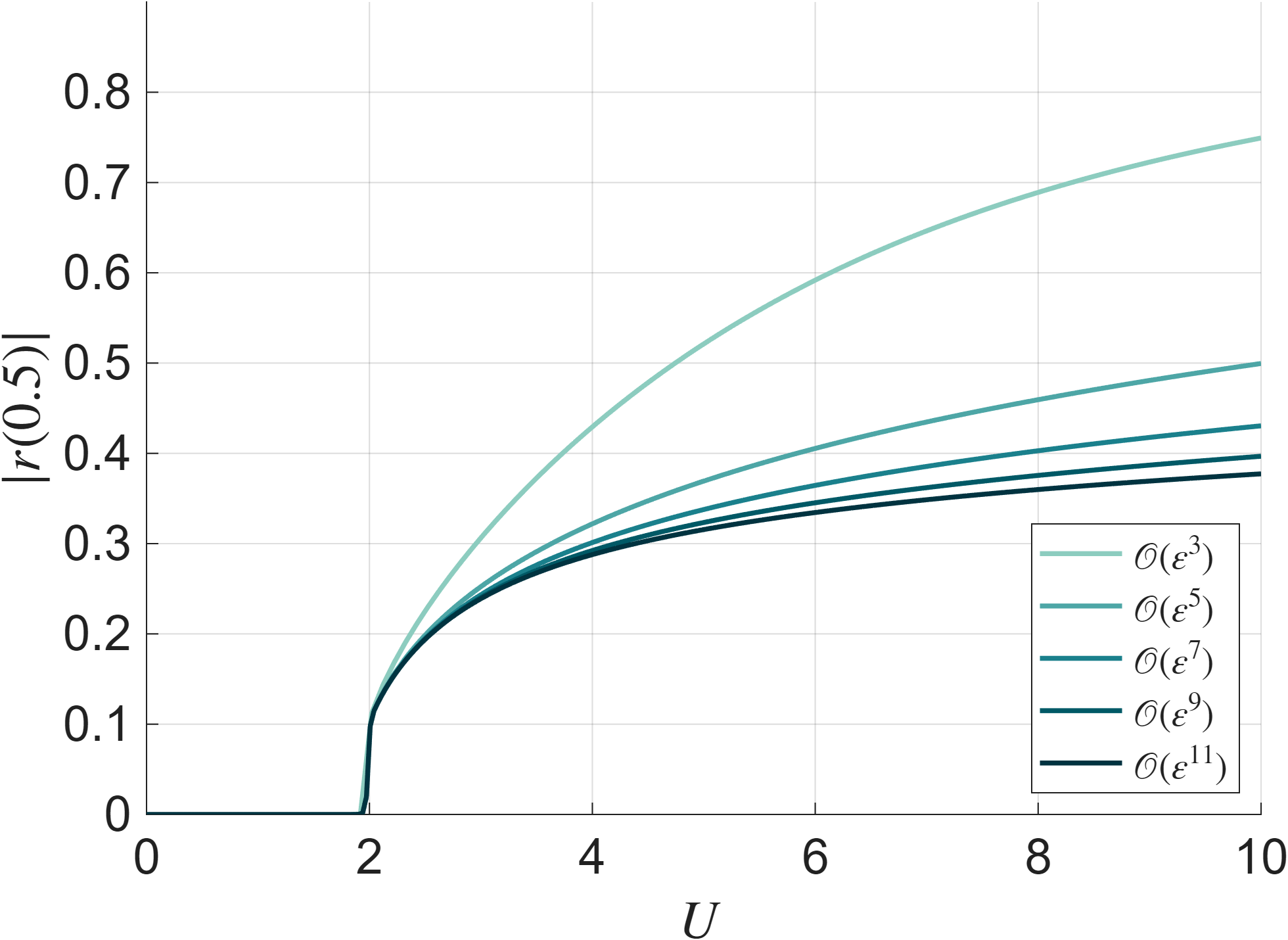}
\caption{Convergence of the steady-state midpoint deflection with the truncation order of the bending curvatures $\eta_{1}$, $\eta_{2}$. Parameters: $\Omega=8$, $\alpha=0.023$, $\sqrt{\beta}=0.536925$.}
\label{fig:truncation-convergence}
\end{figure}

In the equations of motion Eqs.~\eqref{eq:eom-v-rotating}--\eqref{eq:eom-w-rotating}, the bending curvatures were Taylor-expanded only to third order in the displacement gradients,
\begin{equation}
\eta_{1}\approx-w''-\tfrac{1}{2}w'^{2}w'',\qquad
\eta_{2}\approx v''+\tfrac{1}{2}v'^{2}v''+v'w'w''.
\label{eq:eta-3rd}
\end{equation}
This truncation captures the type of the bifurcation correctly but may underestimate the equilibrium amplitude if higher-order geometric stiffening is significant. To assess this, the closed-form curvatures are obtained by substituting the inextensibility constraint Eqs.~\eqref{eq:u-in-terms-of-vw} and the quasi-statically determined twist Eqs.~\eqref{eq:phi-quasistatic} into the kinematic definitions Eqs.~\eqref{eq:eta-euler}:
\begin{equation}
\eta_{1}=-\frac{w''}{\sqrt{1-w'^{2}}},\qquad
\eta_{2}=\frac{v''(1-w'^{2})+v'w'w''}{\sqrt{(1-v'^{2}-w'^{2})(1-w'^{2})}}.
\label{eq:eta-exact}
\end{equation}
The factor $(1-v'^{2}-w'^{2})^{-1/2}$ in $\eta_{2}$ grows rapidly as the slopes increase and provides the strong geometric stiffening that limits the post-critical deflection -- the very feature missed by the cubic truncation. The Taylor expansion to ninth order reads
\begin{equation}
\eta_{1}=-w''\!\left(1+\tfrac{1}{2}w'^{2}+\tfrac{3}{8}w'^{4}+\tfrac{5}{16}w'^{6}+\tfrac{35}{128}w'^{8}\right)+\mathcal{O}(\varepsilon^{11}),
\label{eq:eta1-9th}
\end{equation}
\begin{equation}
\begin{aligned}
\eta_{2}&=v''+\tfrac{1}{2}v''(v'^{2}+2w'^{2})+v'w'w''
+\tfrac{3}{8}v''(v'^{4}+4v'^{2}w'^{2}+3w'^{4})\\
&\quad+\tfrac{1}{2}v'w'w''(v'^{2}+2w'^{2})
+\tfrac{5}{16}v''(v'^{6}+6v'^{4}w'^{2}+10v'^{2}w'^{4}+4w'^{6})\\
&\quad+\tfrac{3}{8}v'w'w''(v'^{4}+4v'^{2}w'^{2}+3w'^{4})\\
&\quad+\tfrac{35}{128}v''(v'^{8}+8v'^{6}w'^{2}+20v'^{4}w'^{4}+16v'^{2}w'^{6}+5w'^{8})\\
&\quad+\tfrac{5}{16}v'w'w''(v'^{6}+6v'^{4}w'^{2}+10v'^{2}w'^{4}+4w'^{6})+\mathcal{O}(\varepsilon^{11}).
\end{aligned}
\label{eq:eta2-9th}
\end{equation}
To evaluate how the truncation of $\eta_{1}, \eta_{2}$ affects the response, the steady-state midpoint deflection $|r(0.5)|$ is computed as a function of flow velocity at the high rotational speed $\Omega=8$, for truncation orders $\mathcal{O}(\varepsilon^{3})$, $\mathcal{O}(\varepsilon^{5})$, $\mathcal{O}(\varepsilon^{7})$, $\mathcal{O}(\varepsilon^{9})$, and $\mathcal{O}(\varepsilon^{11})$ (Fig.~\ref{fig:truncation-convergence}).

Three observations follow. First, the critical flow velocity (the onset of divergence) is identical at all truncation orders; the linear stiffness terms that determine the critical speed are unaffected by the nonlinear truncation, in agreement with the linear-analysis arguments of Section~\ref{subsubsec:linear-stability-map}. Second, the third-order truncation overestimates the post-critical deflection significantly, particularly at large flow velocities, because it underestimates the geometric stiffening provided by the factor $(1-v'^{2}-w'^{2})^{-1/2}$ in the closed-form curvature Eqs.~\eqref{eq:eta-exact}. Third, the $\mathcal{O}(\varepsilon^{9})$ and $\mathcal{O}(\varepsilon^{11})$ curves are very close to each other; the small additional gain at $\mathcal{O}(\varepsilon^{11})$ does not justify the extra computational cost, and the $\mathcal{O}(\varepsilon^{9})$ truncation is adopted for all subsequent comparisons.

\subsection{Inertial- and rotating-frame views of the post-critical response}\label{subsubsec:U4Omega4-forward}

\begin{figure}[htbp]
\centering
\begin{tabular}{cc}
\includegraphics[width=0.5\textwidth]{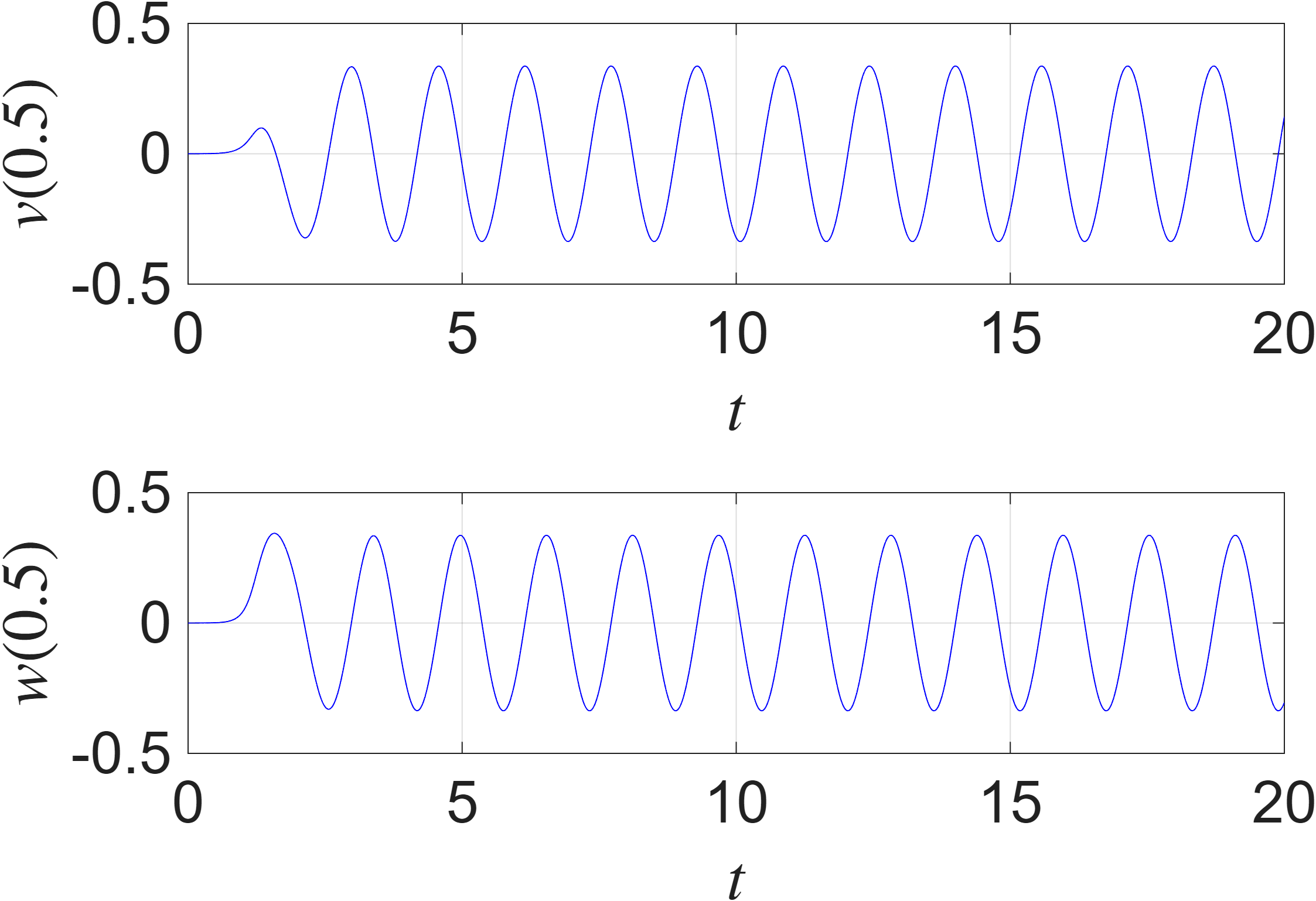} &
\includegraphics[width=0.4\textwidth]{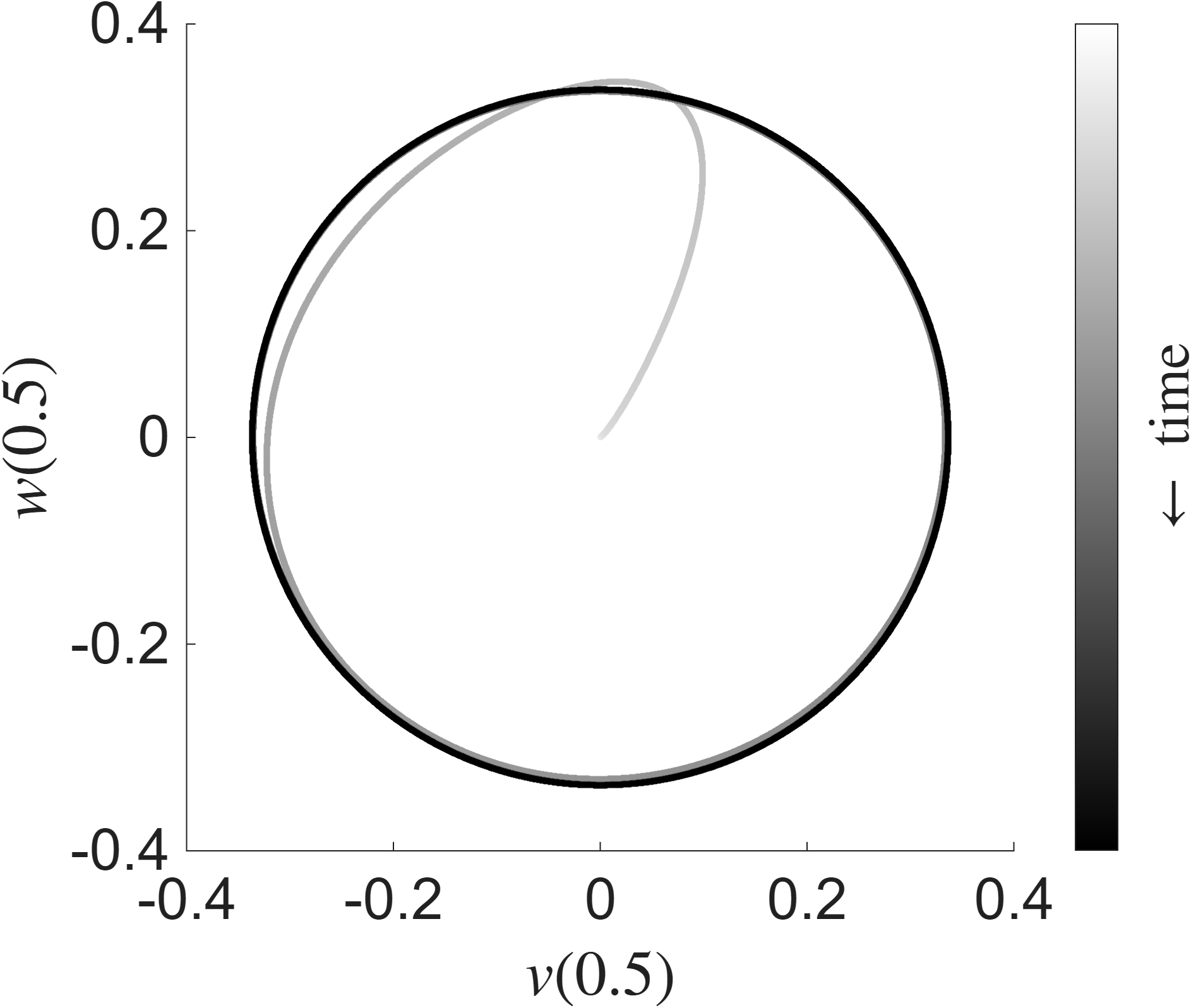} \\
(a) & (b)
\end{tabular}
\caption{Inertial-frame post-critical response at $U=4$, $\Omega=4$: (a)~time histories of the midpoint displacements $v(0.5,t)$ and $w(0.5,t)$; (b)~orbit in the $v$--$w$ plane. Parameters: $\alpha=0.023$, $\sqrt{\beta}=0.536925$. Third-order Taylor truncation.}
\label{fig:U4Omega4-inertial}
\end{figure}

With the $\mathcal{O}(\varepsilon^{9})$ truncation adopted, the post-critical response is examined in detail in both the inertial and rotating frames. At $U=4$, $\Omega=4$ (well outside the ellipse-like stable region), Figure~\ref{fig:U4Omega4-inertial} shows the inertial-frame time history and orbit of the midpoint deflection: an initially small perturbation grows due to the divergence instability and saturates onto a limit cycle whose orbit in the $v$--$w$ plane is a closed circle. The orbit direction confirms that the saturated motion is forward whirling -- the midpoint traces its circle in the same sense as the imposed spin -- consistent with the linear-stability result that only the forward-whirling mode is unstable.

\begin{figure}[htbp]
\centering
\begin{tabular}{cc}
\includegraphics[width=0.5\textwidth]{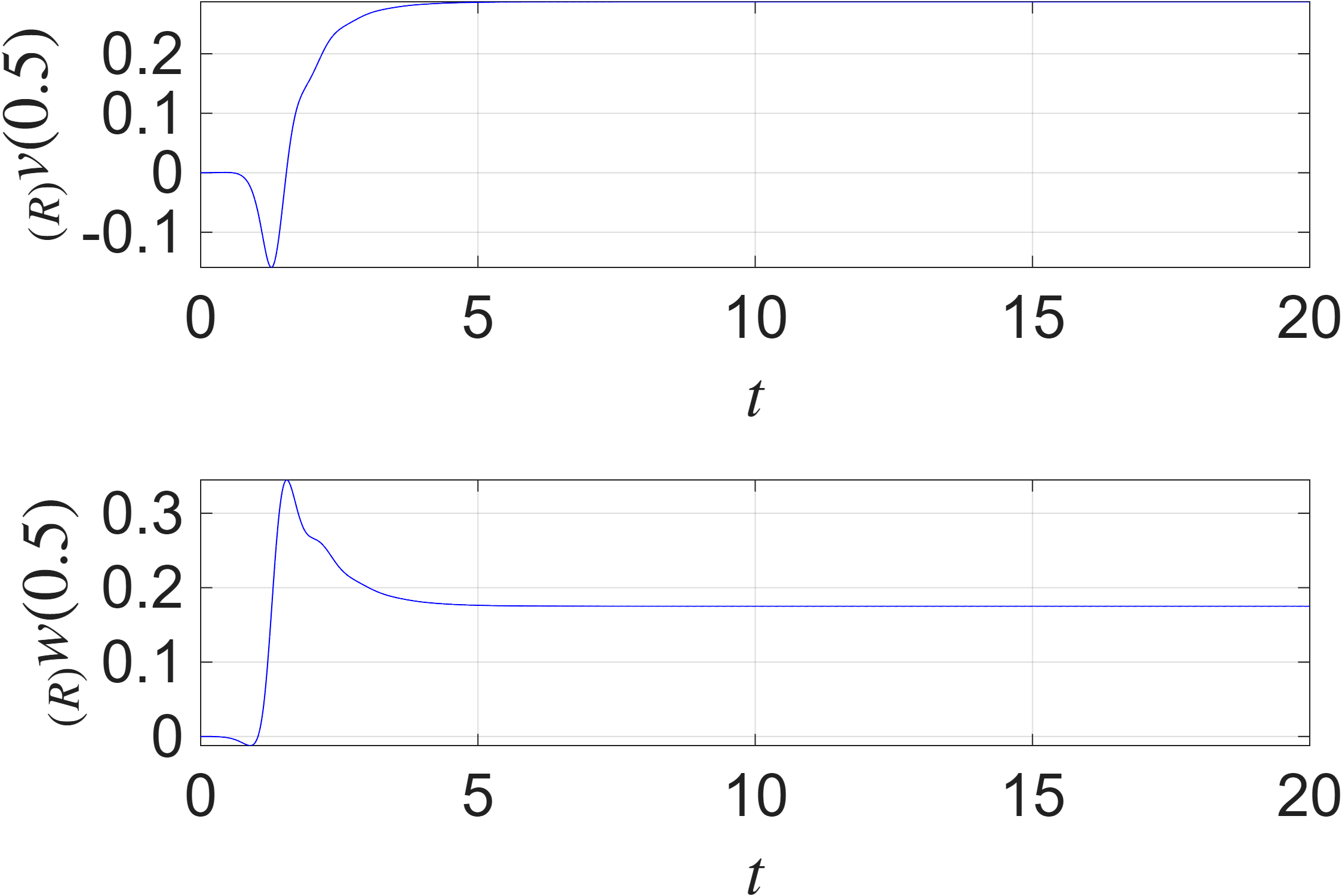} &
\includegraphics[width=0.4\textwidth]{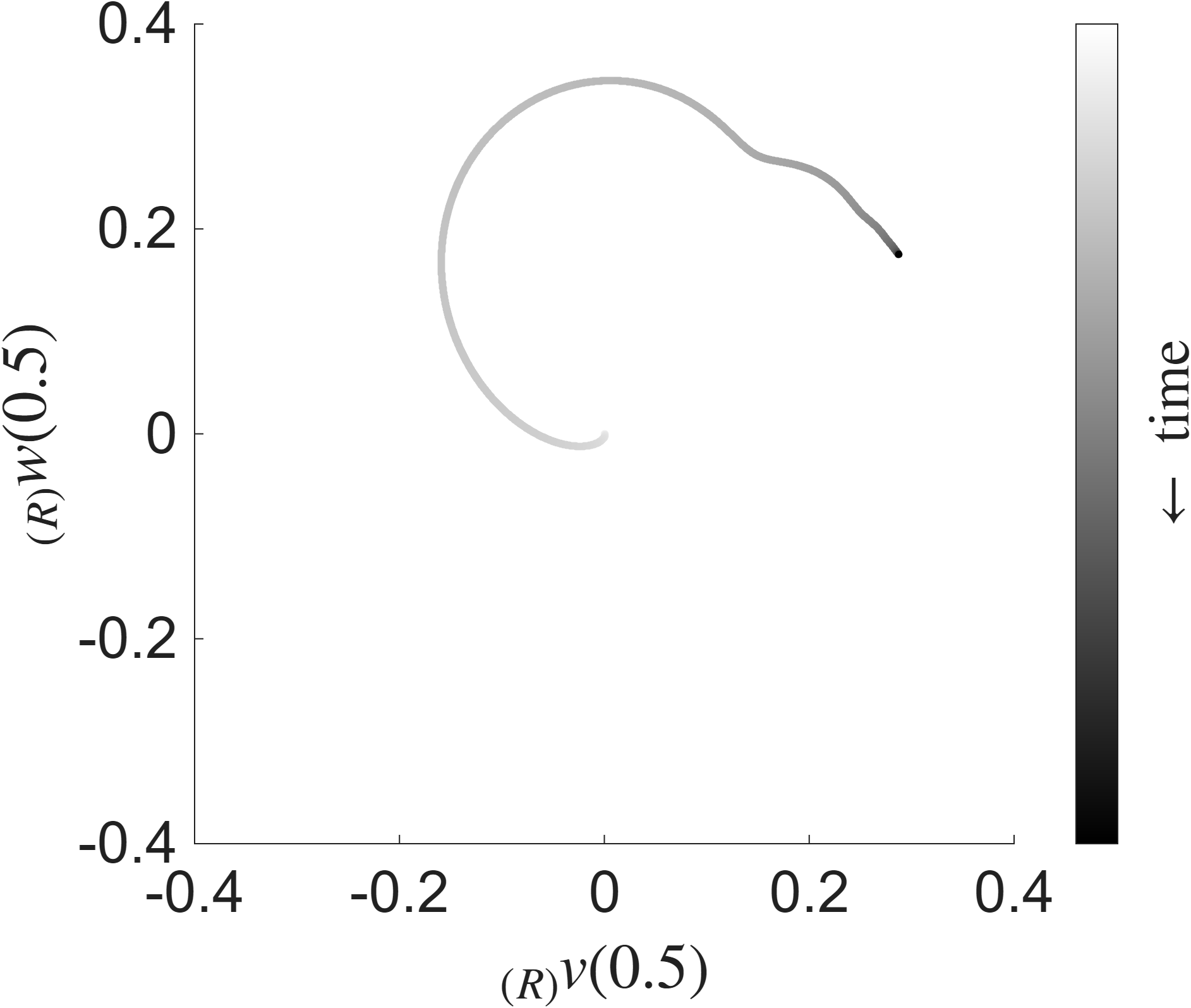} \\
(a) & (b)
\end{tabular}
\caption{Rotating-frame counterpart of Fig.~\ref{fig:U4Omega4-inertial}: (a)~time histories of $\Rom v(0.5,t)$ and $\Rom w(0.5,t)$; (b)~orbit in the $\Rom v$--$\Rom w$ plane, showing convergence to a static equilibrium. Parameters as in Fig.~\ref{fig:U4Omega4-inertial}.}
\label{fig:U4Omega4-rotating}
\end{figure}

The same response viewed in the rotating frame is shown in Fig.~\ref{fig:U4Omega4-rotating}: the deflection components $\Rom v(0.5,t)$ and $\Rom w(0.5,t)$ converge to constant values, corresponding to a single fixed point in the rotating $v$--$w$ plane. The post-critical state is therefore a static divergence in the co-rotating reference frame, exactly as for the pinned--pinned pipe.

\subsection{Global view of the post-critical surface across the \texorpdfstring{$(U,\Omega)$}{(U,Omega)} plane}\label{subsubsec:global-3d}

\begin{figure}[htbp]
\centering
\includegraphics[width=0.82\textwidth]{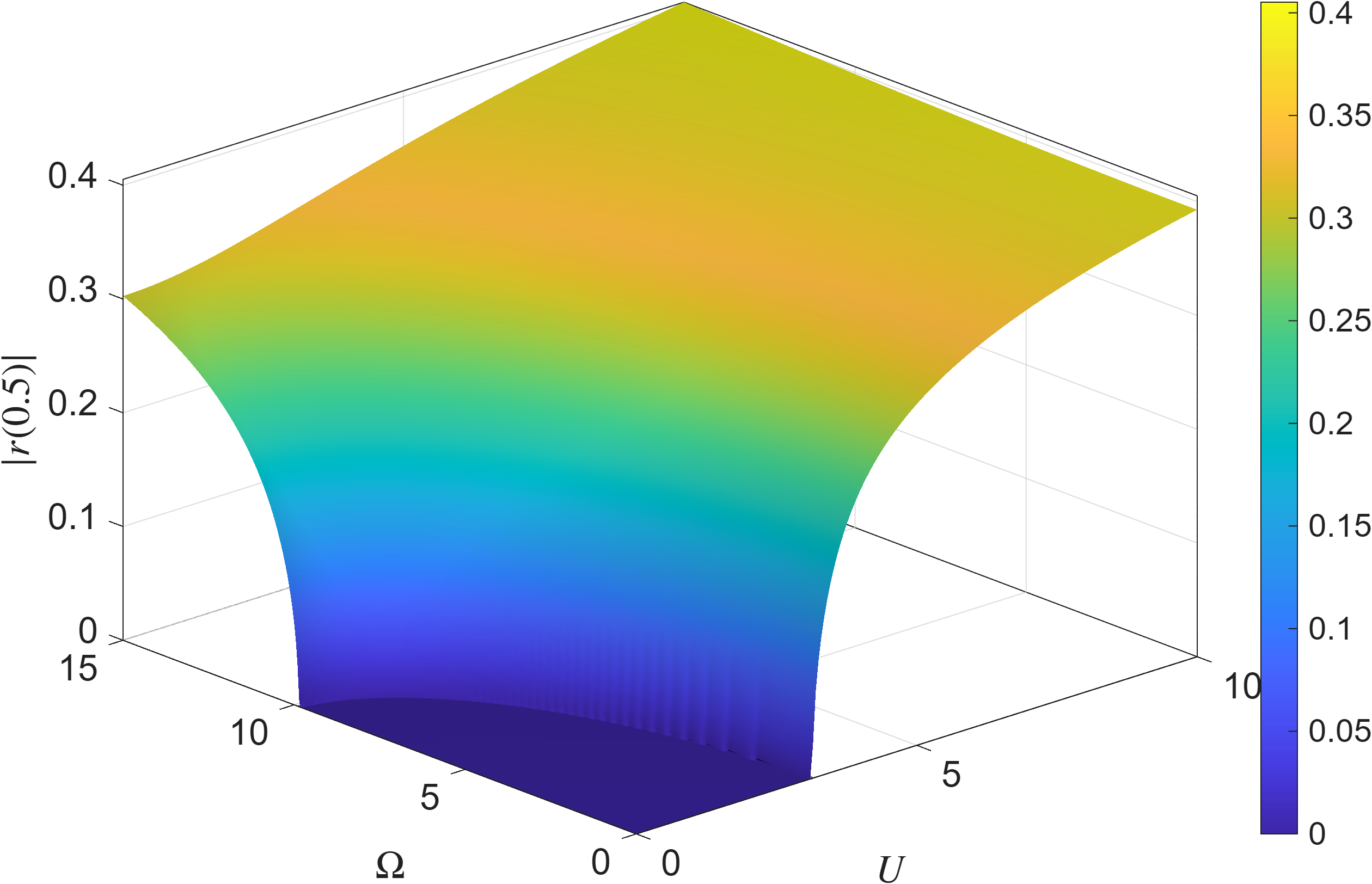}
\caption{Three-dimensional bifurcation surface: steady-state midspan deflection $|r(0.5)|$ over the $(U,\Omega)$ plane, $\mathcal{O}(\varepsilon^{9})$ truncation. Parameters: $\alpha=0.02317$, $\sqrt{\beta}=0.536925$, $\gamma=1$.}
\label{fig:bif-3D-inext}
\end{figure}

To provide a unified picture of the post-critical landscape, the steady-state midspan deflection $|r(0.5)|$ is mapped over the full flow-velocity--rotational-speed plane, $(U,\Omega)\in[0,10]\times[0,15]$. The result is shown in Fig.~\ref{fig:bif-3D-inext}. Inside the ellipse-like stable region the deflection is identically zero, in agreement with the linear-stability prediction; outside it the surface rises smoothly from the boundary, with no discontinuity in either direction -- a signature of the supercritical character of the bifurcation. The surface is qualitatively similar to the corresponding map for the extensible pinned--pinned pipe~\cite{fasihi_nonlinear_2026}, but the deflection magnitudes are larger by more than an order of magnitude across the whole post-critical region.

\begin{figure}[htbp]
\centering
\includegraphics[width=0.82\textwidth]{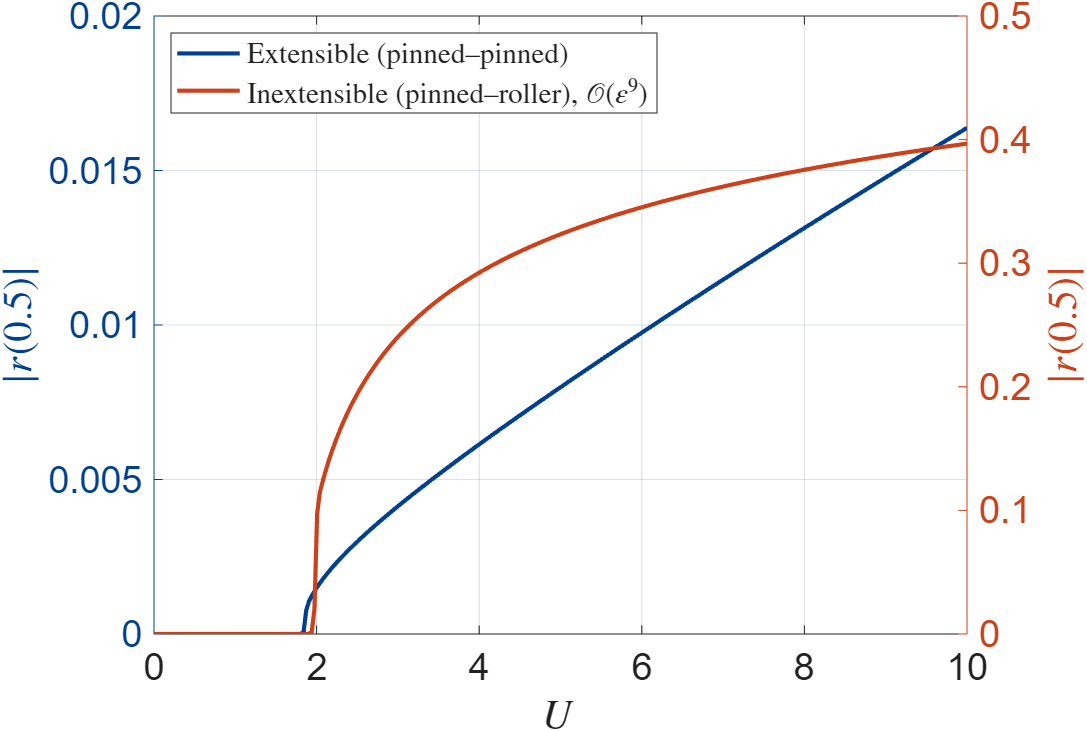}
\caption{Steady-state midpoint deflection versus flow velocity for the extensible pinned--pinned pipe (left axis) and the inextensible pinned--roller pipe (right axis, $\mathcal{O}(\varepsilon^{9})$ truncation). Parameters: $\Omega=8$, $\alpha=0.023$, $\sqrt{\beta}=0.536925$.}
\label{fig:ext-vs-inext}
\end{figure}

This difference is examined more closely using a representative slice. The post-critical bifurcation diagrams of the two configurations are overlaid in Fig.~\ref{fig:ext-vs-inext} for the same physical parameters. A dual vertical axis is used because the post-critical amplitudes differ by more than an order of magnitude.

The inextensible pinned--roller pipe exhibits substantially larger deflections and a steeper bifurcation slope. The difference originates in the axial boundary conditions: in the extensible (pinned--pinned) pipe, axial extension at the supports generates an axial tension that grows quadratically with the lateral deflection, providing a strong nonlinear restoring force that limits the post-critical growth. In the inextensible (pinned--roller) pipe the right support is free to slide axially, so this tension cannot develop, and the only source of geometric stiffening is the bending-curvature nonlinearity~$\eta_{1}^{2}+\eta_{2}^{2}$. The latter is comparatively weak, and the resulting equilibrium amplitudes are correspondingly larger.


\section{Modified Hencky bar-chain model}\label{sec:hencky}

This section develops the modified Hencky bar-chain model as a general-purpose discrete framework for spinning fluid-conveying pipes. The continuous pipe is represented by a chain of rigid links connected at the joints by rotational springs and dampers that localise bending flexibility, with the internal flow modelled as a co-moving incompressible material distributed uniformly along the chain~\cite{wang_hencky_2020}. A first Hencky-type model for the spinning fluid-conveying pipe was developed in the companion paper~\cite{fasihi_nonlinear_2026} using a successive-rotation construction in which each link's orientation is defined relative to its predecessor. Here, that model is modified to a more general formulation in which each link's orientation is expressed directly in the rotating reference frame. The global angular description yields a closed, $n$-independent matrix structure with exact trigonometric kinematics -- directly implementable in any standard programming environment with matrix routines and adaptable to a range of boundary conditions through matrix reductions. It also reduces the computational cost of the symbolic equations of motion by several orders of magnitude (quantified in Section~\ref{subsec:validation}) and handles the pinned--roller boundary condition through a single scalar substitution. The modified description further aligns the discrete formulation with the natural geometry of the articulated pipe model -- the pipe-dynamics counterpart of the Hencky bar-chain, introduced by Benjamin~\cite{benjamin_dynamics_1961} for a planar two-segment cantilever and extended to an arbitrary number of segments by P\"aidoussis and Deksnis~\cite{paidoussis_articulated_1970} -- in which each link's orientation is naturally interpretable as an absolute angle. In this paper, the framework also serves as an independent cross-check of the Galerkin results of Section~\ref{sec:results}.

\subsection{System configuration and dimensionless parameters}\label{subsec:hencky-config}

\begin{figure}[htbp]
\centering
\begin{tabular}{cc}
\includegraphics[width=0.6\textwidth]{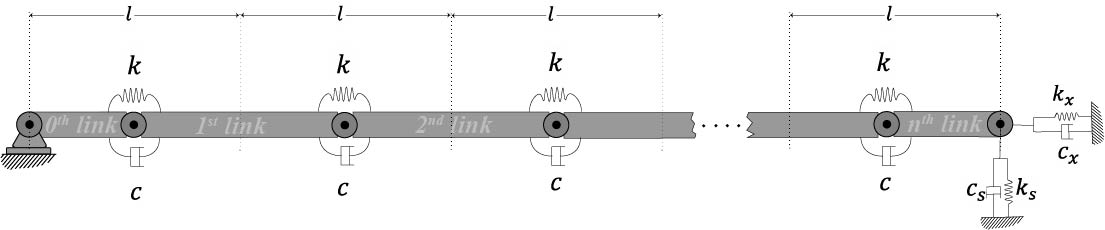} &
\includegraphics[width=0.28\textwidth,height=7cm,keepaspectratio]{SchematicArticulated.jpg} \\
(a) & (b)
\end{tabular}
\caption{Modified Hencky bar-chain model of the spinning fluid-conveying pipe: (a)~rigid-link chain with flexural springs and dampers at the joints; (b)~global angular description -- the orientation of each link is defined directly in the rotating frame $(x,y,z)$ by two successive rotations about the rotating-frame axes.}
\label{fig:hencky-schematic}
\end{figure}

The same continuous pipe introduced in Section~\ref{subsec:system} -- of length~$L$, cross-sectional area~$A$, second moment of area~$I$, mass per unit length~$m$, fluid mass per unit length~$M$, and Kelvin--Voigt damping coefficient~$\eta$ -- is divided into $n+1$ rigid links of equal length $l_i=l$ for $i=1,\ldots,n-1$ with end half-links $l_0=l_n=l/2$, where $l=1/n$ in dimensionless units (Fig.~\ref{fig:hencky-schematic}a). With these end half-links, one massless flexural joint is placed at each interior division point; bending stiffness is concentrated at these $n$ joints. At each interior joint a rotational spring of stiffness $k=1/l$ and a viscous damper of coefficient $c=\alpha/l$ are placed; the link and fluid masses are $m_i=(1-\beta)l_i$ and $M_i=\beta l_i$, respectively. Internal flow at relative velocity~$U$ is treated as a co-moving incompressible material; the corresponding dimensionless flow velocity in the Hencky model is $\tilde{U}=U/\sqrt{\beta}$, following~\cite{paidoussis_fluid-structure_2013}. The same dimensionless groups Eqs.~\eqref{eq:dimensionless-groups} as in the continuum model are used throughout. The left end of the chain ($O_0$) is pinned; the right end ($O_{n+1}$) is supported by transverse and longitudinal springs and dampers of coefficients $k_s$, $c_s$, $k_x=\mu$, $c_x=\alpha\mu$, with the boundary-condition reduction discussed in Section~\ref{subsec:hencky-bc}. The pipe spins about its undeformed longitudinal axis at constant dimensionless angular velocity~$\Omega$.

\subsection{Kinematics with global angular description}\label{subsec:hencky-kinematics}

Three reference frames are introduced: the inertial frame $(X,Y,Z)$ with origin at the left support and unit vectors $(\mathbf{i}_X,\mathbf{i}_Y,\mathbf{i}_Z)$; the rotating frame $(x,y,z)$ that co-rotates with the pipe at angular velocity $\Omega$ about the $X$-axis; and a body-attached frame $(x_i,y_i,z_i)$ for each link~$i=0,1,\ldots,n$, with origin at $O_i$ and unit vectors $(\mathbf{u}_{x_i},\mathbf{u}_{y_i},\mathbf{u}_{z_i})$. The orientation of each link is described by two successive rotations \emph{relative to the rotating frame}: an angle $\theta_i$ about the $y$-axis followed by an angle $\varphi_i$ about the $z$-axis (Fig.~\ref{fig:hencky-schematic}b). The rotation matrix from the link frame to the rotating frame is
\begin{equation}
\mathbf{B}_i=\bigl[\,\Rom\mathbf{u}_{x_i}\;\Rom\mathbf{u}_{y_i}\;\Rom\mathbf{u}_{z_i}\,\bigr]=\mathbf{A}_{i1}\,\mathbf{A}_{i2},
\label{eq:hencky-Bi}
\end{equation}
where $\mathbf{A}_{i1}$ and $\mathbf{A}_{i2}$ are the elementary rotation matrices about the rotating-frame $y$- and $z$-axes, respectively, and $\Rom(\cdot)$ denotes a vector expressed in the rotating frame.

Two consequences follow. The direct expression Eqs.~\eqref{eq:hencky-Bi} makes the generalised coordinates $\theta_i, \varphi_i$ mutually independent, which allows the $\theta$- and $\varphi$-blocks of the equations of motion to be linearly decoupled at equilibrium (Section~\ref{subsec:hencky-linear}). Furthermore, the spin~$\Omega$ is transmitted uniformly to every link regardless of its instantaneous deflection; the joints behave mechanically as constant-velocity (CV) joints rather than universal joints, eliminating a spurious spin-dependent kinematic coupling that would otherwise complicate the linearisation.

The generalised-coordinate vector is grouped by type rather than by link,
\begin{equation}
\mathbf{q}=\begin{bmatrix}\boldsymbol{\theta}\\\boldsymbol{\varphi}\end{bmatrix},\quad
\boldsymbol{\theta}=[\theta_0,\theta_1,\ldots,\theta_n]^{\mathsf{T}},\quad
\boldsymbol{\varphi}=[\varphi_0,\varphi_1,\ldots,\varphi_n]^{\mathsf{T}}.
\label{eq:hencky-q-grouped}
\end{equation}
Joint and mass-centre positions in the rotating frame follow recursively,
\begin{equation}
\Rom\mathbf{r}_{O_{i+1}}=\Rom\mathbf{r}_{O_{i}}+\mathbf{B}_{i}\,{}_{(i)}\mathbf{r}_{O_{i+1}/O_i},\quad
\Rom\mathbf{r}_{C_{i}}=\Rom\mathbf{r}_{O_{i}}+\mathbf{B}_{i}\,{}_{(i)}\mathbf{r}_{C_{i}/O_i},
\label{eq:hencky-positions}
\end{equation}
with ${}_{(i)}\mathbf{r}_{O_{i+1}/O_{i}}=[l_i,0,0]^{\mathsf{T}}$ and ${}_{(i)}\mathbf{r}_{C_{i}/O_{i}}=[l_i/2,0,0]^{\mathsf{T}}$. The mass-centre velocities in the rotating frame are obtained by the transport theorem
\begin{equation}
\Rom\mathbf{v}_{C_i}=\frac{\mathrm{d}}{\mathrm{d}t}\bigl(\Rom\mathbf{r}_{C_i}\bigr)+\Rom\boldsymbol{\omega}_{-1}\times\Rom\mathbf{r}_{C_i},
\label{eq:hencky-velocity-transport}
\end{equation}
where $\Rom\boldsymbol{\omega}_{-1}=[\Omega,0,0]^{\mathsf{T}}$ is the angular velocity of the rotating frame relative to the inertial frame. The fluid velocity in the rotating frame is $\Rom\mathbf{v}_{f_i}=\Rom\mathbf{v}_{C_i}+\mathbf{B}_i\,[U/\sqrt{\beta},0,0]^{\mathsf{T}}$, and the absolute angular velocity of link~$i$, expressed in its body frame, is
\begin{equation}
{}_{(i)}\boldsymbol{\omega}_i=\mathbf{B}_i^{\mathsf{T}}\,\Rom\boldsymbol{\omega}_{-1}+\mathbf{A}_{i2}^{\mathsf{T}}[0,\dot{\theta}_i,0]^{\mathsf{T}}+[0,0,\dot{\varphi}_i]^{\mathsf{T}}.
\label{eq:hencky-omega-link}
\end{equation}
The first term in Eqs.~\eqref{eq:hencky-omega-link} carries the spin: the body angular velocity inherits a component from the rotating-frame spin~$\Omega$, expressed in the local frame via $\mathbf{B}_i^{\mathsf{T}}$. This term is the kinematic origin of all spin-related contributions to the equations of motion derived below.

\subsection{Kinetic, potential, and dissipation contributions}\label{subsec:hencky-energies}

The total kinetic energy is the sum of the translational and rotational contributions of the pipe links and the internal fluid,
\begin{equation}
T=\tfrac{1}{2}\sum_{i=0}^{n}\Bigl[
m_i\,\Rom\mathbf{v}_{C_i}^{\mathsf{T}}\,\Rom\mathbf{v}_{C_i}
+{}_{(i)}\boldsymbol{\omega}_i^{\mathsf{T}}\,{}_{(i)}\mathbf{I}_p\,{}_{(i)}\boldsymbol{\omega}_i
+M_i\,\Rom\mathbf{v}_{f_i}^{\mathsf{T}}\,\Rom\mathbf{v}_{f_i}
+{}_{(i)}\boldsymbol{\omega}_i^{\mathsf{T}}\,{}_{(i)}\mathbf{I}_f\,{}_{(i)}\boldsymbol{\omega}_i
\Bigr],
\label{eq:hencky-T}
\end{equation}
where ${}_{(i)}\mathbf{I}_p$ and ${}_{(i)}\mathbf{I}_f$ are the slender-body inertia tensors of the link and the fluid, respectively, with the polar moment about the link axis neglected and $m_i l_i^2/12$, $M_i l_i^2/12$ on the two transverse diagonals.

The relative angular orientation of two adjacent links connected at joint $O_i$ is quantified by the difference vector in the rotating frame,
\begin{equation}
\Rom\mathbf{s}_i=[s_{ix},s_{iy},s_{iz}]^{\mathsf{T}}=\bigl(\mathbf{B}_i-\mathbf{B}_{i-1}\bigr)[1,0,0]^{\mathsf{T}},
\label{eq:hencky-s}
\end{equation}
with magnitude $|\Rom\mathbf{s}_i|=2\sin(\gamma_i/2)\approx\gamma_i$, where $\gamma_i$ is the angle between links $i-1$ and $i$. The potential energy is then
\begin{equation}
V=\tfrac{1}{2}k\sum_{i=1}^{n}\bigl(s_{ix}^{2}+s_{iy}^{2}+s_{iz}^{2}\bigr)
+\tfrac{1}{2}\,{}_{(I)}\mathbf{r}_{L}^{\mathsf{T}}\,\mathbf{K}_{L}\,{}_{(I)}\mathbf{r}_{L},
\label{eq:hencky-V}
\end{equation}
where $\mathbf{K}_{L}=\mathrm{diag}(\mu,k_s,k_s)$ collects the right-end support stiffnesses and ${}_{(I)}\mathbf{r}_L=\Rom\mathbf{r}_{O_{n+1}}-[1,0,0]^{\mathsf{T}}$ (transformed to the inertial frame, since the support springs are fixed in that frame and do not co-rotate with the pipe). The Rayleigh dissipation function takes the parallel form
\begin{equation}
R=\tfrac{1}{2}c\sum_{i=1}^{n}\bigl(\dot{s}_{ix}^{2}+\dot{s}_{iy}^{2}+\dot{s}_{iz}^{2}\bigr)
+\tfrac{1}{2}\,{}_{(I)}\dot{\mathbf{r}}_{L}^{\mathsf{T}}\,\mathbf{C}_{L}\,{}_{(I)}\dot{\mathbf{r}}_{L},
\label{eq:hencky-R}
\end{equation}
with $\mathbf{C}_{L}=\mathrm{diag}(\alpha\mu,c_s,c_s)$. The spin enters only through the kinetic energy Eqs.~\eqref{eq:hencky-T} via the velocity transport Eqs.~\eqref{eq:hencky-velocity-transport} and the body angular velocity Eqs.~\eqref{eq:hencky-omega-link}; the potential energy Eqs.~\eqref{eq:hencky-V} and the dissipation function Eqs.~\eqref{eq:hencky-R} are independent of $\Omega$.

\subsection{Nonlinear equations of motion in compact matrix form}\label{subsec:hencky-eom}

Substituting Eqs.~\eqref{eq:hencky-T}--\eqref{eq:hencky-R} into Lagrange's equations and collecting the kinetic, potential, and dissipation contributions yields the equations of motion in the compact form
\begin{equation}
\mathbf{f}_T(\mathbf{q},\dot{\mathbf{q}},\ddot{\mathbf{q}})
+\mathbf{f}_V(\mathbf{q})
+\mathbf{f}_R(\mathbf{q},\dot{\mathbf{q}})
=\mathbf{f}_O(\mathbf{q},\dot{\mathbf{q}}),
\label{eq:hencky-eom-compact}
\end{equation}
where $\mathbf{f}_T$, $\mathbf{f}_V$, $\mathbf{f}_R$ are the contributions of kinetic, potential, and dissipation terms, and $\mathbf{f}_O$ is the open-system generalised force arising from the momentum flux of the fluid leaving the downstream end. Each contribution is built from a small set of building-block matrices: the symmetric mass-distribution matrix $\mathbf{D}$, the discrete second-derivative matrix $\mathbf{K}$, the centripetal projection $\mathbf{L}$, the link-length-product matrices $\mathbf{B}_s=\mathbf{b}\mathbf{b}^{\mathsf{T}}$ and the antisymmetric counterpart $\mathbf{B}_a$ (with $\mathbf{b}=[1/(2n),1/n,\ldots,1/n,1/(2n)]^{\mathsf{T}}$), and the configuration-dependent trigonometric vectors and matrices $\mathbf{s}_\theta,\mathbf{c}_\theta,\mathbf{s}_\varphi,\mathbf{c}_\varphi,\mathbf{S}_\theta,\mathbf{C}_\theta$. All terms below -- including the spin-related contributions -- are constructed from this same set. The closed forms of the building blocks are tabulated in \ref{app:hencky}.

Differentiating the kinetic energy and collecting terms produces
\begin{equation}
\mathbf{f}_T=\mathbf{M}\,\ddot{\mathbf{q}}
+\mathbf{N}_{1}\,\dot{\mathbf{q}}^{\,2}
+\mathbf{N}_{2}\,(\dot{\boldsymbol{\theta}}\circ\dot{\boldsymbol{\varphi}})
+\bigl(\sqrt{\beta}\,U\,\mathbf{N}_{TU}+\Omega\,\mathbf{N}_{T\Omega}\bigr)\dot{\mathbf{q}}
+\Omega^{2}\,\mathbf{f}_{T\Omega^{2}}
+\sqrt{\beta}\,U\Omega\,\mathbf{f}_{TU\Omega},
\label{eq:hencky-fT}
\end{equation}
where ``$\circ$'' denotes the Hadamard (element-wise) product. The mass matrix $\mathbf{M}$, the second-order-velocity matrices $\mathbf{N}_1$, $\mathbf{N}_2$, and the flow-Coriolis matrix $\mathbf{N}_{TU}$ are independent of the spin; the latter enters $\mathbf{f}_T$ only through the three explicit $\Omega$- and $U\Omega$-proportional terms shown in Eqs.~\eqref{eq:hencky-fT}: the spin-Coriolis matrix $\mathbf{N}_{T\Omega}$, the centrifugal vector $\mathbf{f}_{T\Omega^{2}}$ proportional to $\Omega^{2}$, and the spin--flow cross-coupling vector $\mathbf{f}_{TU\Omega}$ proportional to $\sqrt{\beta}\,U\Omega$. Their full configuration-dependent forms are reported in \ref{app:hencky}.

The potential energy contributes
\begin{equation}
\mathbf{f}_V=\mathbf{f}_{VF}+k_s\,\mathbf{f}_{Vk_s}+\mu\,\mathbf{f}_{V\mu},
\label{eq:hencky-fV}
\end{equation}
independent of the spin: $\mathbf{f}_{VF}$ collects the flexural-spring forces at the interior joints, $\mathbf{f}_{Vk_s}$ the right-end transverse-spring contribution, and $\mathbf{f}_{V\mu}$ the longitudinal-spring contribution that vanishes for the pinned--roller configuration ($\mu=0$).

Likewise,
\begin{equation}
\mathbf{f}_R=\alpha\,\mathbf{f}_{R\alpha}+c_s\,\mathbf{C}_{c_s}\,\dot{\mathbf{q}}+\alpha\mu\,\mathbf{C}_{\alpha\mu}\,\dot{\mathbf{q}},
\label{eq:hencky-fR}
\end{equation}
The flexural damping $\alpha\mathbf{f}_{R\alpha}$ is the discrete counterpart of the continuum Kelvin--Voigt term $\alpha\dot{v}''''$.

The momentum flux of the fluid leaving the downstream end produces
\begin{equation}
\mathbf{f}_O=U^{2}\,\mathbf{f}_{OU^{2}}+\sqrt{\beta}\,U\Omega\,\mathbf{f}_{OU\Omega}+\sqrt{\beta}\,U\,\mathbf{N}_{OU}\,\dot{\mathbf{q}}.
\label{eq:hencky-fO}
\end{equation}
The first and third terms are independent of the spin; the middle term, proportional to $\sqrt{\beta}\,U\Omega$, is the discrete counterpart of the continuum-model term $-\Omega\,2\beta^{1/2}U\,w'$ that appears in Eqs.~\eqref{eq:eom-v-rotating} after rotating-frame transformation. Collectively, Eqs.~\eqref{eq:hencky-fT}--\eqref{eq:hencky-fO} provide the explicit forms of the four contributions in Eq.~\eqref{eq:hencky-eom-compact}.

\subsection{Linearised equations of motion}\label{subsec:hencky-linear}

Linearising Eqs.~\eqref{eq:hencky-eom-compact} about the straight equilibrium $\mathbf{q}=\mathbf{0}$ yields
\begin{equation}
\mathbf{M}_0\,\ddot{\mathbf{q}}+\mathbf{C}_0\,\dot{\mathbf{q}}+\mathbf{K}_0\,\mathbf{q}=\mathbf{0},
\label{eq:hencky-linearised}
\end{equation}
where the equilibrium reduces all trigonometric vectors and configuration-dependent matrices in Eqs.~\eqref{eq:hencky-fT}--\eqref{eq:hencky-fO} to constants. The mass matrix is
\begin{equation}
\mathbf{M}_{0}=\begin{bmatrix}\mathbf{D}&\mathbf{0}\\\mathbf{0}&\mathbf{D}\end{bmatrix},
\label{eq:hencky-M0}
\end{equation}
with the off-diagonal blocks vanishing at equilibrium and the diagonal blocks reducing to the mass-distribution matrix~$\mathbf{D}$.

The damping matrix takes the form
\begin{equation}
\mathbf{C}_{0}=\alpha\begin{bmatrix}\mathbf{K}&\mathbf{0}\\\mathbf{0}&\mathbf{K}\end{bmatrix}
+c_s\begin{bmatrix}\mathbf{B}_s&\mathbf{0}\\\mathbf{0}&\mathbf{B}_s\end{bmatrix}
+\sqrt{\beta}\,U\begin{bmatrix}\mathbf{B}_a+\mathbf{B}_s&\mathbf{0}\\\mathbf{0}&\mathbf{B}_a+\mathbf{B}_s\end{bmatrix}
+2\Omega\begin{bmatrix}\mathbf{0}&-\mathbf{D}\\\mathbf{D}&\mathbf{0}\end{bmatrix},
\label{eq:hencky-C0}
\end{equation}
and the stiffness matrix takes the form
\begin{equation}
\begin{aligned}
\mathbf{K}_{0}&=\begin{bmatrix}\mathbf{K}&\mathbf{0}\\\mathbf{0}&\mathbf{K}\end{bmatrix}
+k_s\begin{bmatrix}\mathbf{B}_s&\mathbf{0}\\\mathbf{0}&\mathbf{B}_s\end{bmatrix}\\[2pt]
&\quad+\sqrt{\beta}\,U\Omega\begin{bmatrix}\mathbf{0}&-\bigl(\mathbf{B}_a+\mathbf{B}_s\bigr)\\\bigl(\mathbf{B}_a+\mathbf{B}_s\bigr)&\mathbf{0}\end{bmatrix}
-\Omega^{2}\begin{bmatrix}\mathbf{D}&\mathbf{0}\\\mathbf{0}&\mathbf{D}\end{bmatrix}
-U^{2}\begin{bmatrix}\mathbf{L}&\mathbf{0}\\\mathbf{0}&\mathbf{L}\end{bmatrix}.
\end{aligned}
\label{eq:hencky-K0}
\end{equation}
The spin-independent contributions to $\mathbf{C}_0$ are the structural damping ($\alpha$), the transverse-support damping ($c_s$), and the flow-induced Coriolis term ($\sqrt{\beta}U$); the spin-independent contributions to $\mathbf{K}_0$ are the bending stiffness ($\mathbf{K}$), the transverse-support stiffness ($k_s$), and the centrifugal flow term ($-U^{2}\mathbf{L}$). The spin-induced contributions appear as off-diagonal blocks: the $2\Omega\,\mathbf{D}$ block in $\mathbf{C}_0$ couples the $\theta$- and $\varphi$-planes through gyroscopic precession, and the $\sqrt{\beta}\,U\Omega\,(\mathbf{B}_a+\mathbf{B}_s)$ block in $\mathbf{K}_0$ provides the spin--flow cross-stiffness. The diagonal centrifugal contribution $-\Omega^{2}\,\mathbf{D}$ in $\mathbf{K}_0$ acts as a state-independent destiffening of both subsystems.

Two structural consequences follow. \emph{First}, in the absence of spin ($\Omega=0$) the off-diagonal blocks of both $\mathbf{C}_0$ and $\mathbf{K}_0$ vanish, the linearised system decouples into two identical $(n+1)$-dimensional subsystems for $\boldsymbol{\theta}$ and $\boldsymbol{\varphi}$, and the frequency analysis of the bar-chain reduces to a single planar problem. \emph{Second}, when $\Omega\neq0$ the off-diagonal $2\Omega\,\mathbf{D}$ block in $\mathbf{C}_0$ couples the two subsystems through the discrete counterpart of the gyroscopic Coriolis term $\mp 2\Omega\dot{w}/\pm 2\Omega\dot{v}$ in the continuum Eqs.~\eqref{eq:eom-v-rotating}--\eqref{eq:eom-w-rotating}; together with the structural damping in the diagonal $\alpha\mathbf{K}$ block, this off-diagonal coupling is the discrete signature of the rotating-damping destabilisation mechanism. The Hencky model thus reproduces the same spin-mediated stability physics as the continuum model, by a fundamentally different route. The linearised system is recast as the first-order state-space form
\begin{equation}
\dot{\mathbf{y}}=\mathbf{A}\,\mathbf{y},\qquad
\mathbf{y}=\begin{bmatrix}\mathbf{q}\\\dot{\mathbf{q}}\end{bmatrix},\quad
\mathbf{A}=\begin{bmatrix}\mathbf{0}&\mathbf{I}\\-\mathbf{M}_{0}^{-1}\mathbf{K}_{0}&-\mathbf{M}_{0}^{-1}\mathbf{C}_{0}\end{bmatrix},
\label{eq:hencky-state-space}
\end{equation}
whose eigenvalues are computed directly for stability assessment.

\subsection{Boundary-condition reduction}\label{subsec:hencky-bc}

The model accommodates several support configurations through reduction of the system matrices. Two configurations are relevant to this paper.

For the pinned--pinned configuration, the full $2(n+1)$ generalised coordinates of Eqs.~\eqref{eq:hencky-q-grouped} are retained. The longitudinal-spring stiffness $\mu$ enters through the term $\mu\mathbf{f}_{V\mu}$ of $\mathbf{f}_V$ and through the corresponding damping $\alpha\mu\mathbf{C}_{\alpha\mu}\dot{\mathbf{q}}$ of $\mathbf{f}_R$, which together generate the axial-tension nonlinearity that limits post-critical amplitudes in the continuum (extensible) model. The transverse-support stiffness and damping, $k_s$ and $c_s$, are chosen numerically large enough to enforce vanishing transverse displacement at the right end while remaining numerically well-conditioned ($k_s\sim 10^{6}$, $c_s\sim 10^{2}$ in the present computations).

For the pinned--roller configuration, the right end is allowed to translate axially, so the longitudinal-spring contribution is removed by setting
\begin{equation}
\mu=0.
\label{eq:hencky-mu-zero}
\end{equation}
The transverse-support coefficients $k_s$, $c_s$ are retained at the same numerical values to enforce vanishing transverse displacement at the right end. No re-derivation of the matrix structure is required: the pinned--roller pipe is obtained from the general formulation by a single scalar substitution. This minimal adjustment is the principal practical advantage of the modified Hencky formulation for the present application.

Because the $\mu$-proportional terms enter only through the nonlinear forces $\mathbf{f}_V$ and $\mathbf{f}_R$, the linearised Hencky dynamics are independent of $\mu$ and coincide for the two configurations -- consistent with the continuum result of Section~\ref{subsec:rotating-frame}. The two configurations differ only in the post-critical regime, examined in Section~\ref{sec:results}.

\subsection{Application to the extensible pinned--pinned configuration}\label{subsec:validation}

The modified Hencky model is verified against the pinned--pinned (extensible) baseline of~\cite{fasihi_nonlinear_2026} before its application to the pinned--roller pipe. Four checks are performed: a quantitative measure of the computational advantage of the global angular description (Section~\ref{subsec:hencky-kinematics}); a convergence study of the linear-stability boundary with the number of links; a comparison of post-critical bifurcation diagrams against the Galerkin reference; and a time-history comparison under combined flow and rotation.

\begin{figure}[htbp]
\centering
\includegraphics[width=0.72\textwidth]{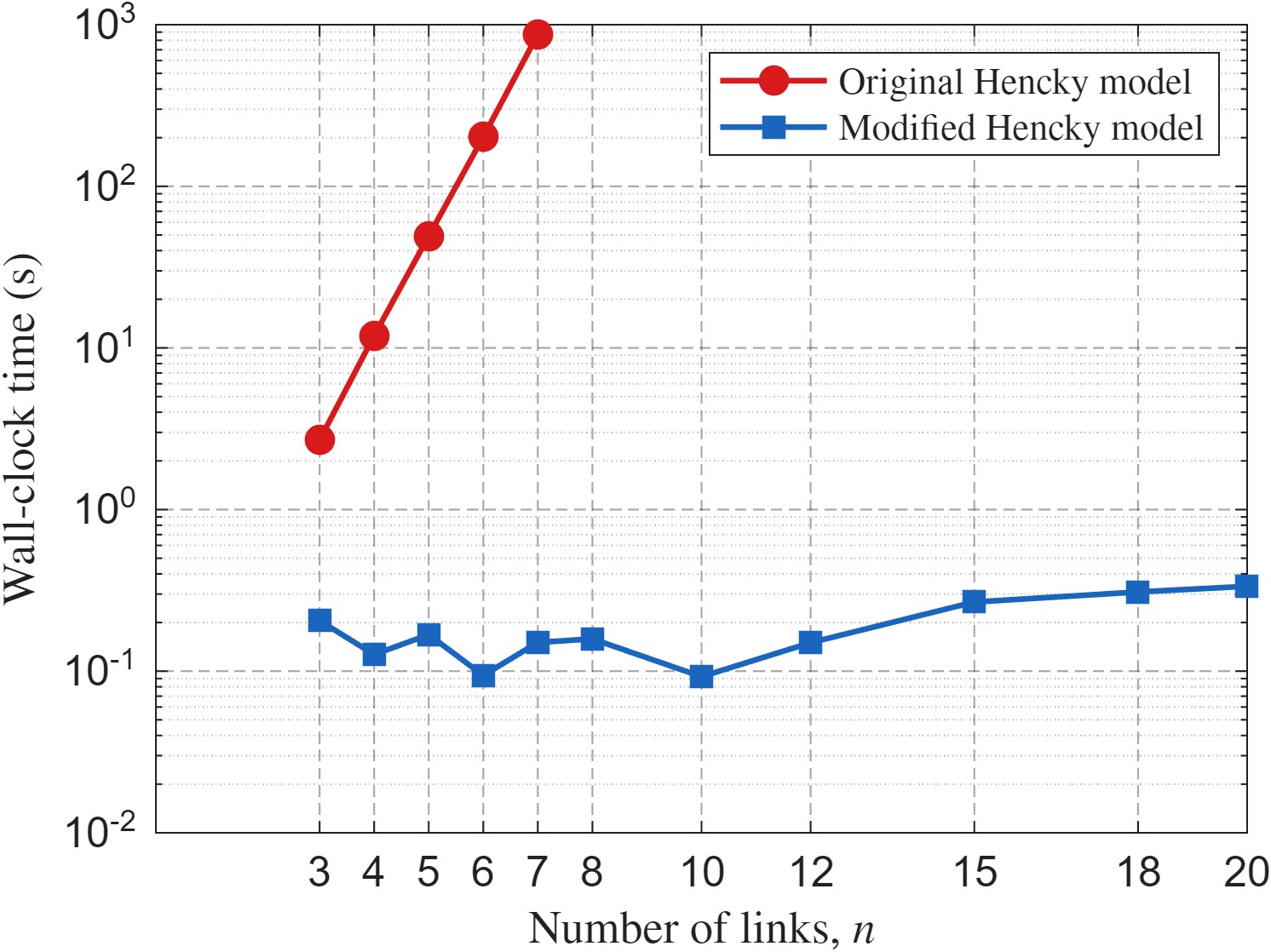}
\caption{Wall-clock computation time for a fixed simulation as a function of the number of links~$n$: the original successive-rotation Hencky formulation of~\cite{fasihi_nonlinear_2026} vs.\ the modified global-angular formulation of this paper. Parameters as in~\cite{fasihi_nonlinear_2026}.}
\label{fig:comp-time}
\end{figure}

The original Hencky formulation used in~\cite{fasihi_nonlinear_2026} chains successive local rotations $\mathbf{B}_i=\mathbf{B}_{i-1}\mathbf{A}_{i1}\mathbf{A}_{i2}$ link by link, so that the symbolic complexity of the equations of motion grows rapidly with the number of links. Wall-clock times for a fixed simulation on the same machine were approximately $2.7$, $11.9$, $49.2$, $202.0$, and $872.3$~s for $n=3,4,5,6,7$, respectively, with no practical extension beyond $n\approx 7$. With the global angular description, where each link is referenced directly to the rotating frame, the same simulation completes in under $0.35$~s for every $n$ tested in the range $3$--$20$. The reduction in computation time at $n=7$ exceeds three orders of magnitude, as shown in Fig.~\ref{fig:comp-time}. The modified formulation thereby brings $n=15$, $n=30$, and beyond into routine reach.

\begin{figure}[htbp]
\centering
\includegraphics[width=0.62\textwidth]{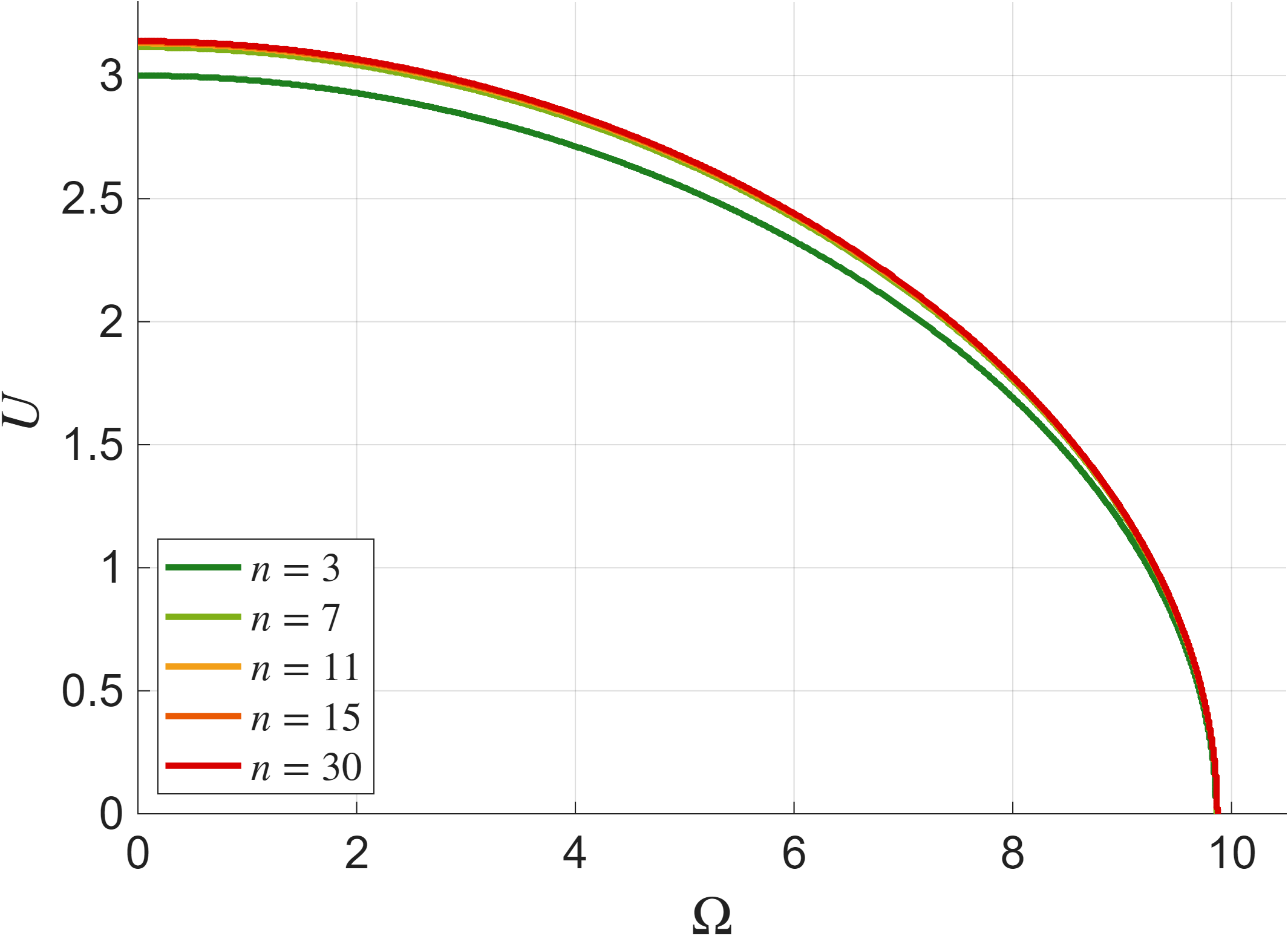}
\caption{Stability boundary in the $U$--$\Omega$ plane from the frequency analysis of the modified Hencky model for $n=3,7,11,15,30$, compared with the continuum ellipse-like reference. Parameters as in Fig.~\ref{fig:comp-time}.}
\label{fig:stability-conv-n}
\end{figure}

The frequency analysis of the linearised modified Hencky system, Eqs.~\eqref{eq:hencky-state-space}, predicts a stability boundary in the $U$--$\Omega$ plane that approaches the continuum reference as~$n$ increases. Figure~\ref{fig:stability-conv-n} shows the boundary for $n=3,7,11,15,30$. The $n=3$ result departs noticeably from the continuum reference, particularly at high flow velocity, while $n=7$ is already close and the curves for $n=11, 15, 30$ are indistinguishable from one another.

\begin{figure}[htbp]
\centering
\begin{tabular}{cc}
\includegraphics[width=0.45\textwidth]{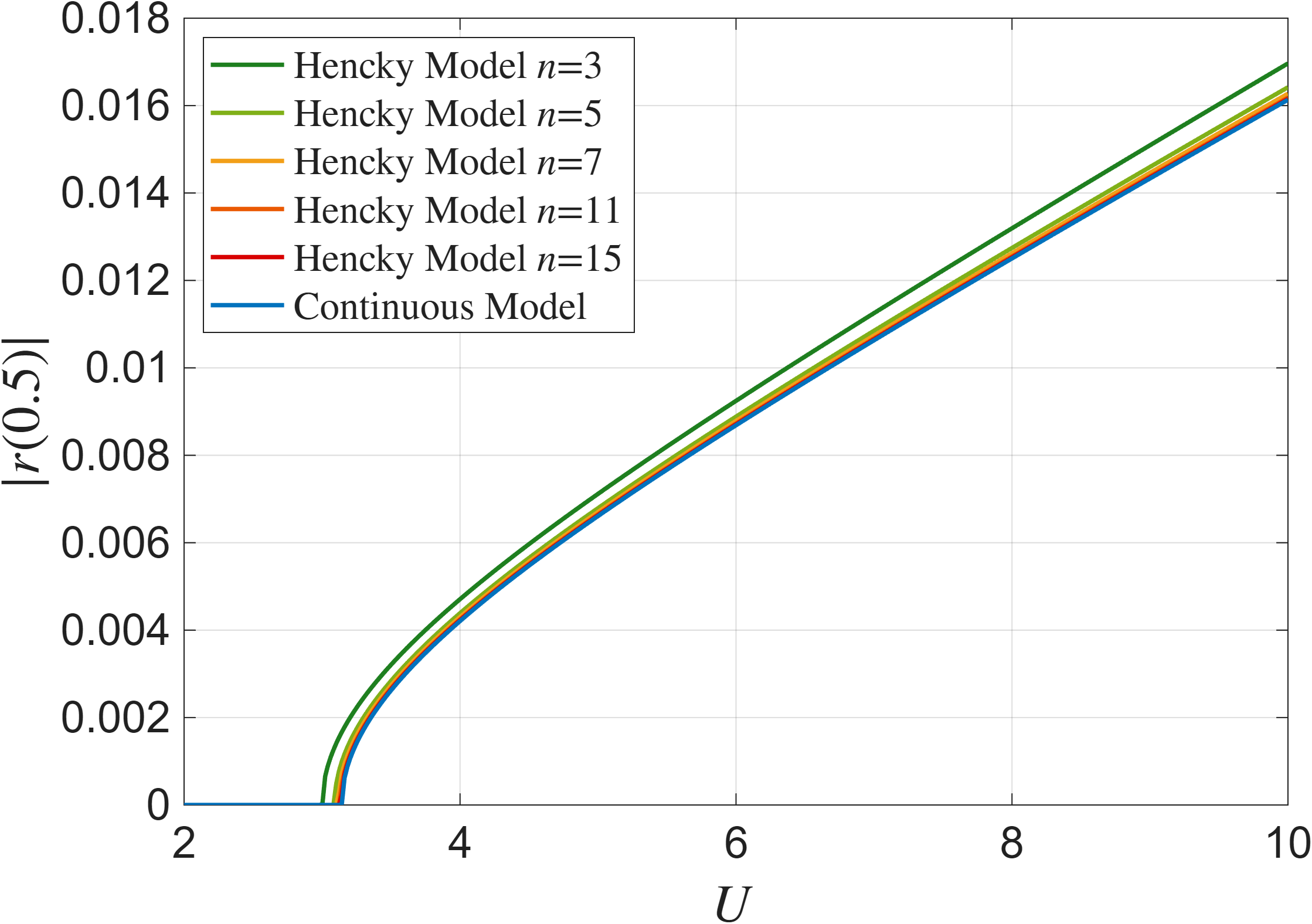} &
\includegraphics[width=0.45\textwidth]{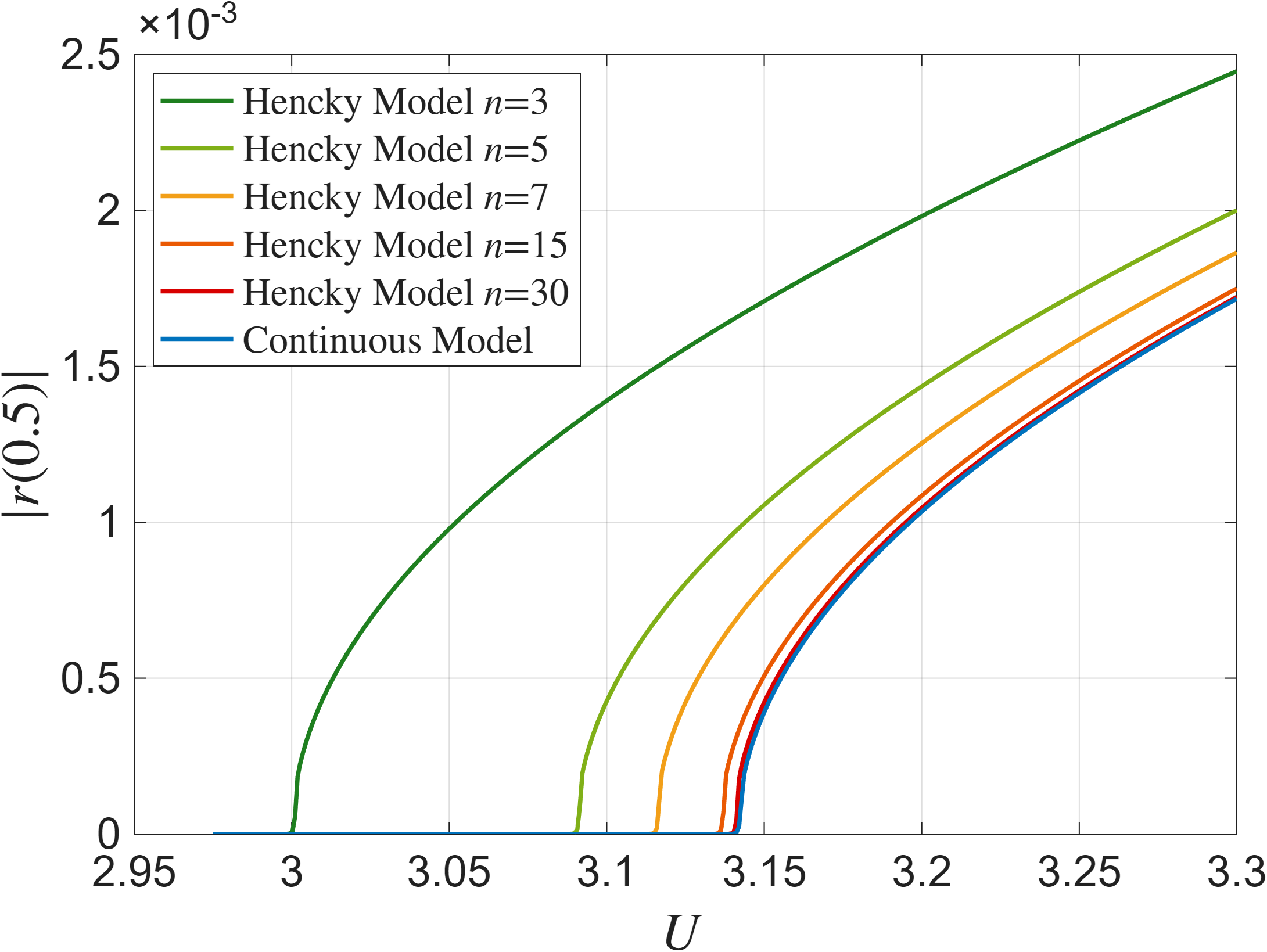} \\
(a) & (b)
\end{tabular}
\caption{Bifurcation diagram of $|r(0.5)|$ versus flow velocity~$U$ at $\Omega=0$: (a)~modified Hencky model at several values of~$n$ vs the Galerkin reference; (b)~magnified view of the onset, including $n=30$. Parameters as in Fig.~\ref{fig:comp-time}.}
\label{fig:bif-U-Omega0}
\end{figure}

The post-critical response is examined via the steady-state midspan deflection magnitude $|r(0.5)|$ as a function of the swept parameter, obtained by direct numerical integration of the nonlinear equations to steady state. Figure~\ref{fig:bif-U-Omega0}(a) shows the bifurcation diagram in flow velocity~$U$ for the modified Hencky model at $n=3,5,7,11,15$, overlaid on the Galerkin reference. The discrete model approaches the continuum result systematically as $n$ increases: by $n=15$ the agreement is excellent across the entire post-critical range. A magnified view of the bifurcation onset (Fig.~\ref{fig:bif-U-Omega0}(b), with $n=30$ included) shows the critical flow velocity converging monotonically to the continuum value $U=\pi$.

\begin{figure}[htbp]
\centering
\includegraphics[width=0.62\textwidth]{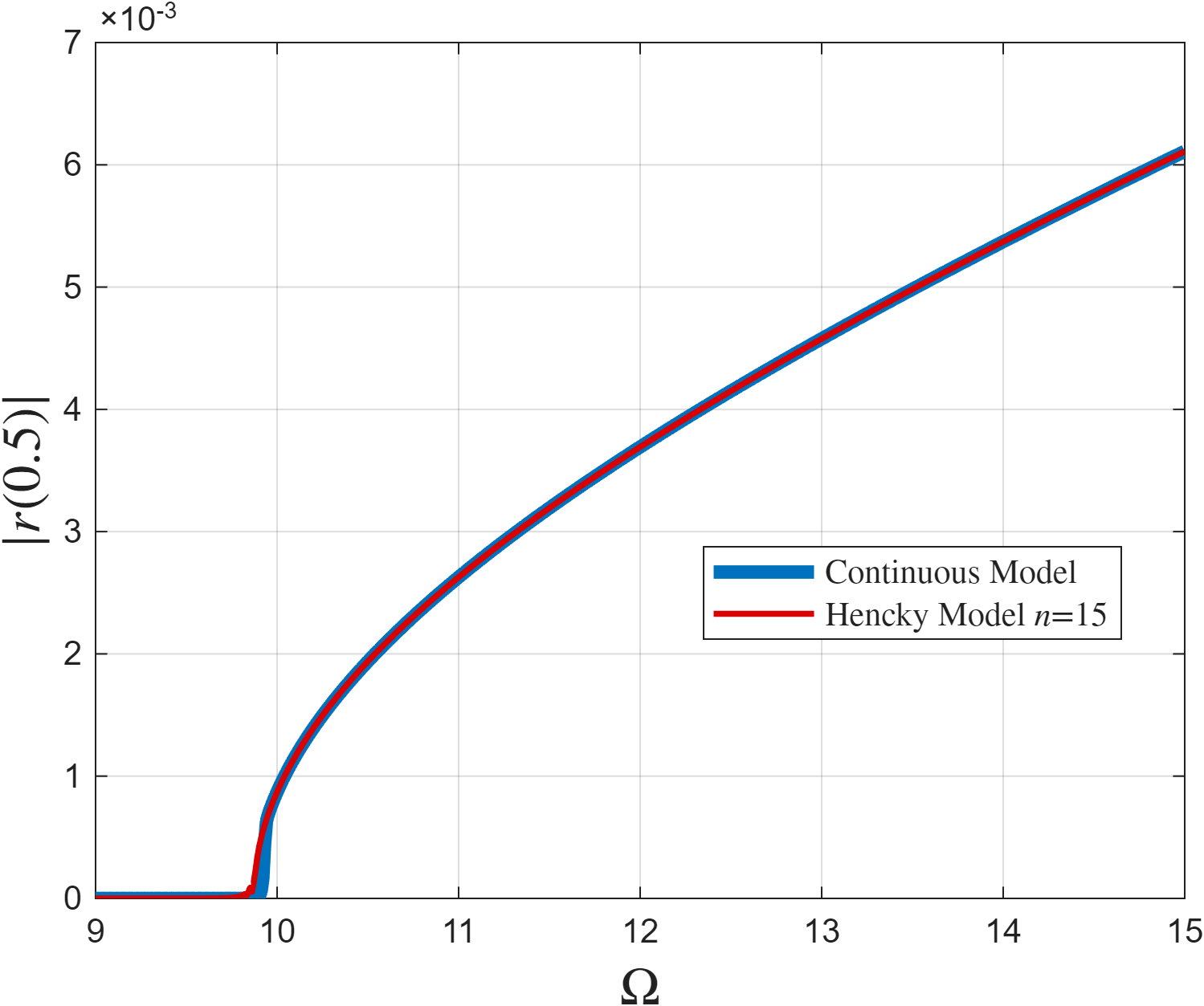}
\caption{Bifurcation diagram of $|r(0.5)|$ versus rotational speed~$\Omega$ at $U=0$, modified Hencky model ($n=15$) vs the Galerkin reference. Parameters as in Fig.~\ref{fig:comp-time}.}
\label{fig:bif-Omega-U0}
\end{figure}

\begin{figure}[htbp]
\centering
\begin{tabular}{cc}
\includegraphics[width=0.45\textwidth]{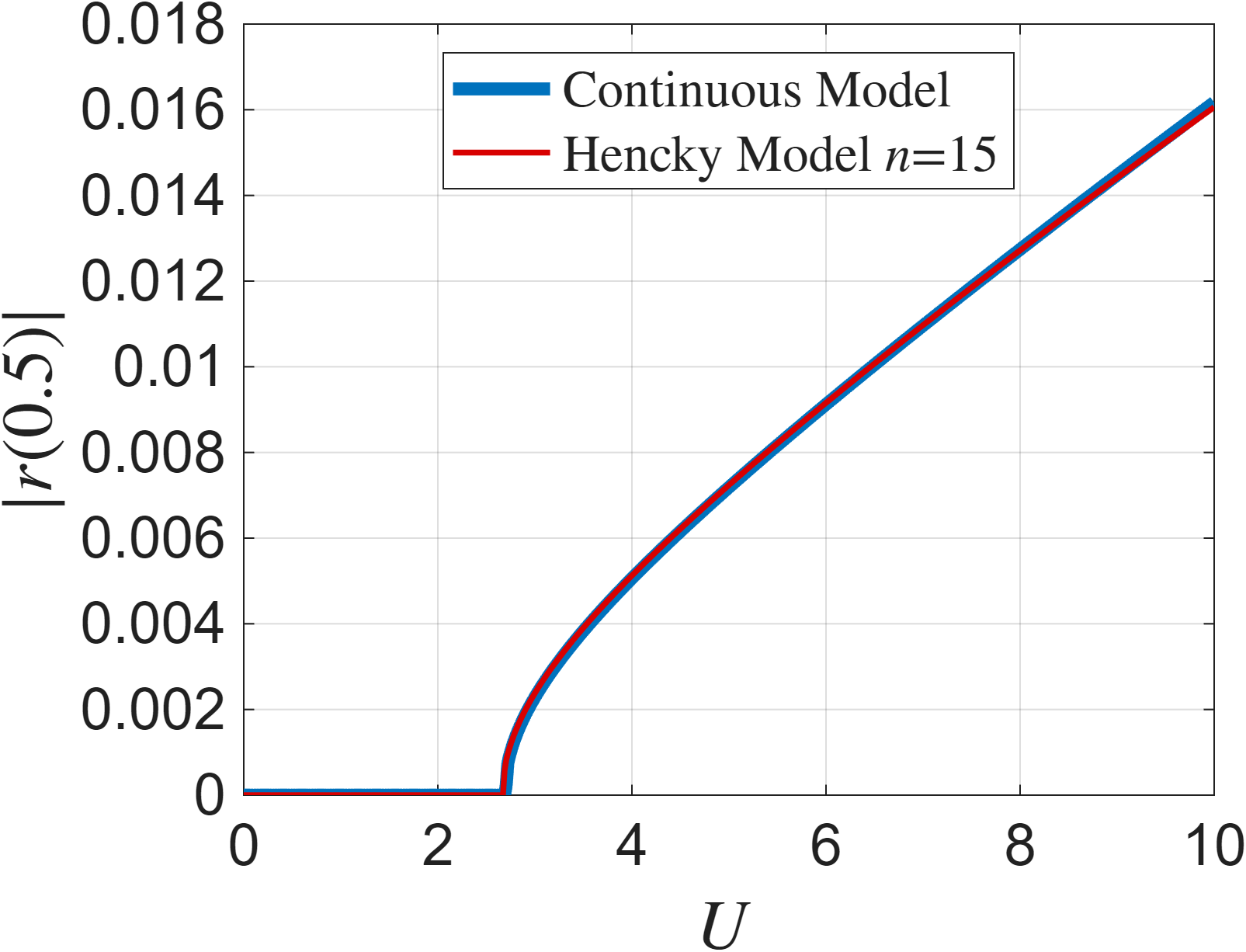} &
\includegraphics[width=0.45\textwidth]{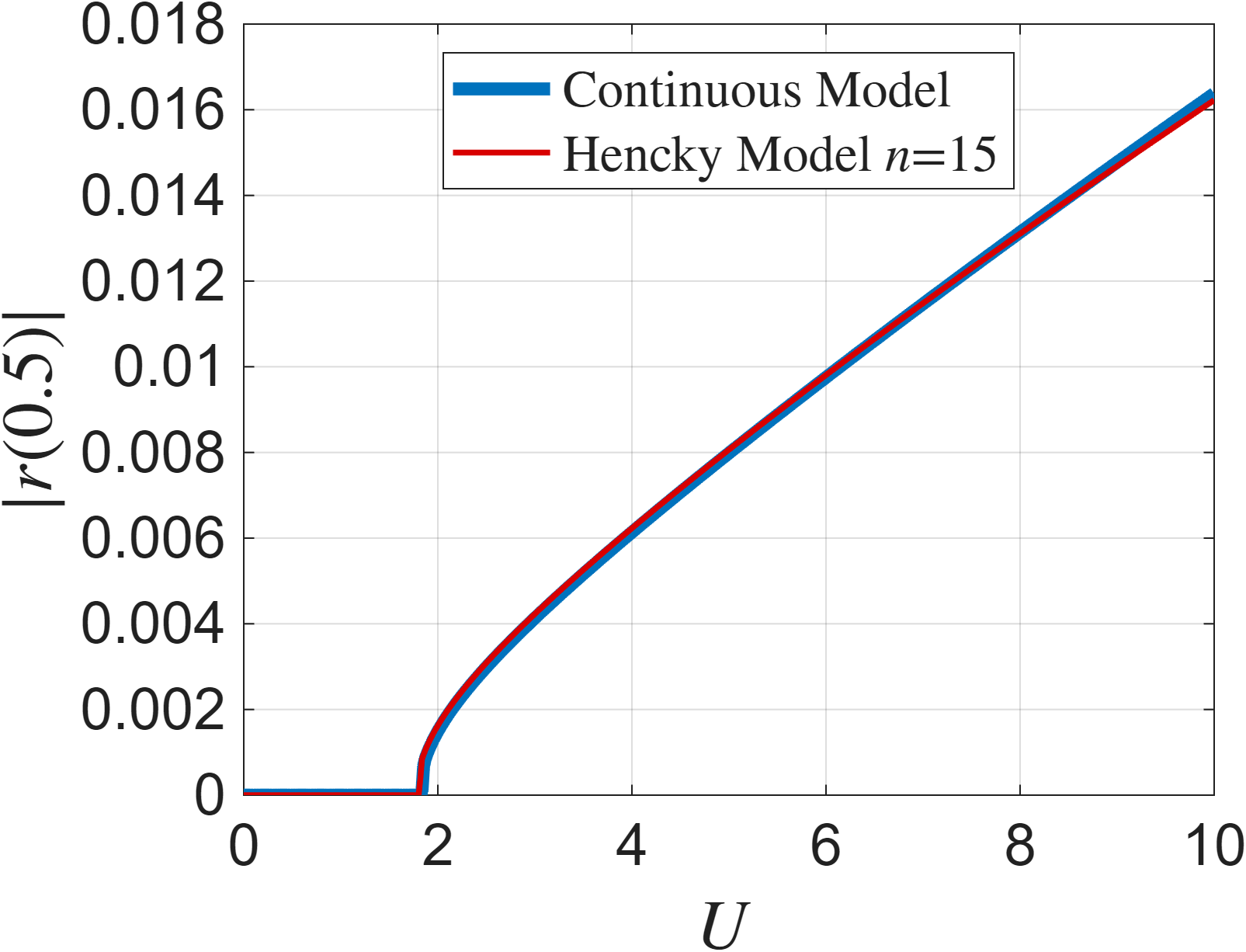} \\
(a) $\Omega=5$ & (b) $\Omega=8$
\end{tabular}
\caption{Bifurcation diagram of $|r(0.5)|$ versus flow velocity~$U$ at non-zero rotational speed: (a)~$\Omega=5$; (b)~$\Omega=8$. Modified Hencky model ($n=15$) vs the Galerkin reference. Parameters as in Fig.~\ref{fig:comp-time}.}
\label{fig:bif-U-Omega58}
\end{figure}

The same comparison on the rotational-speed axis is shown in Fig.~\ref{fig:bif-Omega-U0}, at $U=0$ with $n=15$. The critical rotational speed and the post-critical growth agree with the Galerkin prediction. This was already expected: in the linear-stability figure above, the curves for different~$n$ are very close near the $\Omega$-axis, where rotation alone drives the instability. The comparison is extended to combined flow and rotation in Fig.~\ref{fig:bif-U-Omega58}, for $\Omega=5$ and $\Omega=8$: the critical flow velocity decreases with increasing~$\Omega$ -- the destabilising influence of rotation -- and the agreement with the Galerkin reference is maintained.

\begin{figure}[htbp]
\centering
\begin{tabular}{cc}
\includegraphics[width=0.45\textwidth]{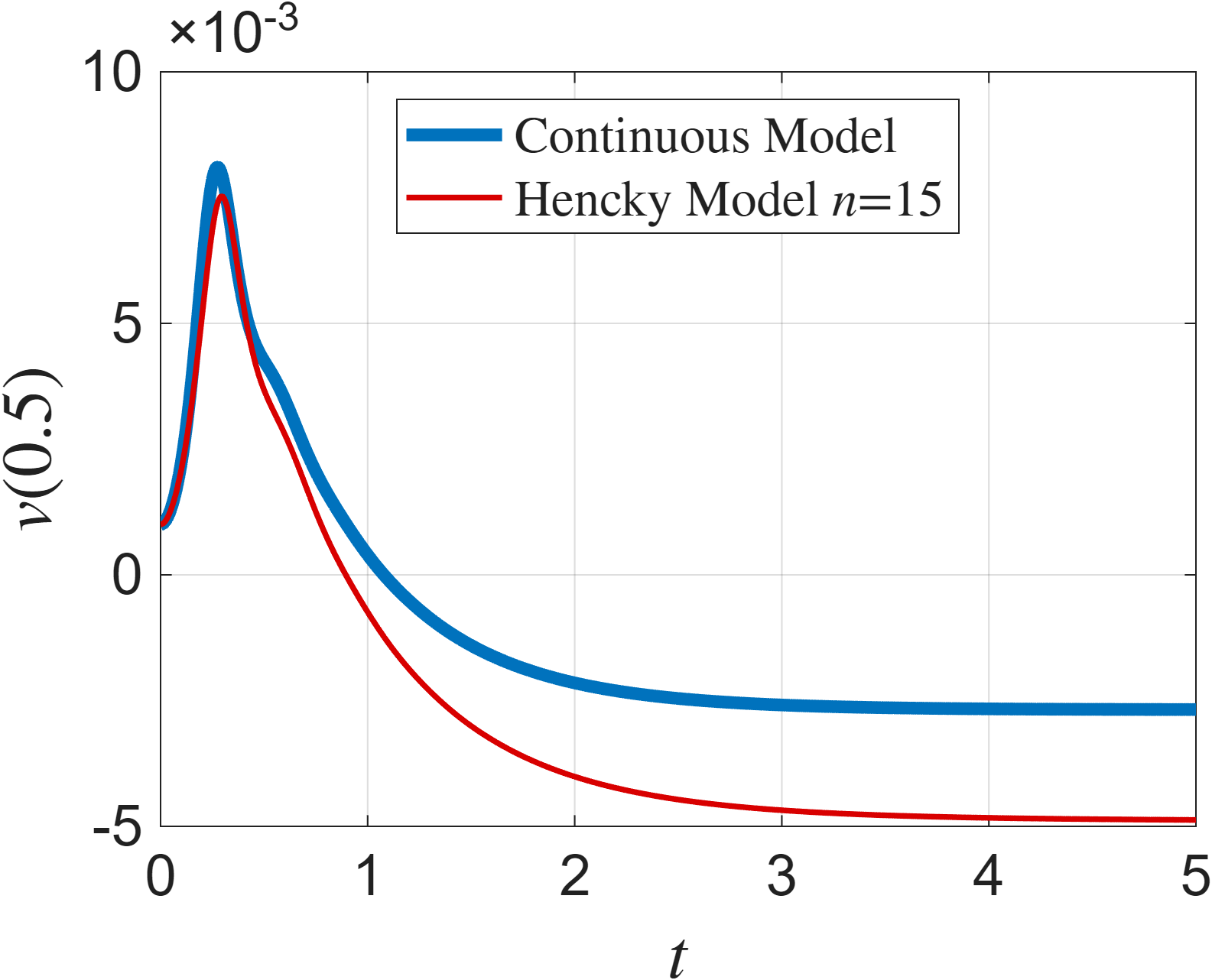} &
\includegraphics[width=0.45\textwidth]{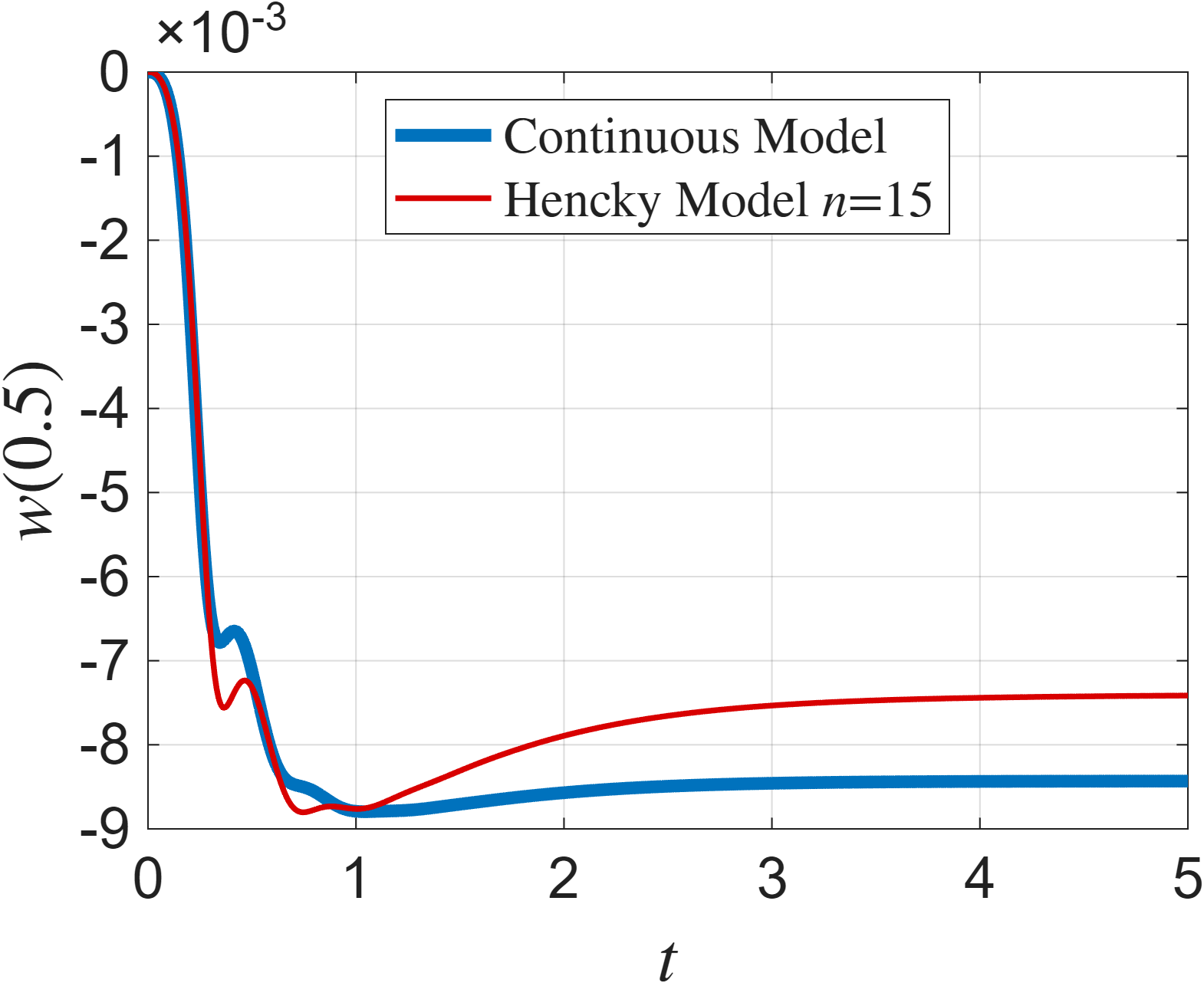} \\
(a) $v(0.5,t)$ & (b) $w(0.5,t)$ \\[6pt]
\multicolumn{2}{c}{\includegraphics[width=0.5\textwidth]{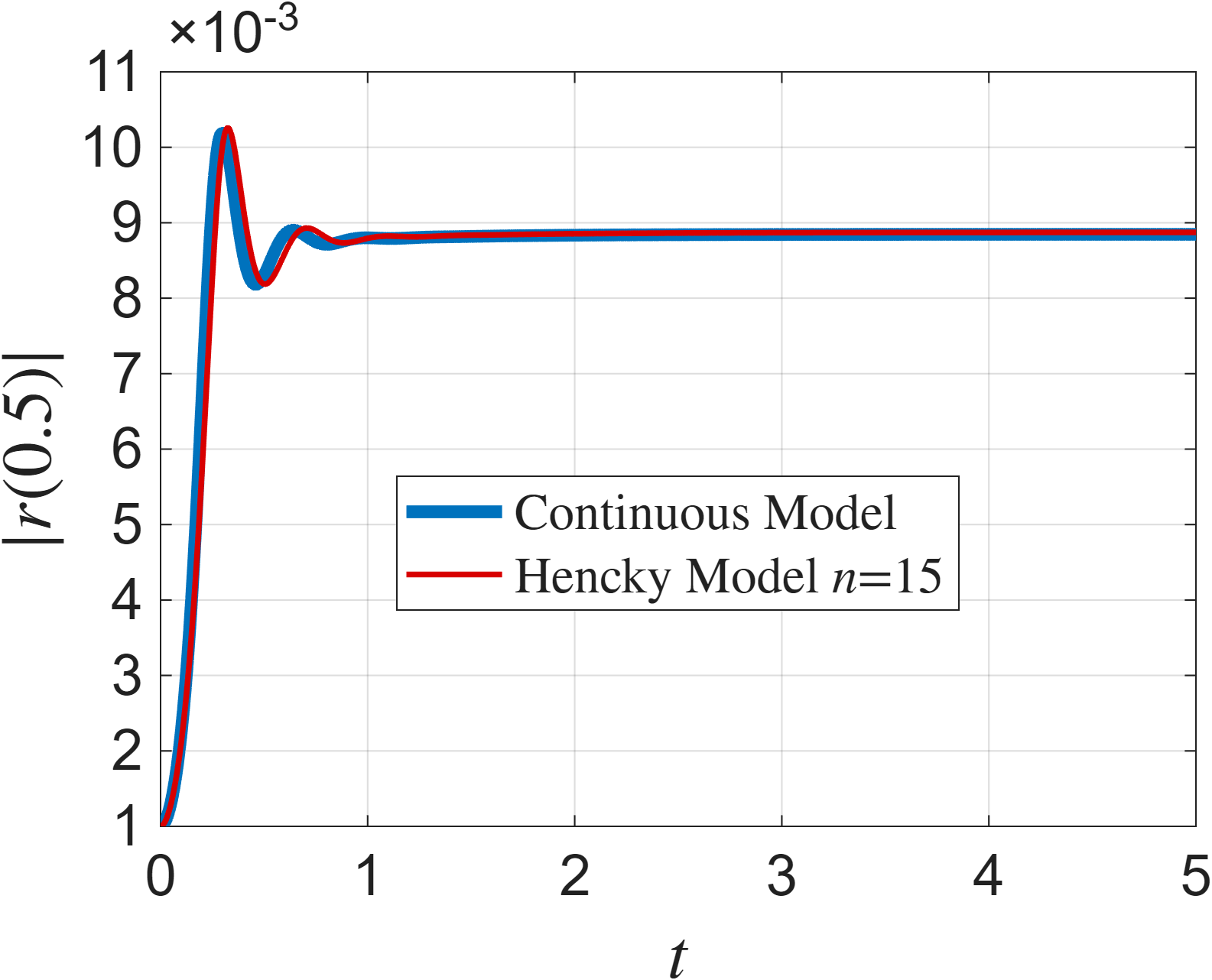}} \\
\multicolumn{2}{c}{(c) $|r(0.5,t)|$}
\end{tabular}
\caption{Time history of the midpoint response under combined flow and rotation: (a)~rotating-frame $v(0.5,t)$; (b)~rotating-frame $w(0.5,t)$; (c)~total deflection magnitude $|r(0.5,t)|$. Modified Hencky model ($n=15$) vs the Galerkin reference. Parameters as in Fig.~\ref{fig:comp-time}.}
\label{fig:time-hencky-vs-galerkin}
\end{figure}

Bifurcation diagrams characterise only the steady-state response. To check that the transient dynamics, including the spin-induced coupling between the $v$- and $w$-planes, are also captured, the two models are compared directly in the time domain for a representative case with both flow and rotation. The initial perturbation is placed entirely in the $v$-direction; motion in $w$ then develops through the spin-induced coupling. Figure~\ref{fig:time-hencky-vs-galerkin}\,(a,b) shows the time histories of the midpoint displacements $v(0.5,t)$ and $w(0.5,t)$, and panel~(c) the total deflection magnitude $|r(0.5,t)|$. The individual components $v$ and $w$ differ slightly between the two formulations -- they project the whirling motion onto the $v$ and $w$ axes in slightly different ways -- but the total deflection magnitude $|r|$ agrees closely throughout the transient and at the post-critical equilibrium. The spin-induced coupling is therefore captured by the Hencky model.

Together with the linear-stability and bifurcation comparisons above, the close time-domain agreement establishes the modified Hencky model as a faithful independent discretisation for the extensible pinned--pinned configuration.


\subsection{Application to the inextensible pinned--roller pipe}\label{subsec:hencky-pr-verification}

Setting $\mu=0$ reduces the modified Hencky model to the pinned--roller configuration, providing an independent test against the inextensible continuum results of Section~\ref{sec:results}. Since the linearised dynamics are unaffected, the comparison targets the post-critical regime.

\begin{figure}[htbp]
\centering
\includegraphics[width=0.62\textwidth]{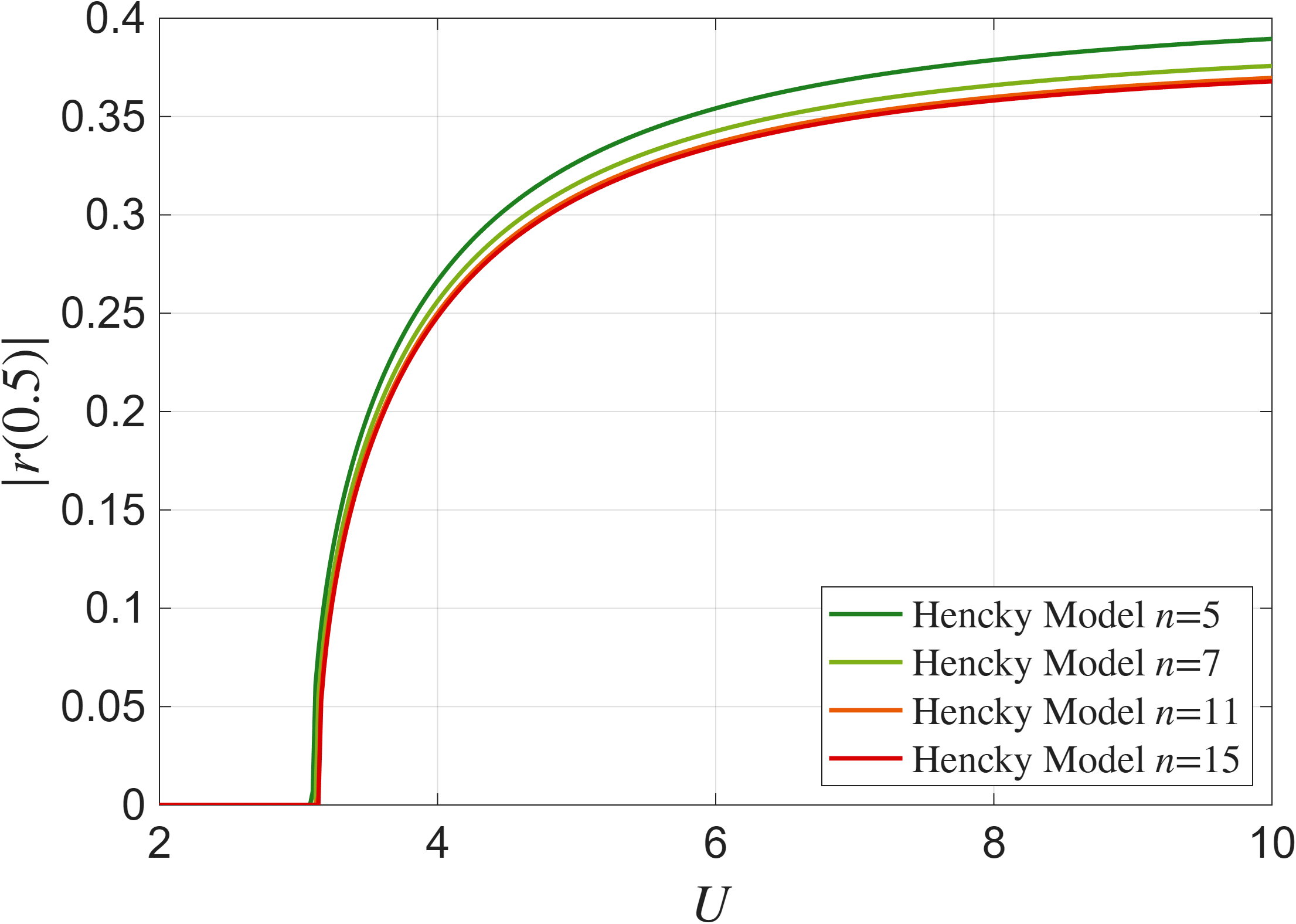}
\caption{Bifurcation diagram of $|r(0.5)|$ versus flow velocity~$U$ at $\Omega=0$ for the pinned--roller pipe; modified Hencky model at $n=5,7,11,15$. Parameters as in Fig.~\ref{fig:comp-time}.}
\label{fig:hencky-pr-U-Omega0}
\end{figure}

Figure~\ref{fig:hencky-pr-U-Omega0} shows the bifurcation diagram in flow velocity at $\Omega=0$ for the modified Hencky model at $n=5, 7, 11, 15$. The discrete curves converge systematically as $n$ increases, with $n=11$ and $n=15$ practically indistinguishable -- $n=15$ is therefore already converged here, in contrast to the pinned--pinned case of Section~\ref{subsec:validation}, where the comparison ran to $n=30$. The $n=3$ case is omitted: the coarse discretisation produced excessive deflections at the post-critical amplitudes characteristic of the inextensible pipe. The critical flow velocity remains at $U\approx\pi$, in agreement with the linearised prediction and with the pinned--pinned result of Section~\ref{subsec:validation}. The post-critical amplitudes, however, are noticeably larger than in the pinned--pinned case, consistent with the order-of-magnitude difference reported in Section~\ref{sec:results}.

\begin{figure}[htbp]
\centering
\includegraphics[width=0.62\textwidth]{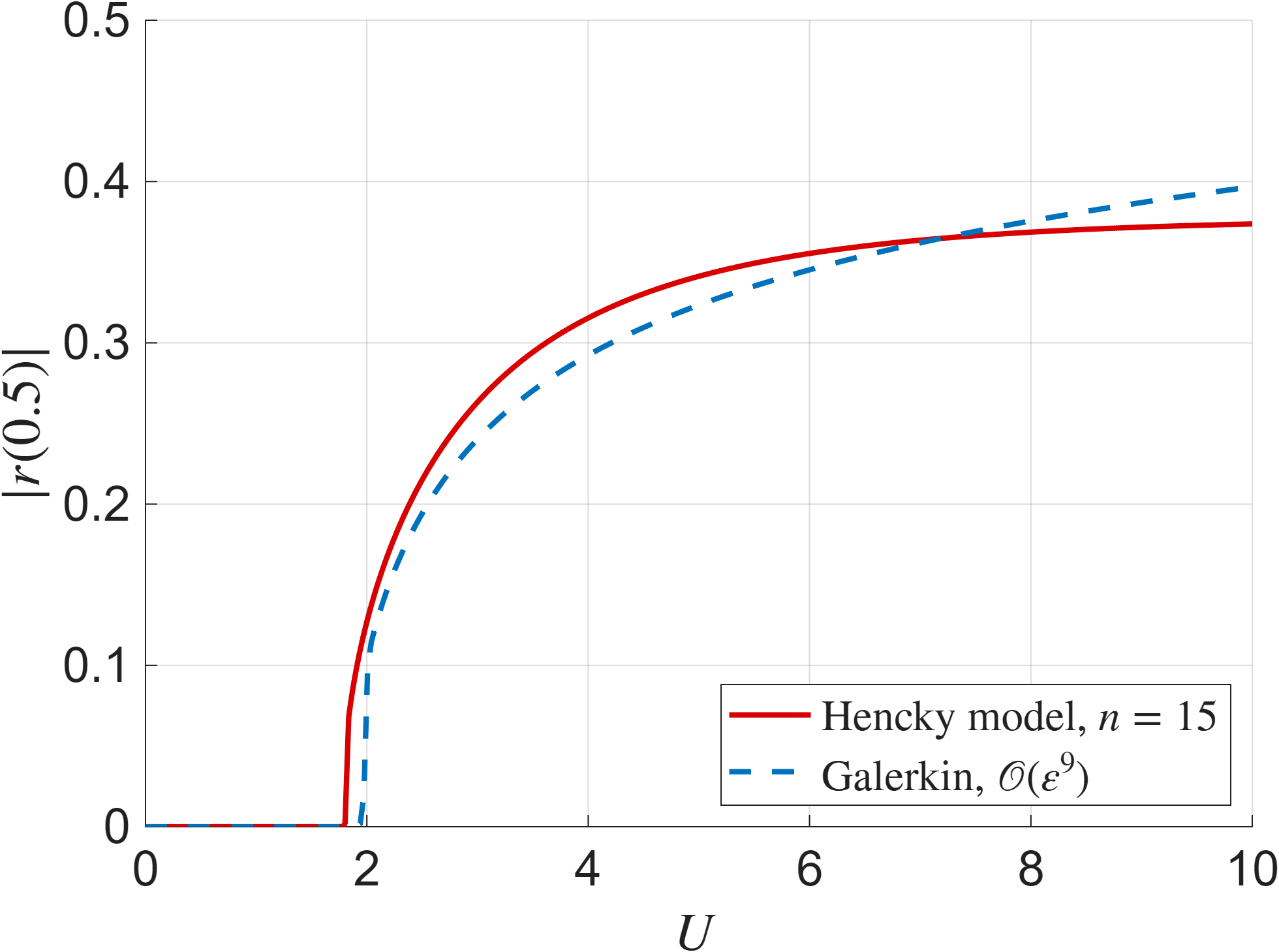}
\caption{Steady-state midpoint deflection $|r(0.5)|$ versus flow velocity at $\Omega=8$ for the pinned--roller pipe: modified Hencky model ($n=15$) vs the Galerkin $\mathcal{O}(\varepsilon^{9})$ reference. Parameters as in Fig.~\ref{fig:comp-time}.}
\label{fig:hencky-pr-vs-galerkin}
\end{figure}

The central validation is the direct comparison of the modified Hencky model with the Galerkin $\mathcal{O}(\varepsilon^{9})$ reference at combined flow and rotation. Figure~\ref{fig:hencky-pr-vs-galerkin} overlays the two predictions at $\Omega=8$. The critical flow velocity is reproduced and the two post-critical curves are in close agreement throughout the sweep, with only small differences in the post-critical amplitudes. This comparison carries two pieces of evidence at once: the ninth-order Taylor truncation of the bending curvatures, established in Section~\ref{sec:results} as the lowest order needed to resolve the post-critical amplitude, is confirmed against an exact-trigonometry reference -- the modified Hencky model uses the closed-form kinematics with no truncation; and $n=15$ is shown to be adequate for the pinned--roller case under combined loading.

\begin{figure}[htbp]
\centering
\begin{tabular}{cc}
\includegraphics[width=0.45\textwidth]{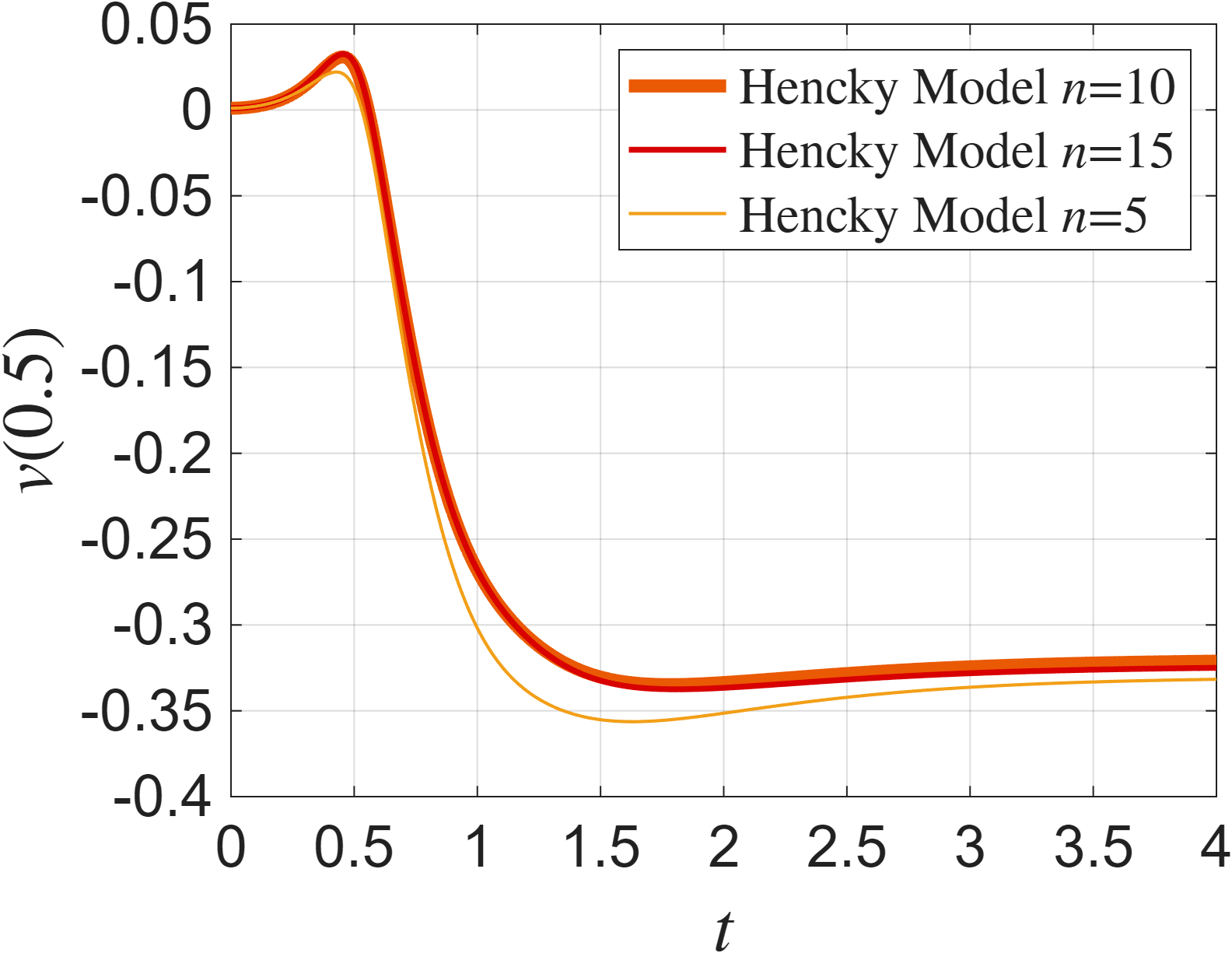} &
\includegraphics[width=0.45\textwidth]{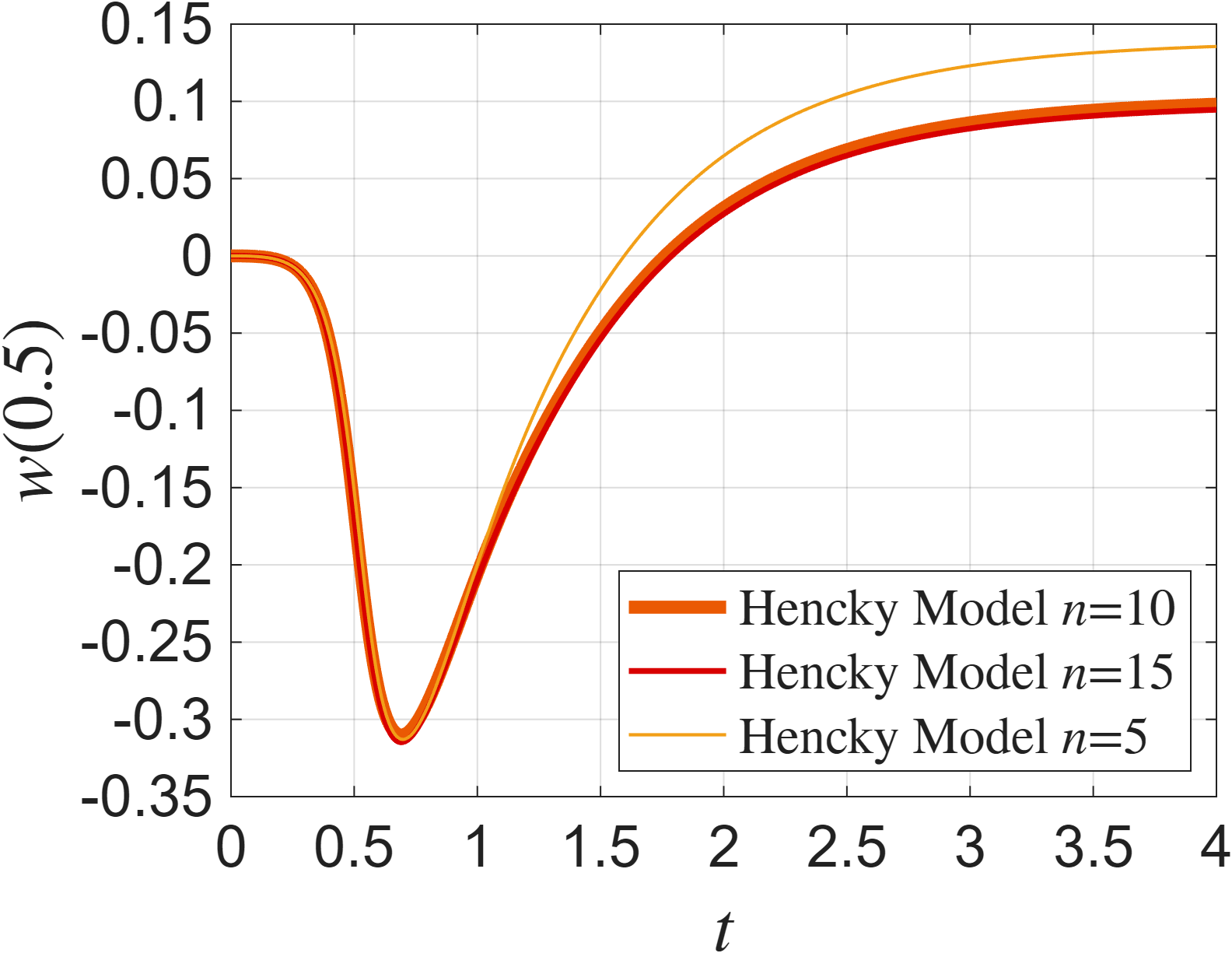} \\
(a) $v(0.5,t)$ & (b) $w(0.5,t)$ \\[6pt]
\multicolumn{2}{c}{\includegraphics[width=0.5\textwidth]{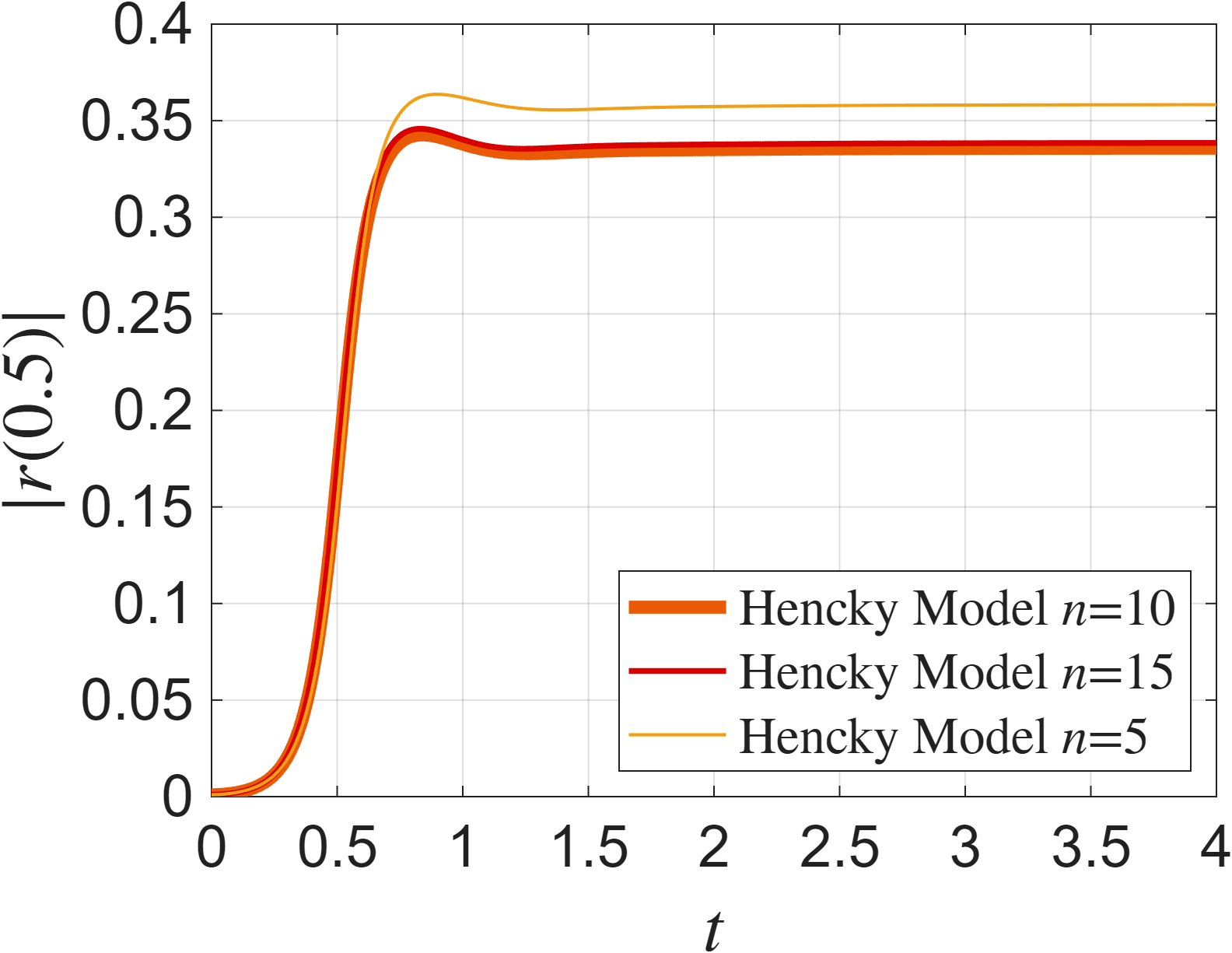}} \\
\multicolumn{2}{c}{(c) $|r(0.5,t)|$}
\end{tabular}
\caption{Time history of the midpoint response under combined flow and rotation for the pinned--roller pipe, modified Hencky model at $n=5,10,15$: (a)~$v(0.5,t)$; (b)~$w(0.5,t)$; (c)~total deflection magnitude $|r(0.5,t)|$. $U=6$, $\Omega=3$. Parameters as in Fig.~\ref{fig:comp-time}.}
\label{fig:hencky-pr-time-history}
\end{figure}

As for the pinned--pinned case, the transient behaviour is checked through a time-history simulation at combined loading. Figure~\ref{fig:hencky-pr-time-history} shows the midpoint response at $U=6$, $\Omega=3$ for the modified Hencky model at $n=5, 10, 15$. The total deflection magnitude $|r(0.5,t)|$ converges with $n$, and the $n=10$ and $n=15$ results are in close agreement; the individual components $v(0.5,t)$ and $w(0.5,t)$ show small differences for the coarsest discretisation but settle by $n=10$. The transient grows from the initial perturbation and reaches a steady whirling regime consistent with the bifurcation predictions above. Together with the bifurcation results, this confirms that the modified Hencky model captures both the steady-state and the transient post-critical response of the pinned--roller pipe, verifying the Galerkin $\mathcal{O}(\varepsilon^{9})$ predictions of Section~\ref{sec:results}. With both the extensible and the inextensible configurations now covered, the modified Hencky model is established as a reliable independent discretisation more generally.

\section{Conclusions}\label{sec:conclusions}

In this paper, the nonlinear dynamics and stability boundaries of an inextensible pinned--roller pipe conveying fluid and spinning about its longitudinal axis were investigated. The equations of motion were derived using a Rayleigh-beam model with Kelvin--Voigt damping in a curvilinear body-attached frame and were discretised by the Galerkin method. The results were then verified by an independent discretisation: a modified Hencky bar-chain model in which each link's orientation is expressed directly in the rotating reference frame.

The linearised dynamics, common to the extensible and inextensible formulations, gave an ellipse-like stability boundary in the flow-velocity--rotational-speed plane with semi-axes $U=\pi$ and $\Omega=\pi^{2}$. Three damping regimes were recovered: the undamped boundary at $\alpha=0$, the damped stable region at low $\Omega$, and the unstable region at high $\Omega$ where the rotating-damping cross-coupling overcomes the structural damping. The boundary depends on the mass ratio $\beta$ and on the flow-profile modification factor $\gamma$; a swirl-number argument based on the slender-pipe scaling showed that pipe rotation in the stable region is too weak to change the flow regime, so $\gamma$ stays fixed. The frequency analysis also revealed that only the forward-whirling mode loses stability across the unstable region.

In the post-critical regime, with the axial-stretching nonlinearity of the extensible case removed by the sliding support, the nonlinear vector splits into five operator groups arising from the inertial, gyroscopic, centripetal, stiffness, and damping contributions. Switching off each group at a representative post-critical point showed that the bending-curvature stiffness group dominates the equilibrium amplitude, with a moderate contribution from the centripetal group; the velocity-dependent groups affect only the transient approach to equilibrium. The bending curvatures themselves were found to require Taylor truncation to ninth order to resolve the post-critical amplitude -- the third-order truncation overestimates the deflection significantly because it misses the geometric stiffening from the inextensibility factor $(1-v'^{2}-w'^{2})^{-1/2}$. The post-critical surface mapped across the $(U,\Omega)$ plane rises smoothly from the stability boundary, with deflection magnitudes larger by more than an order of magnitude than the extensible pinned--pinned counterpart.

The modified Hencky bar-chain model provides a general-purpose discrete framework for spinning fluid-conveying pipes: a closed, $n$-independent matrix formulation with exact trigonometric kinematics, directly implementable in any standard programming environment with matrix routines and adaptable to a range of boundary conditions through matrix reductions. The global angular description reduces the symbolic complexity of the equations of motion by more than three orders of magnitude relative to the standard successive-rotation construction, bringing $n=15$, $n=30$, and beyond into routine reach. Application to the extensible pinned--pinned configuration and to the inextensible pinned--roller pipe (through the single substitution $\mu=0$) gives close agreement with the Galerkin reference across linear-stability, bifurcation, and time-history comparisons, mutually confirming the ninth-order Taylor truncation and establishing the modified Hencky bar-chain as a reliable general-purpose framework for the large-deflection regime of spinning fluid-conveying pipes.

\printcredits

\section*{Declaration of competing interest}
The authors declare that they have no known competing financial interests or personal relationships that could have appeared to influence the work reported in this paper.

\section*{Acknowledgments}
This work has been supported by the Polish National Science Center under grant PRELUDIUM~22  no. 2023/49/N/ST8/01957. For the purpose of Open Access, the authors have applied a CC-BY public copyright license to any Author Accepted Manuscript (AAM) version arising from this submission.

\section*{Data availability}
The MATLAB and Wolfram Mathematica source code, the precomputed
numerical data, and the rendered figure files used to produce every
computational figure of this manuscript are openly available in the
RepOD repository at \url{https://doi.org/10.18150/M8AU3O}.

\appendix
\section{Continuous-model details}\label{app:inext-details}

This appendix collects the explicit forms of (i)~the three coupled partial differential equations obtained from the Hamilton-principle variation of Section~\ref{subsec:eom} and (ii)~the five nonlinear operator groups appearing in the rotating-frame equations of motion Eqs.~\eqref{eq:eom-v-rotating}--\eqref{eq:eom-w-rotating}. The kinematics, energy formulation, and Hamiltonian variational procedure are identical to those of~\cite{fasihi_nonlinear_2026}; only the inextensibility constraint and its consequences are specific to the pinned--roller configuration treated here.

\subsection*{A.1 Three coupled PDEs prior to twist elimination}

Substituting the inextensibility constraint Eqs.~\eqref{eq:u-in-terms-of-vw} together with the third-order Taylor expansions of the kinematic quantities into the Hamiltonian variational principle and collecting the coefficients of $\delta\phi$, $\delta v$, $\delta w$ yields the following three coupled nonlinear partial differential equations:
\begin{align}
& I_x \Big(
       \ddot{v}'\,w'
       + \dot{v}'\,\dot{w}'
       + \ddot{\phi}
\Big) \notag\\
&{}+ D_x \Big(
       v'''\,w'
       + v''\,w''
       + \phi''
\Big) \notag\\
&{}+ C_x \Big(
       -\dot{v}'''\,w'
       -\dot{v}''\,w''
       -v'''\,\dot{w}'
       -v''\,\dot{w}''
       -\dot{\phi}''
\Big) = 0,
\label{eq:eom-phi-full}
\end{align}

\begin{align}
&m\Bigg( v''\!\int_{0}^{s}\!\!\int_{L}^{s}\!\bigl(\dot{v}'^{2}+v'\ddot{v}'+\dot{w}'^{2}+w'\ddot{w}'\bigr)\mathrm{d}s\,\mathrm{d}s \notag\\
&\qquad{}+ v'\!\int_{0}^{s}\!\bigl(\dot{v}'^{2}+v'\ddot{v}'+\dot{w}'^{2}+w'\ddot{w}'\bigr)\mathrm{d}s + \ddot{v}\Bigg) \notag\\
&{}+ M \Bigg(v''\!\int_{0}^{s}\!\!\int_{L}^{s}\!\bigl(\dot{v}'^{2}+v'\ddot{v}'+\dot{w}'^{2}+w'\ddot{w}'\bigr)\mathrm{d}s\,\mathrm{d}s \notag\\
&\qquad{}+ v'\!\int_{0}^{s}\!\bigl(\dot{v}'^{2}+v'\ddot{v}'+\dot{w}'^{2}+w'\ddot{w}'\bigr)\mathrm{d}s
+ 2U v''\!\int_{L}^{s}\!\bigl(v'\dot{v}'+w'\dot{w}'\bigr)\mathrm{d}s \notag\\
&\qquad{}+ 2U v'^{2}\dot{v}' + 2U v'w'\dot{w}'
+ U^{2} v''\!\int_{L}^{s}\!\bigl(v'v''+w'w''\bigr)\mathrm{d}s
+ U^{2} v'^{2} v'' \notag\\
&\qquad{}+ U^{2} v'w'w'' + \ddot{v} + 2U\dot{v}' + U^{2} v''\Bigg) \notag\\
&{}+ I_x \Big(
        -\ddot{v}''\,w'^{2}
        -2\dot{v}''\,w'\dot{w}'
        -2\ddot{v}'\,w'w''
        -2\dot{v}'\,\dot{w}'\,w''
        -2\dot{v}'\,w'\,\dot{w}'' \notag\\
&\qquad{}- \Omega\,\dot{w}''
        - \ddot{\phi}'\,w'
        - \dot{\phi}'\,\dot{w}'
        - \ddot{\phi}\,w''
        - \dot{\phi}\,\dot{w}''\Big) \notag\\
&{}+ I \Big(
      -v''\,\ddot{w}'\,w'
      - v''\,\dot{w}'^{2}
      - v'\,\ddot{w}''\,w'
      -2 v'\,\dot{w}''\,\dot{w}'
      - v'\,\ddot{w}'\,w''
      - \ddot{v}''\Big) \notag\\
&{}+ C_x \Big(
       \dot{v}''''\,w'^{2}
       +4\dot{v}'''\,w'w''
       +2\dot{v}''\,w''^{2}
       +2\dot{v}''\,w'w'''
       +w'''\,v''\,\dot{w}' \notag\\
&\qquad{}+2w''\,v'''\,\dot{w}'
       +2w''\,v''\,\dot{w}''
       +w'\,v''''\,\dot{w}'
       +2w'\,v'''\,\dot{w}''
       +w'\,v''\,\dot{w}''' \notag\\
&\qquad{}+w'''\,\dot{\phi}'
       +2w''\,\dot{\phi}''
       +w'\,\dot{\phi}'''\Big) \notag\\
&{}+ C \Big(
       \dot{v}''\,w''^{2}
       +\dot{v}''''\,v'^{2}
       +v''''\,v'\,\dot{v}'
       +\tfrac{1}{2}\Omega\,w''''\,w'^{2}
       +\dot{v}'\,w'\,w'''' \notag\\
&\qquad{}+\dot{v}''\,w'\,w'''
       +3\dot{v}''\,v''^{2}
       +3v'''\,v''\,\dot{v}'
       +3\dot{v}'\,w''\,w'''
       +3\dot{v}''\,v'''\,v' \notag\\
&\qquad{}+4\dot{v}'''\,v''\,v'
       +v'''\,\dot{w}'\,w''
       +4v''\,\dot{w}''\,w''
       +2v''\,\dot{w}'\,w'''
       +3v'\,\dot{w}'''\,w'' \notag\\
&\qquad{}+3v'\,\dot{w}''\,w'''
       +v'\,\dot{w}'\,w''''
       +v'''\,w'\,\dot{w}''
       +2v''\,w'\,\dot{w}'''
       +v'\,w'\,\dot{w}'''' \notag\\
&\qquad{}+\tfrac{1}{2}\Omega\,w''''\,v'^{2}
       +\Omega\,w'\,w''\,w'''
       +\Omega\,w'''\,v''\,v'
       +\Omega\,w''''
       +\dot{v}''''\Big) \notag\\
&{}+ D_x \Big(
        - v''''\,w'^{2}
        +4v'''\,w'w''
        +2v''\,w''^{2}
        +2v''\,w'w''' \notag\\
&\qquad{}+ w'''\,\phi'
        +2w''\,\phi''
        + w'\,\phi'''\Big) \notag\\
&{}+ D \Big(
       v''^{3}
       +4v'v''v'''
       +v''\,w''^{2}
       +v''\,w'\,w'''
       +v'^{2}\,v'''' \notag\\
&\qquad{}+3v'\,w''\,w'''
       +v'\,w'\,w''''
       +v''''\Big) = 0,
\label{eq:eom-v-full}
\end{align}

\begin{align}
&m \Bigg( w''\!\int_{0}^{s}\!\!\int_{L}^{s}\!\bigl(\dot{v}'^{2}+v'\ddot{v}'+\dot{w}'^{2}+w'\ddot{w}'\bigr)\mathrm{d}s\,\mathrm{d}s \notag\\
&\qquad{}+ w'\!\int_{0}^{s}\!\bigl(\dot{v}'^{2}+v'\ddot{v}'+\dot{w}'^{2}+w'\ddot{w}'\bigr)\mathrm{d}s + \ddot{w} \Bigg) \notag\\
&{}+ M \Bigg(w''\!\int_{0}^{s}\!\!\int_{L}^{s}\!\bigl(\dot{v}'^{2}+v'\ddot{v}'+\dot{w}'^{2}+w'\ddot{w}'\bigr)\mathrm{d}s\,\mathrm{d}s \notag\\
&\qquad{}+ w'\!\int_{0}^{s}\!\bigl(\dot{v}'^{2}+v'\ddot{v}'+\dot{w}'^{2}+w'\ddot{w}'\bigr)\mathrm{d}s
+ 2U w''\!\int_{L}^{s}\!\bigl(v'\dot{v}'+w'\dot{w}'\bigr)\mathrm{d}s \notag\\
&\qquad{}+ 2U w'^{2}\dot{w}' + 2U w'v'\dot{v}'
+ U^{2} w''\!\int_{L}^{s}\!\bigl(v'v''+w'w''\bigr)\mathrm{d}s
+ U^{2} w'^{2} w'' \notag\\
&\qquad{}+ U^{2} w'v'v'' + \ddot{w} + 2U\dot{w}' + U^{2} w''\Bigg) \notag\\
&{}+ I_x \Big(
    2\dot{v}'\,w'\,\dot{v}''
    + \dot{v}'^{2}\,w''
    + \Omega\,\dot{v}''
    + \dot{\phi}'\,\dot{v}'
    + \dot{\phi}\,\dot{v}''\Big) \notag\\
&{}+ I \Big(
      -\ddot{v}''\,v'\,w'
      -2\dot{v}''\,\dot{v}'\,w'
      - \ddot{v}'\,v''\,w'
      - \ddot{v}'\,v'\,w''
      - \dot{v}'^{2}\,w''
      - \ddot{w}''\Big) \notag\\
&{}+ D_x \Big(
       -2v''\,w'\,v'''
       - v''^{2}\,w''
       - v'''\,\phi'
       - v''\,\phi''\Big) \notag\\
&{}+ D \Big(
       v''^{2}\,w''
       + v'\,v'''\,w''
       + w''^{3}
       +4w'\,w''\,w'''
       +3v''\,w'\,v''' \notag\\
&\qquad{}+ v'\,v''''\,w'
       + w'^{2}\,w''''
       + w''''\Big) \notag\\
&{}+ C_x \Big(
       -\dot{v}'''\,w'\,v''
       -\dot{v}''\,w''\,v''
       -\dot{v}''\,w'\,v'''
       -2v''\,\dot{w}'\,v'''
       -v''^{2}\,\dot{w}'' \notag\\
&\qquad{}-v'''\,\dot{\phi}'
       -v''\,\dot{\phi}''\Big) \notag\\
&{}+ C \Big(
       -2\Omega\,v'\,w''\,w'''
       -\Omega\,v'''\,w''\,w'
       -3\Omega\,v'''\,v'\,v''
       +3w'''\,w''\,\dot{w}' \notag\\
&\qquad{}+3\dot{w}''\,w''^{2}
       +\dot{w}''''\,w'^{2}
       +\dot{v}'''\,v'\,w''
       +\dot{v}''\,v''\,w''
       +4\dot{w}'''\,w''\,w' \notag\\
&\qquad{}+3\dot{w}''\,w'''\,w'
       +\dot{v}''''\,v'\,w'
       +2\dot{v}'''\,v''\,w'
       +\dot{v}''\,v'''\,w'
       +w''''\,w'\,\dot{w}' \notag\\
&\qquad{}-\Omega\,v''^{3}
       -\Omega\,v''\,w''^{2}
       -\tfrac{1}{2}\Omega\,v''''\,w'^{2}
       -\tfrac{1}{2}\Omega\,v''''\,v'^{2}
       -\Omega\,v'''' \notag\\
&\qquad{}+\dot{w}''''\Big) = 0.
\label{eq:eom-w-full}
\end{align}

Here, $D_{x}=GJ$ is the torsional stiffness, $C_{x}=\eta_{x}J$ the torsional damping coefficient, and $I_{x}$ the polar mass moment of inertia per unit length of the cross-section. The transverse equations contain rotary-inertia and gyroscopic groups, multiplied by the cross-section moments of inertia $I$, $I_{x}$ and the torsional stiffness and damping coefficients $D_{x}$, $C_{x}$, that couple bending to the twist~$\phi$. Because of the slenderness of the pipe and the moderate spin rates considered here, these contributions are negligible compared with the bending and flow-induced terms; they vanish identically once the twist is eliminated quasi-statically through Eqs.~\eqref{eq:phi-quasistatic} and the rotary inertia is dropped. The reduced two-equation set in $v$ and $w$ that results -- after non-dimensionalisation -- is reported in Section~\ref{subsec:dimensionless} and re-expressed in the rotating frame in Eqs.~\eqref{eq:eom-v-rotating}--\eqref{eq:eom-w-rotating}.

\subsection*{A.2 Nonlinear operator groups}

The five nonlinear operator groups appearing in the decomposition Eqs.~\eqref{eq:N-v-inext-decomp}--\eqref{eq:N-w-inext-decomp} are defined as follows.

\begin{equation}
\begin{aligned}
\mathrm{NMT}_{v}^{(1)}&=v''\!\int_{0}^{s}\!\int_{1}^{s_{1}}\bigl(\dot{v}'^{2}+v'\ddot{v}'+\dot{w}'^{2}+w'\ddot{w}'\bigr)\mathrm{d}s_{2}\,\mathrm{d}s_{1},\\
\mathrm{NMT}_{v}^{(2)}&=v'\!\int_{0}^{s}\bigl(\dot{v}'^{2}+v'\ddot{v}'+\dot{w}'^{2}+w'\ddot{w}'\bigr)\mathrm{d}s_{1},\\
\mathrm{NMT}_{w}^{(1)}&=w''\!\int_{0}^{s}\!\int_{1}^{s_{1}}\bigl(\dot{v}'^{2}+v'\ddot{v}'+\dot{w}'^{2}+w'\ddot{w}'\bigr)\mathrm{d}s_{2}\,\mathrm{d}s_{1},\\
\mathrm{NMT}_{w}^{(2)}&=w'\!\int_{0}^{s}\bigl(\dot{v}'^{2}+v'\ddot{v}'+\dot{w}'^{2}+w'\ddot{w}'\bigr)\mathrm{d}s_{1}.
\end{aligned}
\label{eq:NMT-app}
\end{equation}

\begin{equation}
\begin{aligned}
\mathrm{NGT}_{v}^{(1)}&=v'^{2}\dot{v}'+v'w'\dot{w}',\qquad
\mathrm{NGT}_{v}^{(2)}=v''\!\int_{1}^{s}\bigl(v'\dot{v}'+w'\dot{w}'\bigr)\mathrm{d}s_{1},\\
\mathrm{NGT}_{w}^{(1)}&=w'^{2}\dot{w}'+w'v'\dot{v}',\qquad
\mathrm{NGT}_{w}^{(2)}=w''\!\int_{1}^{s}\bigl(v'\dot{v}'+w'\dot{w}'\bigr)\mathrm{d}s_{1}.
\end{aligned}
\label{eq:NGT-app}
\end{equation}

\begin{equation}
\begin{aligned}
\mathrm{NCT}_{v}^{(1)}&=v'^{2}v''+v'w'w'',\qquad
\mathrm{NCT}_{v}^{(2)}=v''\!\int_{1}^{s}\bigl(v'v''+w'w''\bigr)\mathrm{d}s_{1},\\
\mathrm{NCT}_{w}^{(1)}&=w'^{2}w''+w'v'v'',\qquad
\mathrm{NCT}_{w}^{(2)}=w''\!\int_{1}^{s}\bigl(v'v''+w'w''\bigr)\mathrm{d}s_{1}.
\end{aligned}
\label{eq:NCT-app}
\end{equation}

\begin{equation}
\begin{aligned}
\mathrm{NST}_{v}&=v''^{3}+v''w''^{2}+v'^{2}v''''+v''w'w'''+v'w'w''''+4v'v''v'''+3v'w''w''',\\
\mathrm{NST}_{w}&=w''^{3}+w''v''^{2}+w'^{2}w''''+w''v'v'''+w'v'v''''+4w'w''w'''+3w'v''v'''.
\end{aligned}
\label{eq:NST-app}
\end{equation}
Higher-order forms obtained from the closed-form curvatures Eqs.~\eqref{eq:eta-exact} are given in Section~\ref{subsec:taylor-truncation}.

\begin{equation}
\begin{aligned}
\mathrm{NDT}_{v}&=v''''v'\dot{v}'+\tfrac{1}{2}\Omega w''''w'^{2}+\dot{v}'w'w''''+\dot{v}''w'w'''
+3\dot{v}''v''^{2}+3v'''v''\dot{v}'\\
&\quad+3\dot{v}'w''w'''+3\dot{v}''v'''v'+4\dot{v}'''v''v'+v'''\dot{w}'w''+4v''\dot{w}''w''+2v''\dot{w}'w'''\\
&\quad+3v'\dot{w}'''w''+3v'\dot{w}''w'''+v'\dot{w}'w''''+v'''w'\dot{w}''+2v''w'\dot{w}'''+v'w'\dot{w}''''\\
&\quad+\tfrac{1}{2}\Omega w''''v'^{2}+\Omega w'w''w'''+\Omega w'''v''v'+\dot{v}''w''^{2}+\dot{v}''''v'^{2},\\[3pt]
\mathrm{NDT}_{w}&=-2\Omega v'w''w'''+3w'''w''\dot{w}'+3\dot{w}''w''^{2}+\dot{w}''''w'^{2}\\
&\quad+\dot{v}'''v'w''+\dot{v}''v''w''+4\dot{w}'''w''w'+3\dot{w}''w'''w'\\
&\quad+\dot{v}''''v'w'+2\dot{v}'''v''w'+\dot{v}''v'''w'+w''''w'\dot{w}'\\
&\quad-\Omega v''^{3}-\Omega v''w''^{2}-\tfrac{1}{2}\Omega v''''w'^{2}-\tfrac{1}{2}\Omega v''''v'^{2}\\
&\quad-\Omega v'''w''w'-3\Omega v'''v'v''.
\end{aligned}
\label{eq:NDT-app}
\end{equation}

The two superscripts $(1)$ and $(2)$ of NMT, NGT, NCT distinguish between the local-product contribution and the integral contribution that arises from the inextensibility constraint when expressing the axial displacement~$u$ in terms of the transverse fields. The Galerkin projection of these operators is given in Eqs.~\eqref{eq:nonlinear-projections}.

\section{Detailed expressions of the modified Hencky matrices}\label{app:hencky}

This appendix collects the closed forms of the building-block matrices that enter the modified Hencky model of Section~\ref{sec:hencky}, together with the configuration-dependent expressions of the spin-related contributions $\mathbf{N}_{T\Omega}$, $\mathbf{f}_{T\Omega^{2}}$, $\mathbf{f}_{TU\Omega}$, and $\mathbf{f}_{OU\Omega}$ that are specific to the spinning case. Only the entries needed in the linearised analysis Eqs.~\eqref{eq:hencky-M0}--\eqref{eq:hencky-K0} are reproduced here for completeness. Throughout, $\mathbf{1}=[1,1,\ldots,1]^{\mathsf{T}}$ denotes the $(n+1)$-dimensional vector of ones, and ``$\circ$'' denotes the Hadamard (element-wise) product.

\subsection*{B.1 Mass-distribution matrix $\mathbf{D}$}

The symmetric $(n+1)\times(n+1)$ mass-distribution matrix has entries
\begin{equation}
\begin{aligned}
D_{1,1}&=\frac{3(n-1)+2}{12n^{3}},\quad
D_{n+1,n+1}=\frac{1}{24n^{3}},\quad
D_{i,i}=\frac{6(n+1)-6i-1}{12n^{3}},\quad i=2,\ldots,n,\\
D_{1,j}&=D_{j,1}=\frac{n+1-j}{2n^{3}},\quad j=2,\ldots,n,\quad
D_{1,n+1}=D_{n+1,1}=\frac{1}{16n^{3}},\\
D_{i,j}&=D_{j,i}=\frac{n+1-j}{n^{3}},\quad 2\le i<j\le n,\quad
D_{i,n+1}=D_{n+1,i}=\frac{1}{8n^{3}},\quad i=2,\ldots,n.
\end{aligned}
\label{eq:hencky-D}
\end{equation}

\subsection*{B.2 Discrete second-derivative matrix $\mathbf{K}$}

The tridiagonal $(n+1)\times(n+1)$ matrix
\begin{equation}
\mathbf{K}=n\,\begin{bmatrix}
1&-1&0&\cdots&0&0\\
-1&2&-1&\cdots&0&0\\
0&-1&2&\cdots&0&0\\
\vdots&\vdots&\vdots&\ddots&\vdots&\vdots\\
0&0&0&\cdots&2&-1\\
0&0&0&\cdots&-1&1
\end{bmatrix}
\label{eq:hencky-K}
\end{equation}
is the discrete counterpart of the continuum bending operator $\partial^{2}/\partial s^{2}$, evaluated through Eqs.~\eqref{eq:hencky-s} and its linearisation. Its entries are $K_{1,1}=K_{n+1,n+1}=n$, $K_{i,i}=2n$ for $i=2,\ldots,n$, and $K_{i,i+1}=K_{i+1,i}=-n$.

\subsection*{B.3 Centripetal projection matrix $\mathbf{L}$}

The $(n+1)\times(n+1)$ matrix $\mathbf{L}$ associated with the open-system centrifugal stiffness has entries
\begin{equation}
\begin{aligned}
L_{1,1}&=\tfrac{1}{2n},\quad L_{1,n+1}=-\tfrac{1}{2n},\\
L_{i,i}&=\tfrac{1}{n},\quad L_{i,n+1}=-\tfrac{1}{n},\quad i=2,\ldots,n,
\end{aligned}
\label{eq:hencky-L}
\end{equation}
with all other entries zero.

\subsection*{B.4 Link-length-product matrices $\mathbf{B}_s$, $\mathbf{B}_a$, vector $\mathbf{b}$}

Define $\mathbf{b}=(1/n)[1/2,1,1,\ldots,1,1/2]^{\mathsf{T}}$. The symmetric and antisymmetric link-length-product matrices are
\begin{equation}
\mathbf{B}_s=\mathbf{b}\,\mathbf{b}^{\mathsf{T}},\qquad
B_{a,ij}=\begin{cases}b_i b_j\,\mathrm{sgn}(j-i)& i\neq j,\\0& i=j.\end{cases}
\label{eq:hencky-Bs-Ba}
\end{equation}

\subsection*{B.5 Trigonometric building-block matrices}

The configuration-dependent matrices $\mathbf{S}_\theta$ and $\mathbf{C}_\theta$ have entries
\begin{equation}
S_{\theta,ij}=\sin(\theta_{i-1}-\theta_{j-1}),\qquad
C_{\theta,ij}=\cos(\theta_{i-1}-\theta_{j-1}),\quad i,j=1,\ldots,n+1.
\label{eq:hencky-Stheta-Ctheta}
\end{equation}
$\mathbf{C}_\theta$ is symmetric; $\mathbf{S}_\theta$ is antisymmetric. The trigonometric vectors $\mathbf{s}_\theta,\mathbf{c}_\theta,\mathbf{s}_\varphi,\mathbf{c}_\varphi$ collect the sines and cosines of the generalised coordinates,
\begin{equation}
\mathbf{s}_\theta=[\sin\theta_0,\ldots,\sin\theta_n]^{\mathsf{T}},\quad
\mathbf{c}_\theta=[\cos\theta_0,\ldots,\cos\theta_n]^{\mathsf{T}},
\label{eq:hencky-trig-vectors}
\end{equation}
with analogous definitions for $\mathbf{s}_\varphi$, $\mathbf{c}_\varphi$.

\subsection*{B.6 Spin-related contributions to $\mathbf{f}_T$ and $\mathbf{f}_O$}

The three spin-related contributions to the inertial vector Eqs.~\eqref{eq:hencky-fT} are
\begin{equation}
\mathbf{N}_{T\Omega}=2\begin{bmatrix}\mathbf{D}&\mathbf{D}\\\mathbf{D}&\mathbf{D}\end{bmatrix}\!\circ\!
\begin{bmatrix}
\mathbf{0} & -(\mathbf{c}_\theta\!\circ\!\mathbf{c}_\varphi)\,\mathbf{c}_\varphi^{\mathsf{T}}\\[2pt]
\bigl((\mathbf{c}_\theta\!\circ\!\mathbf{c}_\varphi)\,\mathbf{c}_\varphi^{\mathsf{T}}\bigr)^{\mathsf{T}} &
(\mathbf{s}_\theta\!\circ\!\mathbf{s}_\varphi)\,\mathbf{c}_\varphi^{\mathsf{T}}-\mathbf{c}_\varphi(\mathbf{s}_\theta\!\circ\!\mathbf{s}_\varphi)^{\mathsf{T}}
\end{bmatrix},
\label{eq:hencky-NTOmega-app}
\end{equation}
\begin{equation}
\mathbf{f}_{T\Omega^{2}}=\left(\begin{bmatrix}\mathbf{D}\\\mathbf{D}\end{bmatrix}\circ
\begin{bmatrix}
-(\mathbf{c}_\theta\mathbf{s}_\theta^{\mathsf{T}})\!\circ\!(\mathbf{c}_\varphi\mathbf{c}_\varphi^{\mathsf{T}})\\[2pt]
(\mathbf{s}_\theta\mathbf{s}_\theta^{\mathsf{T}})\!\circ\!(\mathbf{s}_\varphi\mathbf{c}_\varphi^{\mathsf{T}})-(\mathbf{c}_\varphi\mathbf{s}_\varphi^{\mathsf{T}})
\end{bmatrix}\right)\mathbf{1},
\label{eq:hencky-fTOmega2-app}
\end{equation}
\begin{equation}
\mathbf{f}_{TU\Omega}=\left(\begin{bmatrix}\mathbf{B}_a\\\mathbf{B}_a\end{bmatrix}\circ
\begin{bmatrix}
-(\mathbf{c}_\theta\!\circ\!\mathbf{c}_\varphi)\,\mathbf{s}_\varphi^{\mathsf{T}}\\[2pt]
\mathbf{c}_\varphi\,(\mathbf{s}_\theta\!\circ\!\mathbf{c}_\varphi)^{\mathsf{T}}+(\mathbf{s}_\theta\!\circ\!\mathbf{s}_\varphi)\,\mathbf{s}_\varphi^{\mathsf{T}}
\end{bmatrix}\right)\mathbf{1}.
\label{eq:hencky-fTUOmega-app}
\end{equation}
The corresponding spin-related contribution to the open-system vector is
\begin{equation}
\mathbf{f}_{OU\Omega}=\left(\begin{bmatrix}\mathbf{B}_s\\\mathbf{B}_s\end{bmatrix}\circ
\begin{bmatrix}
(\mathbf{c}_\theta\!\circ\!\mathbf{c}_\varphi)\,\mathbf{s}_\varphi^{\mathsf{T}}\\[2pt]
-\mathbf{c}_\varphi(\mathbf{s}_\theta\!\circ\!\mathbf{c}_\varphi)^{\mathsf{T}}-(\mathbf{s}_\theta\!\circ\!\mathbf{s}_\varphi)\,\mathbf{s}_\varphi^{\mathsf{T}}
\end{bmatrix}\right)\mathbf{1}.
\label{eq:hencky-fOUOmega-app}
\end{equation}
At equilibrium $\mathbf{q}=\mathbf{0}$, $\mathbf{N}_{T\Omega}$ reduces to the off-diagonal block $-2\Omega(\mathbf{D}\,\boldsymbol{\sigma})$ shown in Eqs.~\eqref{eq:hencky-C0}, where $\boldsymbol{\sigma}$ is the planar-symplectic permutation between the $\theta$- and $\varphi$-blocks; $\mathbf{f}_{T\Omega^{2}}$ contributes only the $-\Omega^{2}\mathbf{D}$ diagonal terms of Eqs.~\eqref{eq:hencky-K0}; and $\mathbf{f}_{TU\Omega}+\mathbf{f}_{OU\Omega}$ contributes the $\sqrt{\beta}\,U\Omega(\mathbf{B}_a+\mathbf{B}_s)$ off-diagonal block of Eqs.~\eqref{eq:hencky-K0}.

\subsection*{B.7 Element-wise flexural nonlinearities $\mathbf{f}_{VF}$ and $\mathbf{f}_{R\alpha}$}

The flexural-spring restoring vector $\mathbf{f}_{VF}$ and the flexural-damping vector $\mathbf{f}_{R\alpha}$ are partitioned by coordinate type, $\mathbf{f}_{VF}=[\mathbf{f}_{VF\theta}^{\mathsf{T}},\mathbf{f}_{VF\varphi}^{\mathsf{T}}]^{\mathsf{T}}$ and $\mathbf{f}_{R\alpha}=[\mathbf{f}_{R\alpha\theta}^{\mathsf{T}},\mathbf{f}_{R\alpha\varphi}^{\mathsf{T}}]^{\mathsf{T}}$. The element-wise expressions, with the conventions $\theta_{-1}=\theta_0$, $\varphi_{-1}=\varphi_0$, $\theta_{n+1}=\theta_n$, $\varphi_{n+1}=\varphi_n$ and analogous for time derivatives, are
\begin{equation}
\begin{aligned}
f_{VF\theta,i}&=\tfrac{n}{2}\cos\varphi_{i-1}\bigl\{
2\bigl[\cos\theta_{i-2}\cos\varphi_{i-2}+\cos\theta_i\cos\varphi_i\bigr]\sin\theta_{i-1}\\
&\qquad-2\cos\theta_{i-1}\bigl[\cos\varphi_{i-2}\sin\theta_{i-2}+\cos\varphi_i\sin\theta_i\bigr]\bigr\},\\[3pt]
f_{VF\varphi,i}&=n\bigl\{
\bigl[\cos\theta_{i-2}\cos\theta_{i-1}\cos\varphi_{i-2}+\cos\theta_{i-1}\cos\theta_i\cos\varphi_i\\
&\qquad+\sin\theta_{i-1}(\cos\varphi_{i-2}\sin\theta_{i-2}+\cos\varphi_i\sin\theta_i)\bigr]\sin\varphi_{i-1}\\
&\qquad-\cos\varphi_{i-1}(\sin\varphi_{i-2}+\sin\varphi_i)\bigr\},
\end{aligned}
\label{eq:hencky-fVF-elementwise}
\end{equation}
for $i=1,\ldots,n$. The corresponding flexural-damping entries $f_{R\alpha\theta,i}$ and $f_{R\alpha\varphi,i}$ have an analogous structure, with each pair of trigonometric factors replaced by the corresponding bilinear product of the link orientations and angular velocities $\dot{\theta}_j$, $\dot{\varphi}_j$. The full element-wise expressions are obtained directly by differentiating the Rayleigh dissipation function, Eqs.~\eqref{eq:hencky-R}, and are not reproduced here for brevity.

\subsection*{B.8 Right-end support contributions $\mathbf{f}_{Vk_s}$, $\mathbf{f}_{V\mu}$, $\mathbf{C}_{c_s}$, $\mathbf{C}_{\alpha\mu}$}

The vectors and matrices associated with the transverse and axial support spring/damper at the right end take the closed forms:
\begin{equation}
\mathbf{f}_{Vk_s}=\left(\begin{bmatrix}\mathbf{B}_s\\\mathbf{B}_s\end{bmatrix}\circ
\begin{bmatrix}
(\mathbf{c}_\theta\!\circ\!\mathbf{c}_\varphi)(\mathbf{s}_\theta\!\circ\!\mathbf{c}_\varphi)^{\mathsf{T}}\\[2pt]
-(\mathbf{s}_\theta\!\circ\!\mathbf{s}_\varphi)(\mathbf{s}_\theta\!\circ\!\mathbf{c}_\varphi)^{\mathsf{T}}+\mathbf{c}_\varphi\mathbf{s}_\varphi^{\mathsf{T}}
\end{bmatrix}\right)\mathbf{1},
\label{eq:hencky-fVks-app}
\end{equation}
\begin{equation}
\mathbf{f}_{V\mu}=\begin{bmatrix}\mathbf{b}\\\mathbf{b}\end{bmatrix}\!\circ\!
\begin{bmatrix}\mathbf{s}_\theta\!\circ\!\mathbf{c}_\varphi\\\mathbf{c}_\theta\!\circ\!\mathbf{s}_\varphi\end{bmatrix}
-\left(\begin{bmatrix}\mathbf{B}_s\\\mathbf{B}_s\end{bmatrix}\circ
\begin{bmatrix}
(\mathbf{s}_\theta\!\circ\!\mathbf{c}_\varphi)(\mathbf{c}_\theta\!\circ\!\mathbf{c}_\varphi)^{\mathsf{T}}\\[2pt]
(\mathbf{c}_\theta\!\circ\!\mathbf{s}_\varphi)(\mathbf{c}_\theta\!\circ\!\mathbf{c}_\varphi)^{\mathsf{T}}
\end{bmatrix}\right)\mathbf{1}.
\label{eq:hencky-fVmu-app}
\end{equation}
The damping matrices $\mathbf{C}_{c_s}$ and $\mathbf{C}_{\alpha\mu}$ have analogous structure with the same trigonometric building blocks and reduce at equilibrium to the diagonal block contributions $c_s\mathbf{B}_s$ and $\alpha\mu\mathbf{B}_s$ that appear in Eqs.~\eqref{eq:hencky-C0} (the latter dropping out of the pinned--roller configuration).


\end{document}